\documentclass[11pt]{article}
\pdfoutput=1 % if your are submitting a pdflatex (i.e. if you have
\usepackage{jcappub} % for details on the use of the package, please see the JCAP-author-manual
\usepackage{multirow}    
\usepackage{comment}
\usepackage{siunitx,booktabs}
                 
\usepackage[normalem]{ulem}                     
\newcommand{\GeV}{\text{GeV}}

\newcommand{\pc}{\text{pc}}
\newcommand{\kpc}{\text{kpc}}
\newcommand{\cm}{\text{cm}}
\newcommand{\kms}{{\text{km}}~{\text{s}}^{-1}}

\newcommand{\muas}{\mu \text{as}}
\newcommand{\muasy}{\mu \text{as}~\text{y}^{-1}}
\newcommand{\muasyy}{\mu \text{as}~\text{y}^{-2}}

\newcommand{\dd}{{\rm d}}
\newcommand{\betahat}{\hat{\vect{\beta}}}
\newcommand{\vhat}{\hat{\vect{v}}}

\definecolor{deepblue}{rgb}{0.2,0.2,0.8}

\definecolor{deepred}{rgb}{0.8,0.2,0.2}

\usepackage{bm}
\newcommand{\vect}[1]{\boldsymbol{\mathbf{#1}}}

\usepackage[T1]{fontenc} % if needed

\title{Halometry from Astrometry}

\author[a,b]{Ken Van Tilburg,}
\author[a]{Anna-Maria Taki,}
\author[a,c]{Neal Weiner}

\affiliation[a]{Center for Cosmology and Particle Physics, Department of Physics, New York University, New York, NY 10003, USA}
\affiliation[b]{School of Natural Sciences, Institute for Advanced Study, Princeton, NJ 08540, USA}
\affiliation[c]{Center for Computational Astrophysics, Flatiron Institute, New York, NY 10010, USA}
\date{\today}

\emailAdd{kenvt@nyu.edu}
\emailAdd{amt543@nyu.edu}
\emailAdd{neal.weiner@nyu.edu}

\abstract{
Halometry---mapping out the spectrum, location, and kinematics of nonluminous structures inside the Galactic halo---can be realized via variable weak gravitational lensing of the apparent motions of stars and other luminous background sources. Modern astrometric surveys provide unprecedented positional precision along with a leap in the number of cataloged objects. Astrometry thus offers a new and sensitive probe of collapsed dark matter structures over a wide mass range, from one millionth to several million solar masses. It opens up a window into the spectrum of primordial curvature fluctuations with comoving wavenumbers between $5~\text{Mpc}^{-1}$ and $10^5~\text{Mpc}^{-1}$, scales hitherto poorly constrained. We outline detection strategies based on three classes of observables---multi-blips, templates, and correlations---that take advantage of correlated effects in the motion of many background light sources that are produced through time-domain gravitational lensing. While existing techniques based on single-source observables such as outliers and mono-blips are best suited for point-like lens targets, our methods offer parametric improvements for extended lens targets such as dark matter subhalos. Multi-blip lensing events may also unveil the existence, location, and mass of planets in the outer reaches of the Solar System, where they would likely have escaped detection by direct imaging.  
}

\begin{document}
\maketitle

%\flushbottom
%%%%%%%%%%%%%%%%%%%%%%%%%%%%%%%%%%%%%%%%%%%%%%%%%%%%%%%%%%%%%%%%%%%

\section{Introduction}\label{sec:intro}
The nature of the one quarter of the Universe that is composed of dark matter has been an open and important question since it was first discovered. A wide range of proposed models has motivated and shaped direct searches for dark matter particles at colliders and small-scale experiments, as well as indirect searches in astrophysical and cosmological contexts. At present, there is unfortunately no robust indication for any nongravitational interactions of dark matter. Given the plethora of hypothetical sectors with cosmologically stable particles, general effects caused by the dark matter's irreducible gravitational coupling have become more central, as they may narrow down the list of plausible dark matter candidates.

The wealth of information on dark matter garnered from astronomical and cosmological observations---such as the number of light species, nucleosynthesis, galaxy morphology, cluster mergers, and more---have helped form an idea of what it can and cannot be. To date, all observations are consistent with dark matter being a cold, pressureless fluid without interactions (besides gravity) that permeates the entire Universe. Any sign of a departure from this behavior, e.g.~evidence for a self-interaction, for cored halos, or for ultra-compact clumps, would be a major step in pinning down a model that could explain the missing mass.

A large fraction of the body of evidence for dark matter comes from its density fluctuations and their evolution as a function of time. At the largest of observable scales, anisotropies in the cosmic microwave background point to a nearly scale-invariant spectrum of primordial curvature fluctuations, which are also imprinted onto the density fluctuations of a component more abundant than baryons or photons---the dark matter. According to the standard theory of structure formation, it is those tiny fluctuations that grew and eventually collapsed to form the seeds of large-scale structures and galaxies, including the Milky Way.

Many studies have focused on understanding the density structure of the Universe on smaller scales, which today would form a part of galaxies such as our own. These structures would reflect the properties of the dark matter at much smaller length scales, and also of its clustering at much earlier times, when sound waves of the photon-baryon fluid entered the horizon. They would be the end product of primordial fluctuations at small scales, produced at later times in inflationary theories. Unfortunately, extracting quantitative information at these small scales is plagued by myriad complications. Studies of Lyman-$\alpha$ clouds require sophisticated modeling of nonlinear systems. Small galaxies are highly sensitive to baryonic feedback, which can affect many of their properties. Halos on subgalactic scales would have collapsed at high redshifts, endowing them with a relatively high density that would have allowed them to survive the process of galaxy formation and their assembly into larger host halos. These subhalos are thought to have little ordinary matter, sidestepping baryonic feedback effects, but also making them---almost---invisible.

Gravitational lensing has long been an important tool to measure the mass and morphology of large objects such as galaxy clusters and massive galaxies, but lensing can also reveal interesting features about the constituents of galaxies, including the Milky Way halo. Photometric microlensing provided some of the first indications that dark matter was not made of baryonic constituents~\cite{alcock2000macho,tisserand2007limits}. 
Flux-ratio inconsistencies in the multiple images of strongly lensed quasi-stellar objects have been attributed to the substructure of the lenses~\cite{mao1998evidence, metcalf2001compound, chiba2002probing, dalal2002direct, metcalf2002flux, kochanek2004tests}, and can put the standard subhalo spectrum to the test~\cite{mao2004anomalous, amara2006simulations, maccio2006radial, maccio2006effect, xu2009effects, xu2010substructure}. Dark matter substructure could explain perturbations in the positions of lensed images~\cite{koopmans20022016+, chen2007astrometric, williams2008lensed, more2009role} as well as their relative time delays~\cite{keeton2009new, congdon2010identifying}. Lensing by extended extragalactic substructure has been considered in refs.~\cite{inoue2005three, inoue2005extended, koopmans2005gravitational, vegetti2009bayesian, vegetti2009statistics,vegetti2014density}. Strong gravitational lensing data can determine the density power spectrum inside galaxies, and thus pin down the spectrum of dark matter subhalos~\cite{hezaveh2016measuring, hezaveh2013dark, hezaveh2016detection}. 
Besides the above studies that rely on gravitational lensing by extragalactic substructure, other techniques have been proposed to detect substructure within the Milky Way: measuring Shapiro time delay effects in pulsar timing observations~\cite{siegel2007probing}, tracing gaps created by dark matter subhalos in stellar streams~\cite{erkal2015forensics, erkal2016number}, and monitoring the characteristic trails that a subhalo would imprint on the phase-space distribution of surrounding halo stars~\cite{buschmann2017stellar}. 
    
The rise of precision astrometry, led by the groundbreaking optical space-based observatory \textit{Gaia}~\cite{prusti2016gaia} and the continually improving long-baseline radio interferometers, raises the prospect that astrometric weak lensing might provide more clues about the nature of dark matter.
Previous works have pointed out the potential to detect dark compact objects using astrometric lensing of the strong~\cite{boden1998astrometric} and weak~\cite{dominik2000astrometric} kind. The \textit{Gaia} mission should have the capability to outperform photometric lensing searches, and to detect even a small fraction of the dark matter abundance if it consists of compact objects~\cite{belokurov2002astrometric}. Searches for dilute subhalos of dark matter are more difficult because their size suppresses the lensing deflection angle. Ref.~\cite{erickcek2011astrometric} concluded that subhalos with standard shapes and densities are unlikely to produce detectable astrometric lensing events of individual background sources, unless they are extremely cuspy at their centers, in which case they can give rise to observable signals akin to those produced by compact objects~\cite{li2012new}. Recent simulations of minihalos, however, disfavor such steep inner density profiles~\cite{delos2018ultracompact}. The question remains: are there techniques that do not rely on the dramatic lensing signatures from point-like objects and ultra-cuspy halos, but instead, can tease out the subtle lensing effects from more dilute subhalos?

In this paper, we introduce several methods that leverage the high statistics of current and planned astrometric surveys, potentially enabling the discovery of extended nonluminous structures inside the Milky Way. Our methods might even reveal the presence of massive planets far beyond the Kuiper Belt in our own Solar System, through the correlated lensing effect they produce as they transit past many background stars. In section~\ref{sec:summary}, we give an executive summary of the basic physical effects and a preview of one of our main results, as well as recommendations for the data products and observational strategies of ongoing and future astrometric missions. In section~\ref{sec:lenstargets}, we review the properties of the lens targets that we aim to detect. Section~\ref{sec:signal} introduces new classes of signal observables. Their noise contributions from instrumental as well as intrinsic origins are discussed in section~\ref{sec:background}. Section~\ref{sec:sensitivity} contains our sensitivity projections for the lens targets of section~\ref{sec:lenstargets}. We conclude with our future outlook in section~\ref{sec:conclusions}.
Readers primarily interested in the search for outer Solar System planets may skip many parts of the main text, and follow only sections~\ref{sec:summary}, \ref{sec:planets}, \ref{sec:outliers}, \ref{sec:instrument}, \ref{sec:planetnine}, and \ref{sec:conclusions} for a self-contained analysis.

\section{Summary}\label{sec:summary}

We will investigate how high-statistics, time-domain astrometry can yield insights into the structure of dark matter in the Milky Way. Before presenting our detailed study in later sections, we condense the basic intuition of the physics and our analysis methods here. 

\begin{figure}[t]
%\centering 
\includegraphics[width=.99\textwidth,origin=0,angle=0]{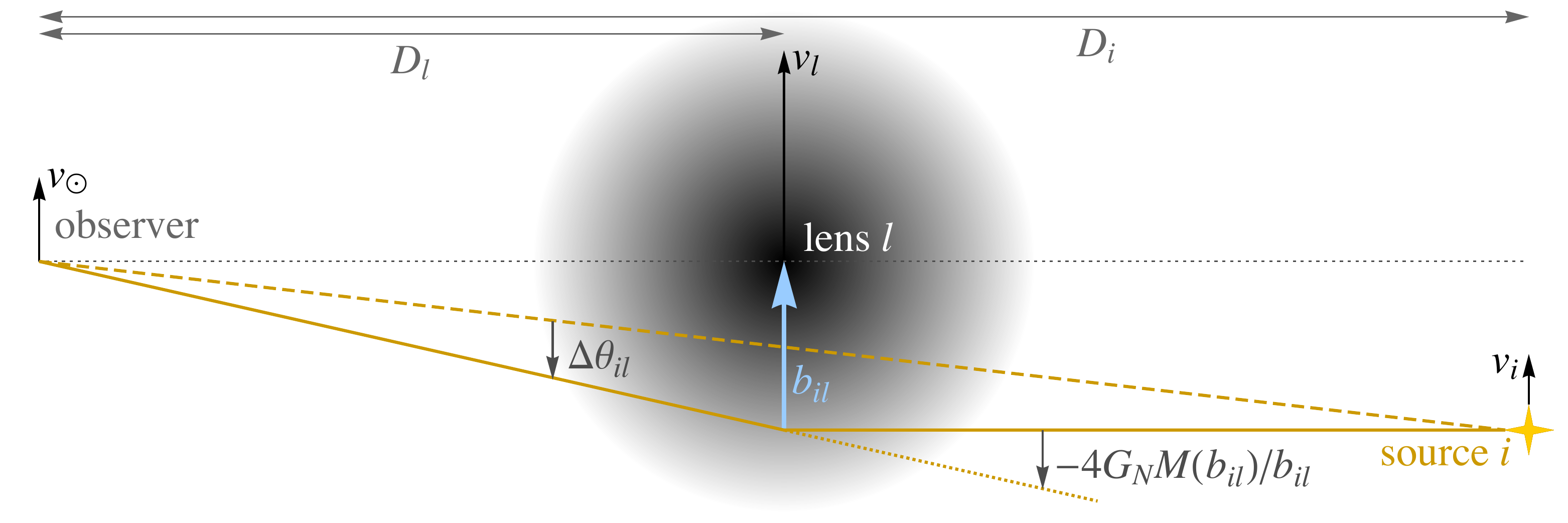}
\caption{Astrometric weak lensing geometry. A lens $l$ comprised of an extended, spherically symmetric mass distribution at a line-of-sight distance $D_l$ bends the light path (dark yellow) of a background source $i$ (yellow four-pointed star) at an impact parameter $\vect{b}_{il}$ (light blue) and distance $D_i$. The deflection angle $\Delta \vect{\theta}_{il}$ is proportional to $1/b_{il}$ as well as the mass $M(b_{il})$ enclosed inside the cylinder with radius equal to $b_{il}$. The lensing deflection evolves nonlinearly in time because the impact parameter $\vect{b}_{il}$ changes due to the peculiar velocities of the lens ($\vect{v}_l$), observer ($\vect{v}_\odot$), and source ($\vect{v}_i$). We ignore velocity components along the line of sight (that is, $D_l$ and $D_i$ are fixed), but it should be understood that $\Delta \vect{\theta}_{il}$, $\vect{b}_{il}$, $\vect{v}_l$, $\vect{v}_\odot$, and $\vect{v}_i$ are 2D vectors perpendicular to the line of sight.} \label{fig:lensing}
\end{figure}

\paragraph{Physical effects}
We portray in figure~\ref{fig:lensing} how the angular deflection vector $\Delta \vect{\theta}_{il}$ arises due to weak gravitational lensing of a background light source $i$ at a distance $D_i$ by a lens $l$ at a distance $D_l < D_i$. The impact parameter of the light path with the lens is defined as $\vect{b}_{il}$. If the lens has a spherically symmetric mass density distribution $\rho(r)$, then the effective mass that contributes to the lensing angle is the enclosed lens mass inside a cylinder parallel to the line of sight with radius equal to the impact parameter: $M(b_{il}) = 2\pi \int_{-\infty}^{+\infty} \dd x \int_{0}^{b_{il}} \dd b' \, b' \rho(\sqrt{x^2 + b^{\prime 2}})$. The angular deflection vector is:
\begin{align}
\Delta \vect{\theta}_{il} = - \left(1-\frac{D_l}{D_i}\right) \frac{4 G_N M(b_{il})}{b_{il}} \hat{\vect{b}}_{il}, \label{eq:lensangle}
\end{align}
where $G_N$ is Newton's gravitational constant.
As a point of reference, a characteristic impact parameter between a typical lens inside the Milky Way and a random background source is 10~kpc (roughly the size of the Galaxy itself); this then gives a typical lensing angle of
\begin{align}
\Delta \theta_{il}^\text{typical} \sim \frac{4G_N M(b_{il}^\text{typical})}{b_{il}^\text{typical}} \approx 4~\muas\left[\frac{M(b_{il}^\text{typical})}{10^6~M_\odot}\right]\left[\frac{10~\kpc}{b_{il}^\text{typical}}\right]. \label{eq:lensangletyp}
\end{align}
As the resolution of astrometric surveys is steadily driven down to the microarcsecond scale, and the effect of eq.~\ref{eq:lensangletyp} can be much larger for smaller-than-typical impact parameters, why have DM structures lighter than $10^7~M_\odot$ not been discovered yet through gravitational lensing?

The first obstacle is that the true angular position $\vect{\theta}_i$ of a background source (corresponding to the dashed line of figure~\ref{fig:lensing}) is not known a priori, so any lens correction $\Delta \vect{\theta}_{il}$, however large, is not observable for a single source. The lens distortion of eq.~\ref{eq:lensangle} reduces the angular number density of background sources behind the lens by a fractional amount of $\mathcal{O}(\Delta \theta_{il})$. However, noting that $1~\muas \approx 5 \times 10^{-12}$, the apparent density reduction is minuscule. Statistically, one can expect to see e.g.~a $10^{-10}$ fractional density fluctuation only for a population of $10^{20}$ or more background objects. The situation is even more hopeless in terms of systematics, as large density variations occur naturally all throughout the local Universe.

Once one enters the time domain, new possibilities arise. When the impact parameter changes by a large fractional amount over the course of a survey, i.e.~$\Delta b_{il} \sim b_{il}$, there can be nonrepeating anomalies in the apparent motion of a star~\cite{dominik2000astrometric,belokurov2002astrometric,erickcek2011astrometric,wyrzykowski2016black}. 
As objects in the Galactic halo move at speeds of $\mathcal{O}(10^{-3}c)$, $\Delta b_{il} \sim b_{il}$ requires $b_{il} \sim 10^{-3}~\pc$ for a survey of a few years. The closest expected approach between any one lens and any one star, out of $N_0$ stars at a characteristic distance $D_i$, is
\begin{equation}
\left\langle \min_{i,l} b_{il} \right\rangle \simeq \sqrt{\frac{M_l}{\pi \rho_l D_i N_0}} \approx 2\times 10^{-5}~\pc \sqrt{ \left[\frac{5~\kpc}{D_i} \right]\left[\frac{10^{-2} \rho_{\text{DM},\odot}}{\rho_l}\right] \left[\frac{M_{l}}{M_\odot }\right] \left[\frac{10^9}{N_0}\right] },	
\end{equation}
where $\rho_l$ is the energy density of these lenses with masses $M_l$. We take the local DM energy density to total $\rho_{\text{DM},\odot} \approx 0.4~\text{GeV}/\text{cm}^{3}$.  Close approaches are rare but do occur when the dark objects are copious in number.

When the dark objects are less numerous, such transits do not occur and $\Delta b_{il}/b_{il}$ is small. In this case, we can simplify our approach by focusing on the instantaneous \emph{time derivatives}, $\Delta \dot{\vect{\theta}}_{il}$ and $\Delta \ddot{\vect{\theta}}_{il}$, of the lensing deflection, due to the rate of change in impact parameter:
\begin{align}
\vect{v}_{il} \equiv \dot{\vect{b}}_{il} = \vect{v}_l - \left(1-\frac{D_l}{D_i}\right) \vect{v}_\odot - \frac{D_l}{D_i} \vect{v}_i,
\end{align}
where the velocities are defined as in figure~\ref{fig:lensing}. These derivatives will be suppressed by additional factors of $v_{il}/b_{il}$, which can be quite small, making the effects difficult to observe. Moreover, the ``fixed stars'' do of course have intrinsic proper motion noise both that is stochastic, such as velocity dispersion and acceleration in binary systems, and that is systematic, such as rotational velocities in large, gravitationally bound systems. However, all of the above noise contributions are suppressed by the large line-of-sight distance $D_i$, and the systematic ones can be modeled and subtracted.

A second obstacle is that the characteristic size $r_s$ of a lens suppresses lensing effects at small impact parameters. Let us assume a mass density profile
\begin{align}
\rho(r) = \frac{2^{3-\gamma} \rho_s}{\left(\frac{r}{r_s}\right)^\gamma \left(1 + \frac{r}{r_s} \right)^{3-\gamma}}
\end{align}
for $0 \le \gamma < 3$.
%$\rho(r) \propto r^{-3}$ or faster for $r>r_s$, and as $\rho(r) \propto r^{-n}$ with $n>0$ for $r<r_s$. 
Ignoring logarithmic factors, we get that the enclosed mass $M(b_{il})$ is roughly the total lens mass $M_s$ for $b_{il} > r_s$. For impact parameters smaller than the size of the lens ($b_{il} < r_s$), we have that $M(b_{il}) \sim M_s (b_{il}/r_s)^{2}$ for $0 \le \gamma \le 1$, and that $M(b_{il}) \sim M_s (b_{il}/r_s)^{3-\gamma}$ for $2 < \gamma < 3$. Even for the relatively cuspy NFW profile~\cite{navarro1997universal}, which has $\gamma=1$, this means that $M(b_{il})/b_{il}$ actually reaches a maximum at $b_{il} \sim r_s$, so that the lensing angle is suppressed for smaller impact parameters. Relative to the angular position on the sky, angular velocities and accelerations come with one and two ``powers'' of $\partial_t \sim \dot{b}_{il} \partial_{b_{il}} \sim b_{il}/v_{il}$, respectively.  The lens-induced corrections are parametrically equal to: 
\begin{align}
\Delta \dot{\theta}_{il} \sim \frac{4G_N M_s v_{il}}{b_{il}^2} \min\left[1,\left(\frac{b_{il}}{r_s}\right)^{\min\left[2,3-\gamma \right]}\right]; \quad
\Delta \ddot{\theta}_{il} \sim \frac{4G_N M_s v_{il}^2}{b_{il}^3} \min\left[1,\left(\frac{b_{il}}{r_s}\right)^{\min\left[2,3-\gamma \right]}\right]. \label{eq:sizeeffects}
\end{align}
The magnitudes of the observable effects at small $b_{il}$ for a finite-size lens are always suppressed by the factor of $(b_{il}/r_s)^{\min\left[2,3-\gamma \right]}$ relative to those of a point-like lens ($r_s = 0$) of the same mass.

To gain a sense of the scales involved, we can estimate the sizes of the effects for a point mass object, assuming a characteristic relative velocity between source and lens $v_{il} \sim 10^{-3}c$: 
\begin{align}
\Delta \dot{\theta}_{il} \sim 0.1~\muasy \left[\frac{M_l}{M_\odot}\right] %\left[\frac{v_{il}}{10^{-3} c}\right]
\left[\frac{10^{-2}~\pc}{b_{il} }\right]^2; \quad 
\Delta \ddot{\theta}_{il} \sim 4 \times 10^{-3}~\muasyy \left[\frac{M_l}{M_\odot}\right] %\left[\frac{v_{il}}{10^{-3} c}\right]^2
\left[\frac{10^{-2}~\pc}{b_{il}} \right]^3. \label{eq:sizeeffectsestpoint} 
\end{align}
Likewise, we can estimate that for an impact parameter $b_{il} \sim r_s$ with an NFW halo, we have:
\begin{align}
\Delta \dot{\theta}_{il} \sim 10^{-4}~\muasy \left[\frac{M_s}{10^7~M_\odot}\right] %\left[\frac{v_{il}}{10^{-3} c}\right]
\left[\frac{1~\kpc}{r_s }\right]^2; \quad 
\Delta \ddot{\theta}_{il} \sim 4 \times 10^{-11}~\muasyy \left[\frac{M_s}{10^7~M_\odot}\right]%\left[\frac{v_{il}}{10^{-3} c}\right]^2
\left[\frac{1~\kpc}{r_s} \right]^3. \label{eq:sizeeffectsesthalo}
\end{align}
As we shall see, with $\muas$ accuracy, there is hope to have sensitivity to point masses from observations of individual luminous sources. In contrast, the lens-induced signals from extended objects will be tiny, and we can only hope to detect them in \emph{aggregate}. Modern astrometric catalogs, with their enormous numbers of objects, will allow us to lever up the signal to noise considerably.

A second fact that will aid us comes from the rapid improvement in the measurement precision of the time-derivative quantities of eq.~\ref{eq:sizeeffects}. Astrometric surveys take time series data of the angular position of a star $i$; if any one measurement has a precision $\sigma_{\delta \theta,i}$ in both longitude and latitude, and measurements are made at a repetition rate of $f_\text{rep}$, then the instrumental precision after a mission time $\tau$ is:\footnote{The numerical prefactors come from a calculation of the expected noise of a parabolic least-squares fit to regularly sampled, independent observations. We ignored covariances and the need to fit for parallax, which will moderately change the numerical prefactors but not the parametric behavior of eq.~\ref{eq:noiseeffects}.}
\begin{align}
\sigma_{\mu,i} = \sqrt{192} \frac{\sigma_{\delta \theta,i}}{\tau \sqrt{f_\text{rep} \tau}} \approx 3.5~\muasy; \quad
\sigma_{\alpha,i} = \sqrt{720} \frac{\sigma_{\delta \theta,i}}{\tau^2 \sqrt{f_\text{rep} \tau}}\approx 1.4~\muasyy. \label{eq:noiseeffects}
\end{align}
In the above numerical estimates, we have plugged in numerical estimates for the precision on angular velocity ($\sigma_\mu$) and angular acceleration ($\sigma_\alpha$) for bright stars in \textit{Gaia}'s data set, after the nominal 5-year mission time.
Inspection of eqs.~\ref{eq:sizeeffects}~and~\ref{eq:noiseeffects} reveals that for every extra time derivative, one loses in the ratio of signal to instrumental noise by the factor of $v_{il} \tau / b_{il}$. Hence, in terms of instrumental precision only, angular-velocity observables are more sensitive than angular-acceleration ones. On the other hand, angular velocities generally have much larger intrinsic noise of both the stochastic (dispersion) and systematic (rotational velocities) kind, which should be added in quadrature with the instrumental noise, whereas intrinsic acceleration noise is typically below the instrumental precision. We shall see that there are applications for both angular velocities and accelerations.

\begin{figure}[ht]
%\centering 
\includegraphics[width=.325\textwidth,origin=0,angle=0]{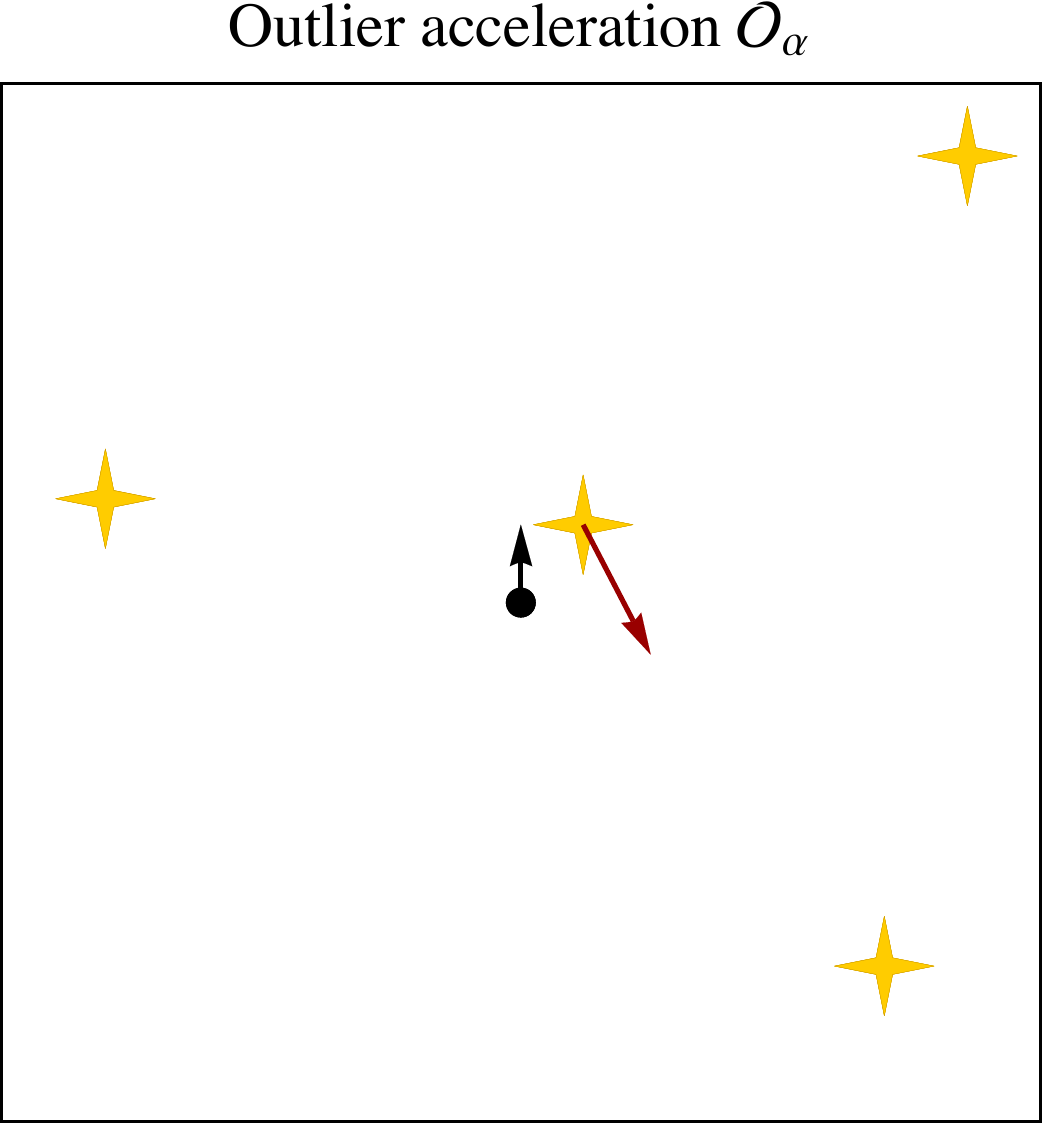}
\includegraphics[width=.325\textwidth,origin=0,angle=0]{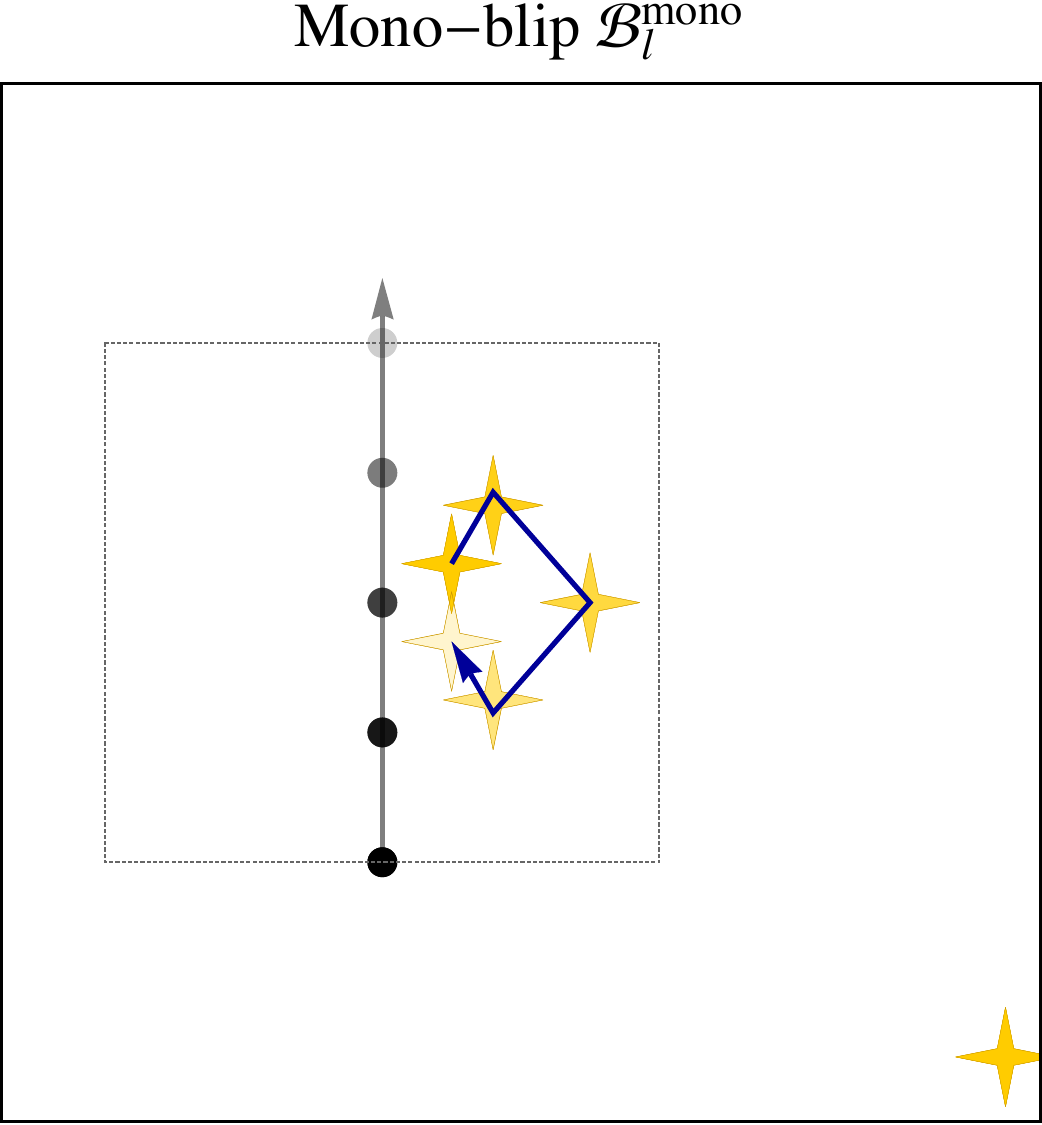}
\includegraphics[width=.325\textwidth,origin=0,angle=0]{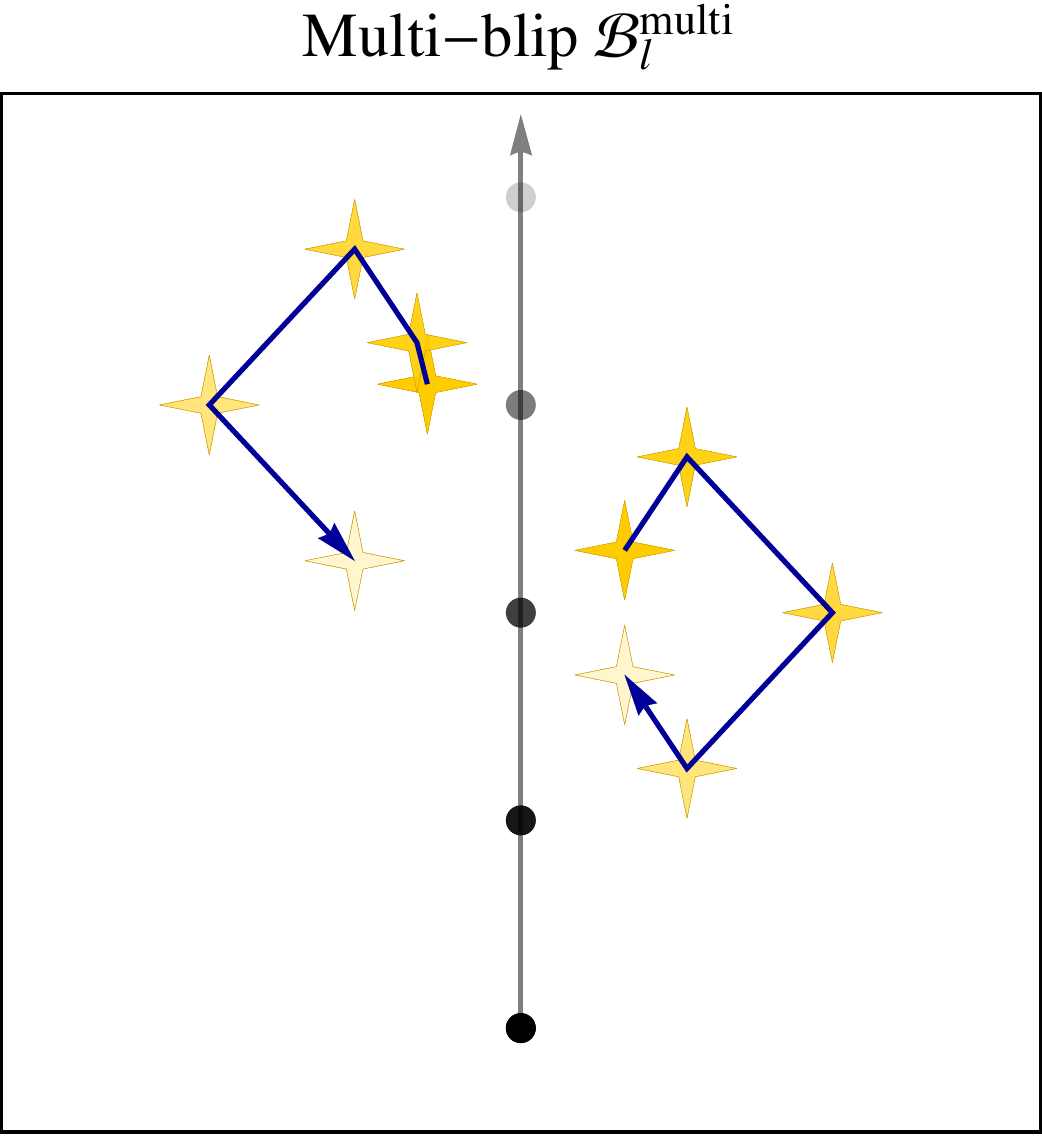}
\includegraphics[width=.325\textwidth,origin=0,angle=0]{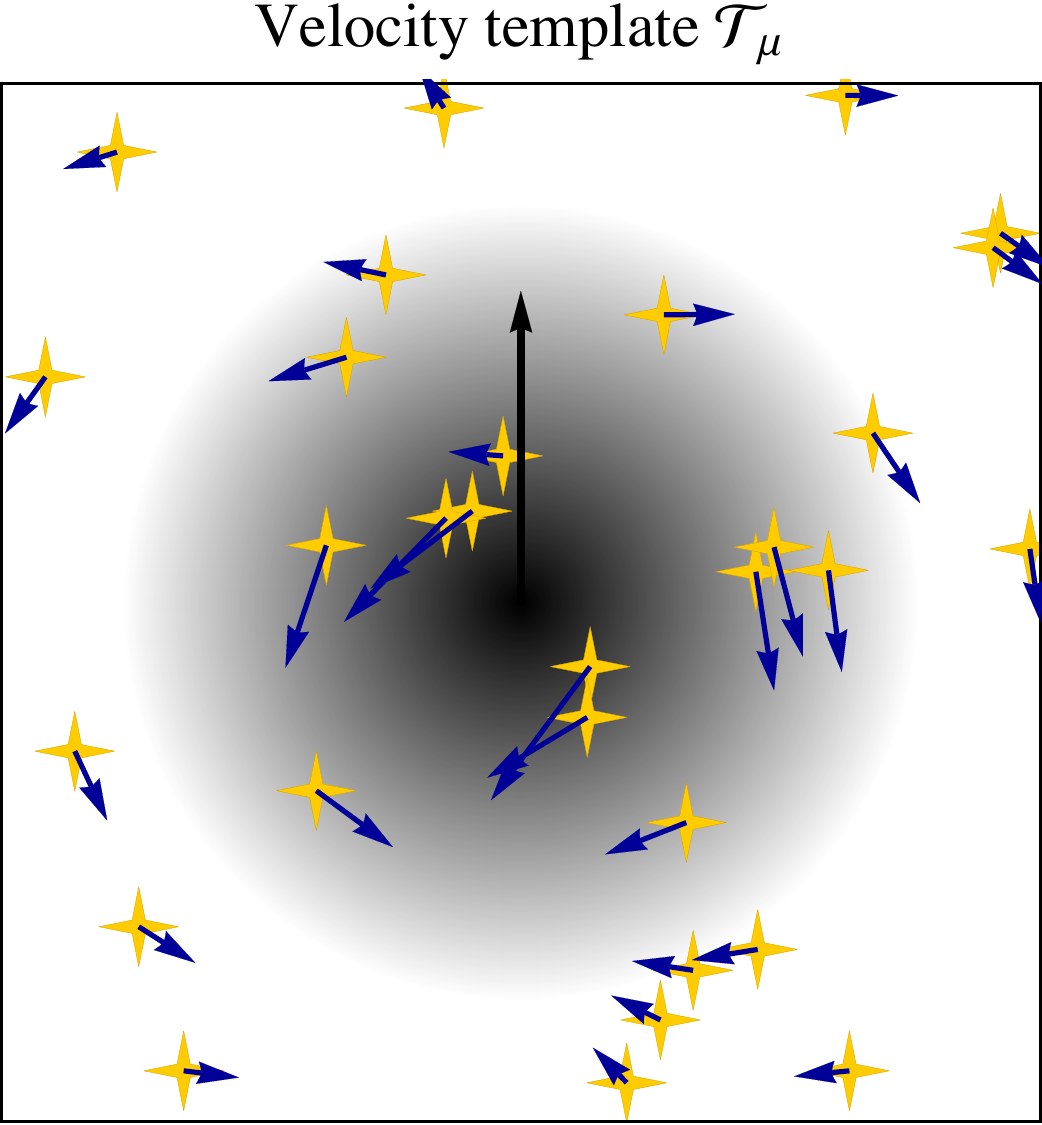}\hspace{0.01\textwidth}
\includegraphics[width=.65\textwidth,origin=0,angle=0]{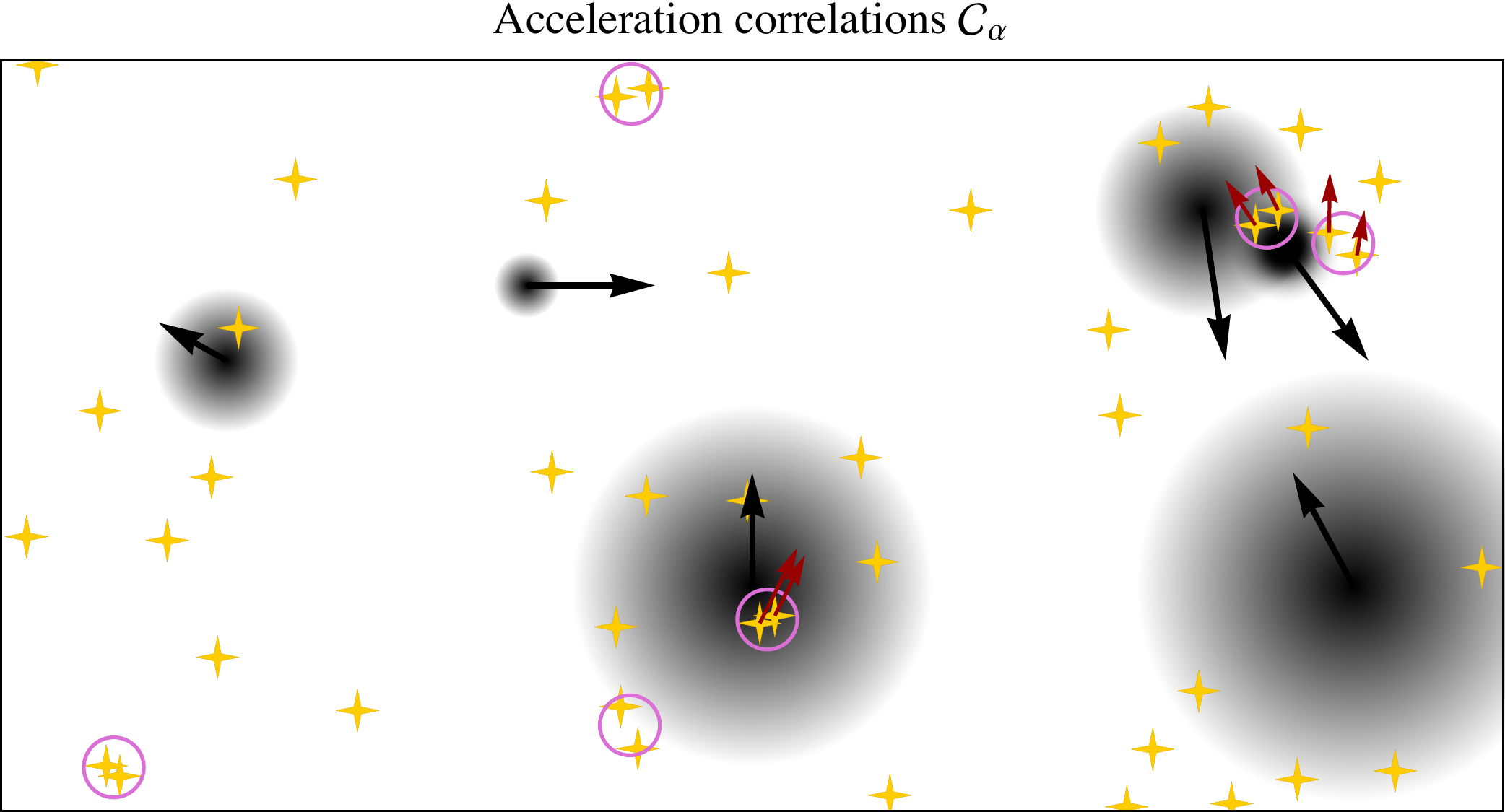}
\caption{Illustration of signal observables. Lenses are indicated by black points (top row) or extended halos (bottom row) with black velocity vectors or paths. Background sources are depicted as four-pointed stars, with blue velocity and red acceleration vectors. Axes are celestial coordinates. \textbf{Top left:} Accidentally small impact parameters between source-lens pairs can produce anomalously large angular velocities or accelerations.  \textbf{Top middle:} If the change in impact parameter of a source-lens pair over the observation time is large, the relative motion scans over the nonlinear part of the lensing deflection, causing a nonrepeating anomaly in the proper motion of a single source---a ``mono-blip''. \textbf{Top right:} If the change in impact parameter is larger than the typical separation between sources (usually when the lens has a small line-of-sight distance), blips in the proper motion of several background sources will occur---a ``multi-blip''. \textbf{Bottom left:} An extended subhalo moving in front of a dense field of background sources will produce correlated, lens-induced velocity shifts according to a pattern determined by the subhalo's density profile, to which a velocity template can be fitted. \textbf{Bottom right:} An ensemble of subhalos (or point masses) moving in front of background sources induces global correlations in the acceleration of nearby star pairs (circled in pink).} \label{fig:signals}
\end{figure}

\paragraph{Classes of targets and observables}
Given the wide range of masses and densities that could in principle be exhibited by clumps of dark matter, there is no single approach that will work for all objects in question. We will discuss three major classes in section~\ref{sec:signal}:% single-source observables, multiple-source observables, and global observables. A sample of the proposed observables is illustrated in figure~\ref{fig:signals}.

The first class of observables involves \emph{rare events}. This class includes outlier velocities, outlier accelerations, and nonrepeating mono-blips in the proper motion of \emph{individual} background sources. Mono-blip observables are useful when the impact parameter changes by a large fractional amount, i.e. $v_{il} \tau / b_{il} \gtrsim 1$, and are essentially the same as techniques proposed previously in e.g.~ref.~\cite{erickcek2011astrometric}. Outliers extend the sensitivity to cases where $v_{il} \tau / b_{il} \lesssim 1$. We begin with this class of observables because they are optimal for detecting and simultaneously localizing point-like lens targets with small proper motions, and also because they provide a foil for comparison against the novel techniques which are more suitable for extended lens targets, and for ones that traverse large arcs on the sky.

The second class is that of \emph{local} multi-source observables, which take advantage of the inherently large number of background sources behind extended lenses or lenses with significant proper motion or parallax. This class is useful for objects which extend over large angular areas, such as large diffuse halos. The class includes velocity and acceleration templates, which pick up correlated effects on the motion of the background sources that are ``eclipsed'' by a lens of finite size subtending a large angle on the sky. Specifically, for the lens-induced velocity shift of eq.~\ref{eq:sizeeffects} with an NFW-like cusp ($\gamma=1$), most of the information that can be extracted is actually from background sources with $b_{il} \sim r_s$ instead of those with $b_{il} \ll r_s$, simply because there are a lot more of the former. This remains true for other profiles, as long as $\gamma < 2$. 
This class also includes multi-blip observables designed to detect candidate lenses that transit past a large number of sources. Again, the cumulative effect of many transits boosts the signal to noise ratio, and may be leveraged to look for compact objects that traverse a large angle on the sky, such as distant planets in our own Solar System.  It can be proven that templates and multi-blips are optimal local observables, in the sense that they are matched filters.

The third class consists of \emph{global} observables aimed at distinguishing whether or not a field of stellar motions is lensed at any characteristic angular scale. This class can be useful for searching for objects which are too diffuse to give the dramatic rare events described above,  not extended or strong enough to show up in the local multi-object observables, but still copious enough to provide some aggregate signal. Small halos are an example of something that could be tested in this fashion.
While not providing localization information, they offer the capability of measuring the spectrum---the size and mass function---of the lens population if a positive signal were to be seen. The correlation observables  measure the two-point function of the angular velocity and acceleration fields of background sources. In many cases, such as the velocity field of extragalactic quasars or the acceleration field of Galactic disk stars, there should be practically no two-point correlations at all, except for those induced by time-variable lensing. Again, we see that this is an observable that exploits the fact that any one lens may affect the motion of several light sources simultaneously. The global correlation observables may turn out to be more sensitive to lower-mass (and thus potentially more numerous) lens targets, when any one lens does not provide a large enough local signal.

\begin{figure}[t]
\centering 
\includegraphics[width=.99\textwidth,origin=0,angle=0]{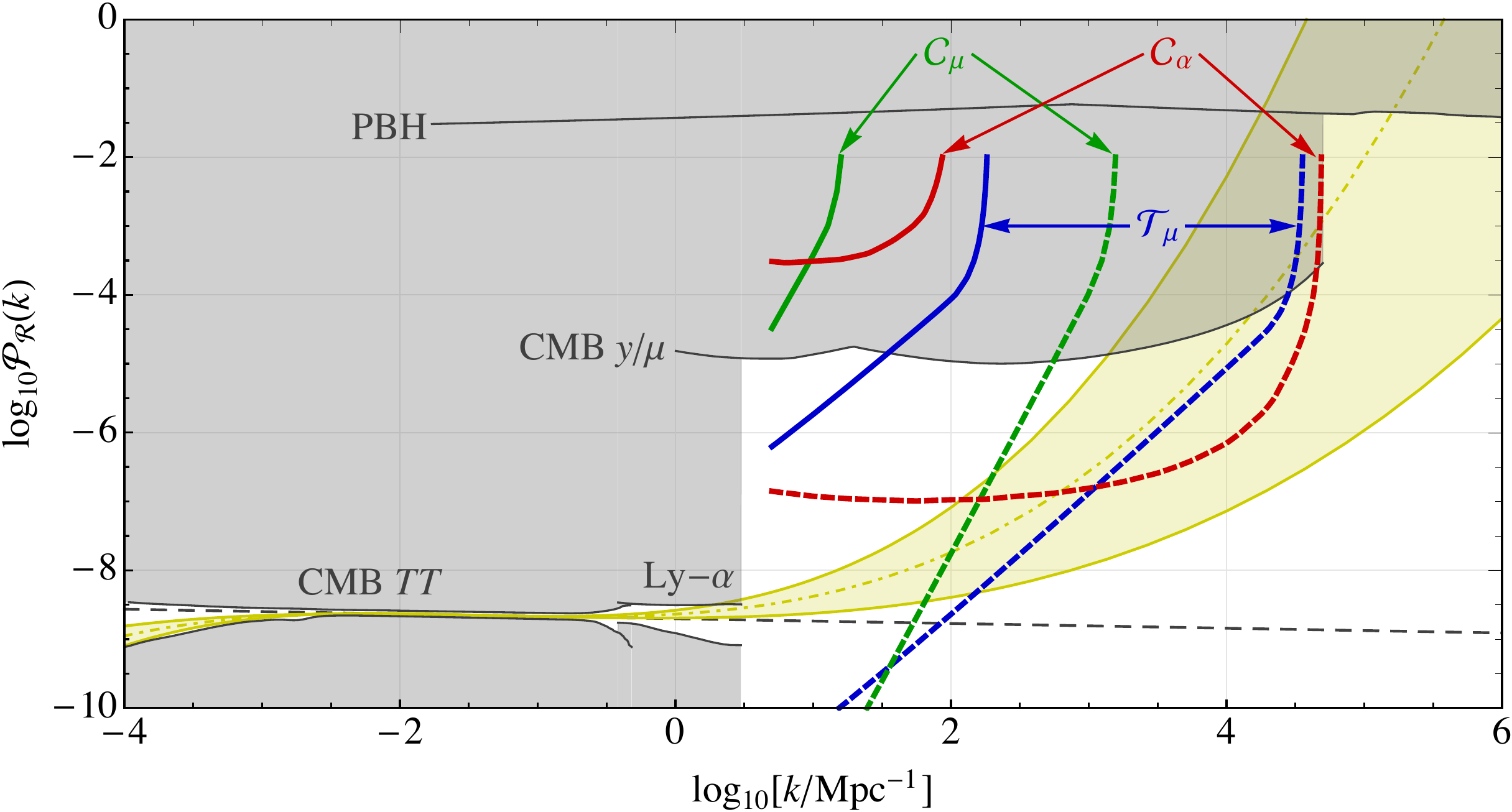}
\caption{Sensitivity projections and constraints on the primordial curvature power spectrum $\mathcal{P}_{\mathcal{R}}$ as a function of comoving wavenumber $k$ (in units of $\text{Mpc}^{-1}$). Forecasts for where on-going and future astrometric surveys can reach unit signal to noise ratio are shown by solid and dashed lines respectively, using velocity templates $\mathcal{T}_\mu$ (blue), velocity correlations $\mathcal{C}_\mu$ (green), and acceleration correlations $\mathcal{C}_\alpha$ (red), for the same parameters as in figure~\ref{fig:halosens}. Gray regions are excluded at 95\% CL by temperature anisotropies in the cosmic microwave background (CMB $TT$), Lyman-$\alpha$ observations, nondetection of spectral distortions of $y-$ and $\mu$-type in the CMB, and limits on primordial black holes (PBH). The black dashed line is the best fit to the \textit{Planck} CMB data assuming a constant spectral tilt $n_s$, while the yellow band indicates the parameter space where $\dd n_s / \dd \ln k$ and $\dd^2 n_s / (\dd \ln k)^2$ were allowed to float by $1\sigma$ from their best fit values (dot-dashed yellow). We refer to sections~\ref{sec:highdensityhalos}~and~\ref{sec:senshalos} for more details.
} \label{fig:prim}
\end{figure}

\paragraph{Preview of results}
Taken together, these observables will offer discovery potential for a wide variety of dark matter substructure: compact objects, high-density clumps, and, if sufficient accuracy is achieved in future surveys, the ability to probe NFW subhalos not much denser---but potentially much smaller---than those of dwarf galaxies. Many of these objects can originate as a direct result of a spectrum of primordial curvature fluctuations that is enhanced at small scales, which can occur in many inflationary models. Compact objects will be dominantly constrained by rare events, large halos by aggregate velocity signals, and acceleration signals will contain high density halos over a range of scales. To preview some of these results in a more familiar context, we display our sensitivity forecasts in terms of primordial perturbations, for current (solid lines) and future (dashed lines) astrometric surveys, in figure~\ref{fig:prim}. In later sections, we will explain in detail how we arrived at this result, and show more sensitivity projections for NFW subhalos, compact objects, and hypothetical planets far beyond the Kuiper Belt in our own Solar System.

\paragraph{Recommendations for astrometric surveys}
Astrometric missions such as \textit{Gaia} typically release only derived data products, such as the distance $D_i$, angular position $\vect{\theta}_i$, and angular velocity $\dot{\vect{\theta}}_i$ of each light source. Searches for gravitational lensing by dark matter substructure that are based on average velocities can already use this data, but it does not contain enough information for searches using acceleration-based observables, or mono- or multi-blip events. 
We encourage the Gaia Data Processing and Analysis Consortium and future astrometric missions to release the full time series
of astrometric position data or transit timing for each source, or residual information that would make it possible to reconstruct the full astrometric time series. This will allow for maximum flexibility for data use and re-use, as those data can be
fitted to nonlinear trajectories to search for lens-induced accelerations as well as mono- and multi-blip lens events.
%We encourage the collaborations to release the full time series data for each light source as soon as possible. This will allow for maximum flexibility, as those data can be fitted to nonlinear trajectories, to search for lens-induced accelerations as well as mono- and multi-blip lens events.

Our analysis reveals that the best source targets are generally those that have higher angular number density, higher apparent brightness, larger line-of-sight distance, and smaller intrinsic proper motion. The relevant figure of merit to be minimized for each proposed observable is defined in eqs.~\ref{eq:FOMO},~\ref{eq:FOMB},~\ref{eq:FOMTmu},~and \ref{eq:FOMC}. These figures of merit warrant deep surveys toward targets such as the Galactic Center and Disk, the Magellanic Clouds and other bright but distant galaxies, and quasars. We suggest these considerations be factored in the observational strategies of future astrometric surveys such as \textit{Theia} and SKA.

%%%%%%%%%%%%%%%%%%%%%%%%%%%%%%%%%%%%%%%%%%%%%%%%%%%%%%%%%%%%%%%%%%%
\section{Lensing targets}\label{sec:lenstargets}
Our aim is to develop astrometric lensing techniques to look for nonluminous objects populating the Milky Way. The long-term goal for such searches would be the subhalos naturally present as a consequence of the hierarchical formation that gave rise to the structures of the Universe. 
As we shall see, next-generation experiments are needed to robustly probe the Milky Way's DM substructure if the primordial curvature power spectrum remains at the $10^{-9}$ level at scales smaller than those probed in Lyman-$\alpha$ observations and CMB experiments. Short of this, there are a variety of motivated objects and scenarios that could be discovered in the shorter term. Examples include:
\begin{itemize}
\item higher-density subhalos from an enhanced primordial power spectrum (subsection~\ref{sec:highdensityhalos});
\item exotic, point-like objects such as primordial black holes and dark stars, or more extended exotic structures that can form from rich DM microphysics, such as dissipation mechanisms and phase transitions (subsection~\ref{sec:exoticobjects}); 
\item new planets in our own Solar System (subsection~\ref{sec:planets}).
\end{itemize}
We start by reviewing the standard spectrum of dark matter subhalos in subsection~\ref{sec:subhalos}, along with a basic model of the Milky Way's own dark matter halo and baryonic disk.

\subsection{Standard subhalos}\label{sec:subhalos}
The ``holy grail'' of this approach would be to measure dark matter substructure in the Milky Way. If the primordial power spectrum is not enhanced at small scales (e.g.~if it is given by the black dashed line in figure~\ref{fig:prim}), then this substructure is expected to consist of a broad, approximately scale-invariant spectrum of subhalos contained within the Milky Way's large halo. This substructure is a natural consequence of the hierarchical structure formation process associated with cold dark matter. These expectations are borne out in numerical $N$-body simulations of collisionless, cold dark matter with only gravitational interactions, starting from a set of initial conditions extrapolated from those inferred from the curvature perturbation spectrum measured in the cosmic microwave background. 

The clumpiness of dark matter is expected to consist of halos and subhalos (smaller halos within larger halos) whose shape can be approximated by the spherically symmetric, NFW density profile~\cite{navarro1997universal}:
\begin{align}
\rho(r) = \frac{4 \rho_s}{\frac{r}{r_s} \left( 1+\frac{r}{r_s} \right) ^2}, \label{eq:rhoNFW}
\end{align}
where $r_s$ is known as the scale radius, and $\rho_s = \rho(r_s)$ is the density at the scale radius. The enclosed mass within a radius $r$ is
\begin{align}
m(r) = \int_0^r dr' \, 4\pi r^{\prime 2} \rho(r') = 16 \pi \rho_s r_s^3 f(r/r_s); \quad f(x) = \ln(1+x) - \frac{x}{1+x}.
\end{align}
The virial radius $r_{200}$ is defined as the radius within which the mean halo density is 200 times the critical density of the universe $\rho_c = 3 H_0^2 / 8\pi G_N$, where $H_0$ is the Hubble constant. The virial mass $m_{200}$ is the mass inside this radius, namely $m_{200} = m(r_{200}) = (4\pi/3) 200 \rho_c r_{200}^3$. 
The concentration parameter $c_{200} = r_{200} / r_s$ is a measure of the halo's compactness. The core mass $M_s$ is defined as the mass contained within the scale radius $r_s$:
\begin{align}
M_s = m(r_s) = 16 \pi \rho_s r_s^3 f(1) = \frac{4\pi}{3} 200 \rho_c r_s^3 c_{200}^3 \frac{f(1)}{f(c_{200})}. \label{eq:Ms}
\end{align}
Note that $M_s \propto r_s^3$ at fixed concentration parameter $c_{200}$. Approximate scale invariance implies that that there is only a weak variation of $c_{200}$ with $r_s$, so to the extent that this approximation holds, we can expect collapsed (sub-)halos to have roughly the same mean density inside their scale radii.

The halo of the MW can also be modeled by the NFW density profile, with $\rho_s^\text{MW} \approx (3.0 \pm 1.5)\times 10^{-3} ~M_\odot ~\pc^{-3}$ and $r_s^\text{MW}\approx 18.0 \pm 4.3~\kpc$ as best-fit parameters, with our Solar System located at a distance $R_\odot^\text{MW} \approx 8.29 \pm 0.16~\kpc$~\cite{mcmillan2011mass}. Using the value $H_0 = 67.8~\kms~\text{Mpc}^{-1}$ for the Hubble constant~\cite{ade2016planck}, those mean values translate to a concentration of $c^\text{MW}_{200} \approx 13$, a virial radius $r_{200}^\text{MW} \approx 240~\kpc$, and a virial mass $m_{200}^\text{MW} \approx 1.5~\times 10^{12} ~M_\odot$, with large (and correlated) uncertainties. For later reference, we also mention that the baryonic disk can be modeled by the density profile:
\begin{align}
\rho_d(R,z) = \frac{\Sigma_{d,0}}{2 z_d} \exp\left\lbrace - \frac{|z|}{z_d} - \frac{R}{R_d}\right\rbrace
\end{align}
as a function of radius $R$ from the center, and height $z$ above (or below) the disk plane. The best-fit values are $\Sigma_{d,0} \approx 741\pm123 ~M_\odot~\pc^{-2}$ for the surface mass density, and $R_d \approx 3.00\pm 0.22 ~\kpc$ for the radial scale, given a fiducial vertical scale height of the thin disk of $z_d = 300~\pc$~\cite{mcmillan2011mass}. (We will ignore the sub-dominant ``thick-disk'' component.)

\begin{figure}[t]
\centering 
\includegraphics[width=.49\textwidth,origin=0,angle=0]{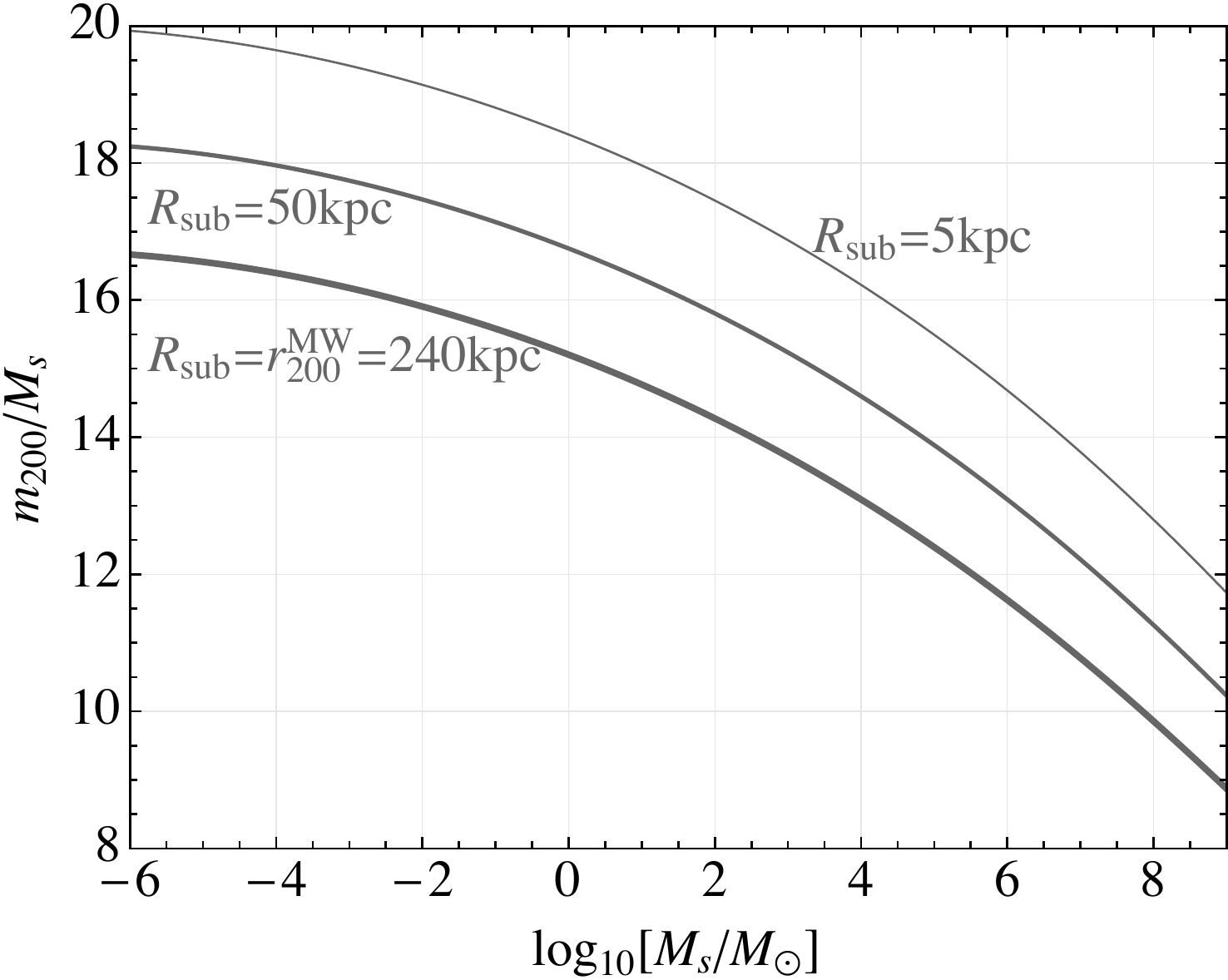}
\includegraphics[width=.49\textwidth,origin=0,angle=0]{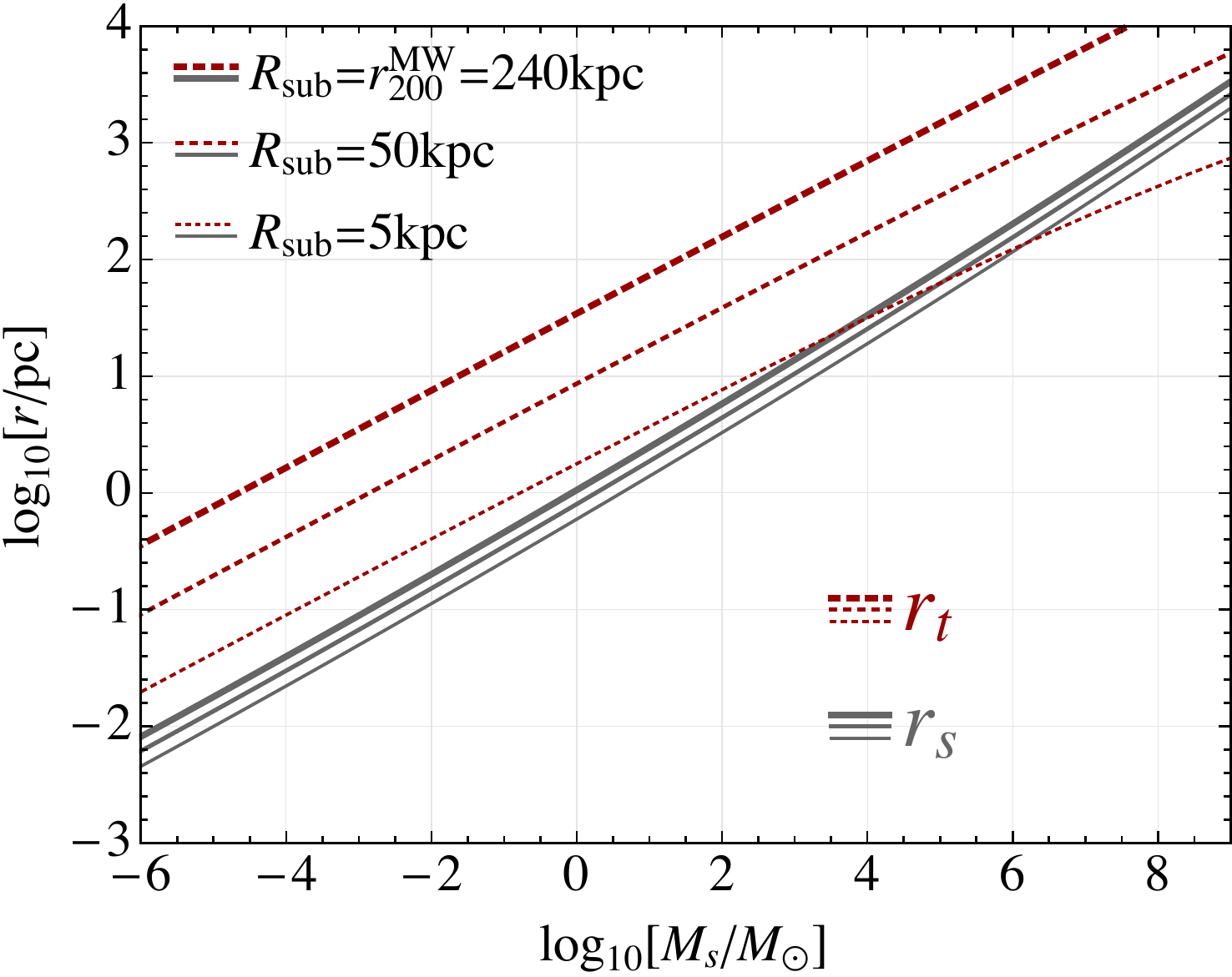}
\caption{\textbf{Left:} Ratio of the virial mass $m_{200}$ and the core mass $M_s$ as a function of $M_s$ in solar-mass units, for three values of subhalo distance $R_\text{sub}$ from the Galactic Center. \textbf{Right:} Scale radius $r_s$ (gray, solid) and tidal radius (red, dotted) as a function of $M_s$. For all but the largest $r_s$ and smallest $R_\text{sub}$, we have that $r_t > r_s$ so that the inner core of the subhalo survives tidal stripping, even though there is often significant mass loss ($m_t < m_{200}$) due to mass stripping at the outer edges of subhalos.}\label{fig:NFWs}
\end{figure}

For subhalos of the Milky Way, we quote results of ref.~\cite{moline2017characterization}, who characterized the properties of subhalos from high-resolution simulations of Milky-Way-type host halos. The properties of the large subhalos are median results from the dark-matter-only simulations VL~II~\cite{diemand2008clumps} and ELVIS~\cite{garrison2014elvis}, which themselves used initial conditions from WMAP's 3-year and 7-year data releases. Their differences among each other and relative to initial conditions extrapolated from \textit{Planck} data should not be large over the range of subhalo masses simulated ($m_{200} \gtrsim 10^6~M_\odot$)~\cite{maccio2007concentration}. At the low-mass end, the fits of ref.~\cite{moline2017characterization} were calibrated against the simulations of micro-subhalos ($10^{-6}~M_\odot \lesssim m_{200} \lesssim 10^{-3}~M_\odot$) of ref.~\cite{ishiyama2014hierarchical}.
The concentration parameter $c_{200}$ of the simulated subhalos with virial mass $m_{200}$ at radius $R_\text{sub}$ from the center of a MW-type  host halo was empirically described by:
\begin{align} 
c_{200} = c_0 \left\lbrace 1 + \sum_{i=1}^3 \left[ a_i \log_{10} \frac{m_{200}}{10^8 h^{-1} M_\odot} \right]^i\right\rbrace \left(1 + b \log_{10} \frac{R_\text{sub}}{r_{200}^\text{MW}}\right) \label{eq:c200moline}
\end{align}
with $c_0 = 19.9$, $a_i = \lbrace-0.195,0.089,0.089\rbrace$, $b = -0.54$, and $h = H_0 / (100~\kms~\text{Mpc}^{-1})$~\cite{moline2017characterization}.
The subhalo-to-subhalo scatter about these median results follows approximately a log-normal distribution with standard deviation $0.13$ and $0.11$ for the resolved subhalos in VL-II and ELVIS, respectively. In the left panel of figure~\ref{fig:NFWs}, we plot the resulting median relation between $M_s$ and $m_{200}$ for subhalos, for three values of $R_\text{sub}$. The virial mass $m_{200}$ is the original mass of the subhalo (before tidal stripping). In the right panel of figure~\ref{fig:NFWs}, we plot the median relation between $M_s$ and $r_s$ in solid gray, for the same three values of $R_\text{sub}$. Due to tidal effects and formation bias, subhalos are more compact than host halos of the same mass, and become increasingly concentrated closer to the MW's center (smaller $R_\text{sub}$).

Tidal gravitational fields generally distort the density profile of extended dark matter subhalos. We will adopt a tidal approximation wherein the subhalo's density $\rho(r)$ of eq.~\ref{eq:rhoNFW} is unperturbed up to a tidal radius $r_t$, and vanishes for $r > r_t$. A physically motivated value for this quantity is the radius at which the subhalo's gravitational force is $3/2$ times the tidal gravitational forces from the perturbing objects, which we will take to be the Milky Way's disk and dark matter halo. (The factor of $3/2$ instead of $1$ is to account for kinetic energy inside the subhalo.) This then allows for the following implicit definition of the tidal radius $r_t$ of a subhalo at a radius $R_\text{sub}$ from the MW center, in the limit $r_t \ll R_\text{sub}$ and $r_t \ll R_d$:
\begin{align}
\frac{m(r_t)}{r_t^2} =\frac{3}{2} \max \left\lbrace m^\text{MW}(R_\text{sub})\frac{2 r_t}{R_\text{sub}^3}, \Sigma_{d,0} e^{-R_\text{sub}/R_d} \left(1-e^{-r_t/2z_d} \right) \right\rbrace.
\end{align}
This equation is approximately valid for circular orbits through our Galaxy; for elliptical orbits $R_\text{sub}$ should be replaced by the distance at periapsis in the halo term, and by the smallest radius at which the subhalo crosses the disk for the disk term. For the fiducial values for the MW halo and disk adopted above, the tidal radius of all subhalos is set by the tidal field of the halo, not that of the disk. The mass $m_t \equiv m(r_t)$ should be regarded as the present-day, physical mass of the subhalo after tidal stripping; $m_{200}$ can be regarded as a fitting parameter, or as the original mass of the subhalo near the time of accretion but before tidal stripping in the host halo. In the right panel of figure~\ref{fig:NFWs}, we plot the tidal radius $r_t$ as a function of $M_s$ in dotted red. Over most of the mass range considered, tidal forces should not dramatically disrupt the profile of the subhalo core, as $r_t > r_s$. These analytic estimates are quite coarse, but recent studies of tidal disruption indicate that substructure at these small scales should survive, although obtaining accurate numerical convergence in simulations of substructure disruption is notoriously difficult~\cite{van2017disruption,van2018dark}.

Most models predict a broad spectrum of substructure over a wide range of scales. We parametrize this by the subhalo mass function:
\begin{align}
\frac{\dd N}{\dd m_t} \approx \frac{A}{m^\text{MW}_{200}} \left(\frac{m_t}{m^\text{MW}_{200}} \right)^{n_m} B[m_t; m_-, m_+],
\end{align}
where $n_m$ is the spectral index, the dimensionless coefficient $A$ sets the overall normalization of the substructure fraction, and $m_-$ and $m_+$ set the minimum and maximum subhalo mass, respectively. The boxcar function $B[m_t; m_-, m_+]$ is taken to be 1 for $m_- < m_t < m_+$ and 0 otherwise. The maximum subhalo mass $m_+$ after tidal stripping is not to exceed the MW virial mass $m_{200}^\text{MW}$. The minimum subhalo mass $m_-$ depends on DM microphysics (such as the mass or self-interactions of the DM particle), or its production mechanism in the early Universe. Strictly speaking, the halo mass function depends also on the location inside the MW halo, a complication we ignore here. For our purposes, the halo mass function refers to the one for which $5~\kpc \lesssim R_\text{sub} \lesssim 50~\kpc$. In a nearly scale-invariant Universe, one expects $n_m \approx -2$, which yields a nearly scale-invariant subhalo spectrum with energy density at all scales, i.e.~$m_t\dd \log( N)/\dd \log(m_t) \simeq \text{constant}$. For example, if one takes $n_m = 2$, $m_- = 10^{-6}~M_\odot$, $m_+ = m_{200}^\text{MW}$, and $A = 0.012$, then one gets that 50\% of the MW's energy density is composed out of substructure, as $\int_{m_-}^{m_+} \dd m_t\, m_t \dd  N/\dd m_t \approx 0.5 m_{200}^\text{MW}$.  Restricting to the mass range resolvable by simulations, $m_- = 10^{6}~M_\odot$, one finds $\sim$ 10\% in substructure. These values are consistent with those found in ref.~\cite{moline2017characterization}, which did start with nearly scale-invariant initial conditions; they should however be taken with a grain of salt when extrapolated outside of the window of simulated subhalo masses.

\begin{figure}[t]
\centering 
\includegraphics[width=.45\textwidth,origin=0,angle=0]{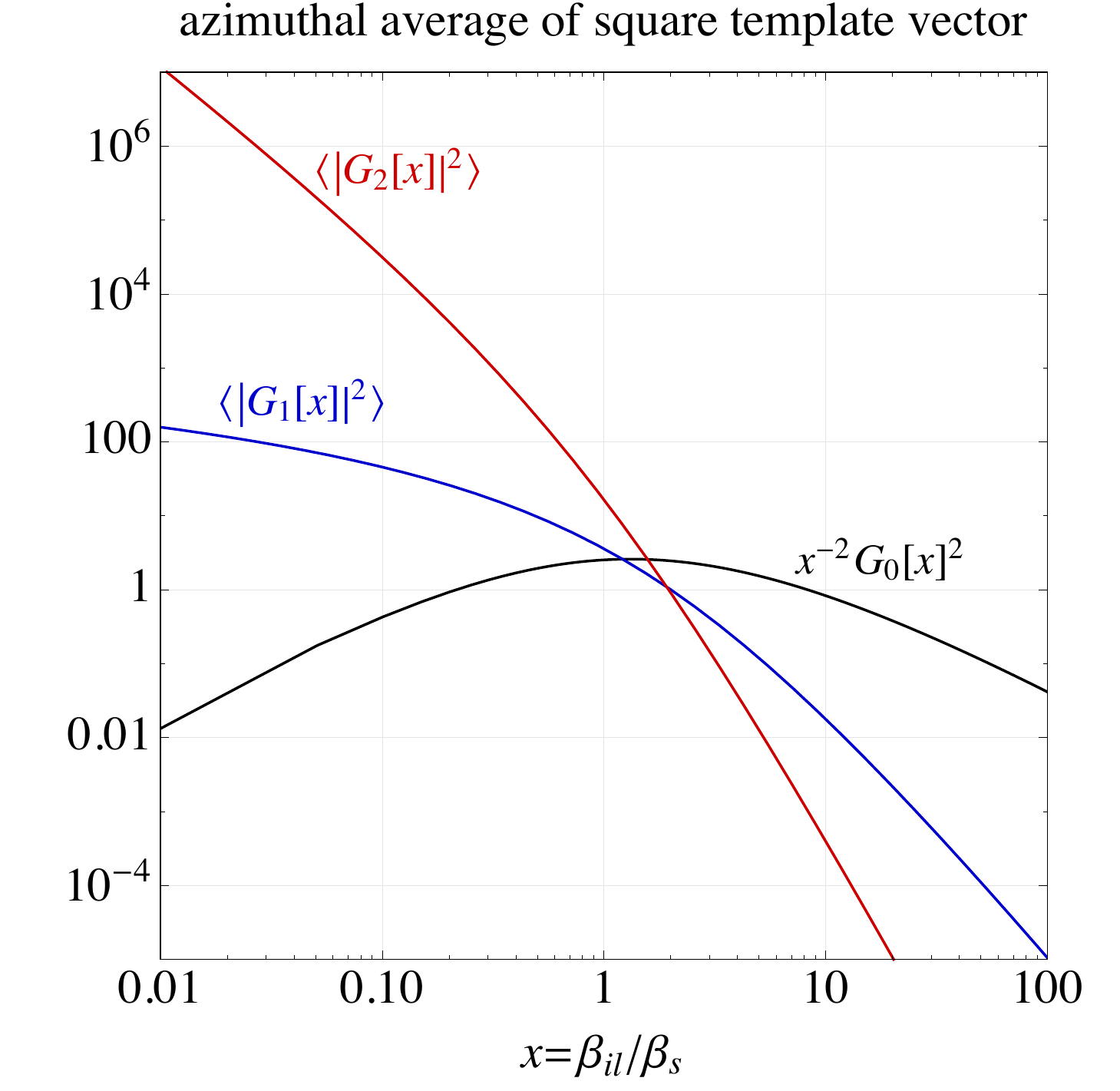}
\includegraphics[width=.45\textwidth,origin=0,angle=0]{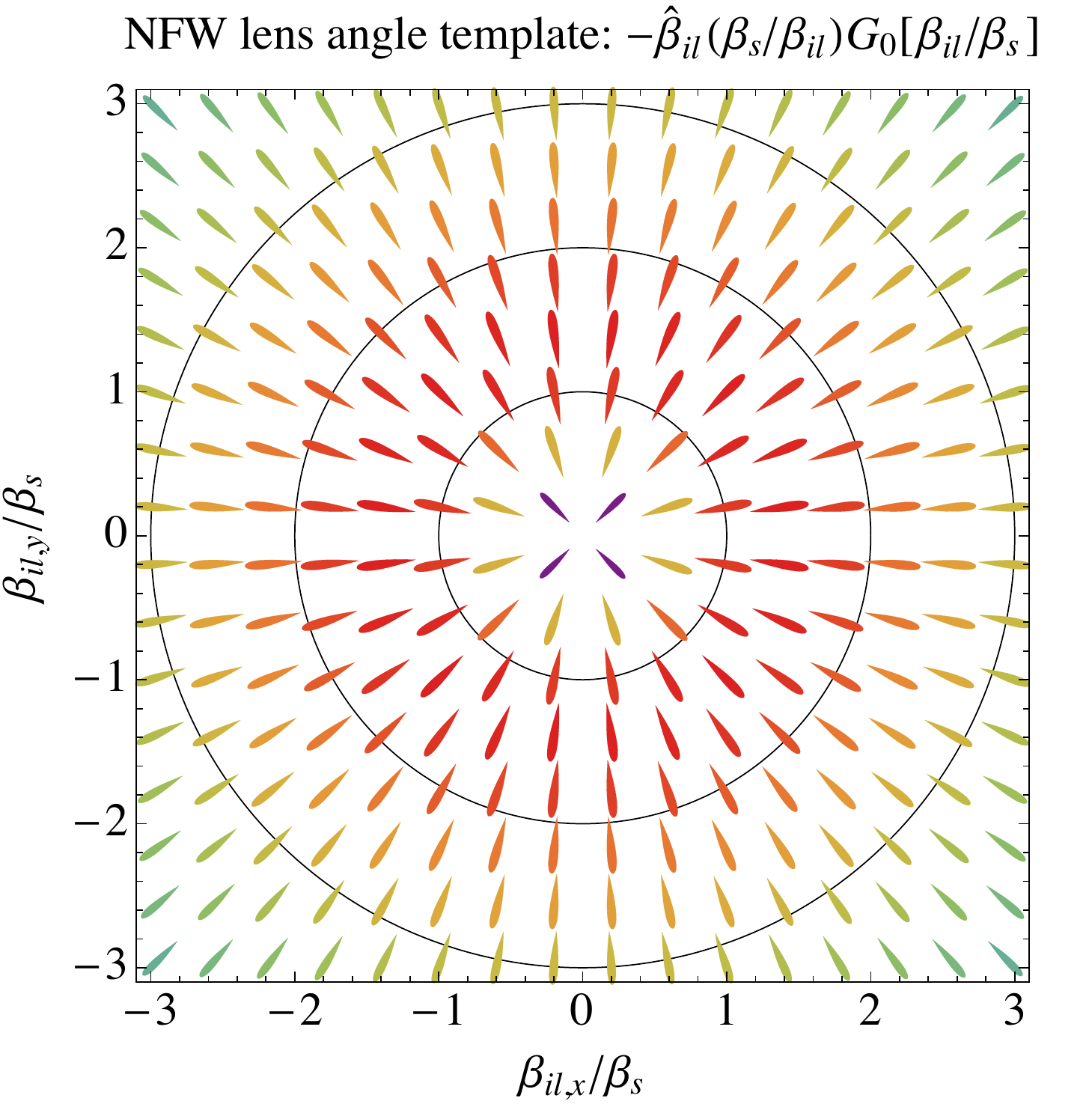}
%\vspace{2em}\\
\includegraphics[width=.45\textwidth,origin=0,angle=0]{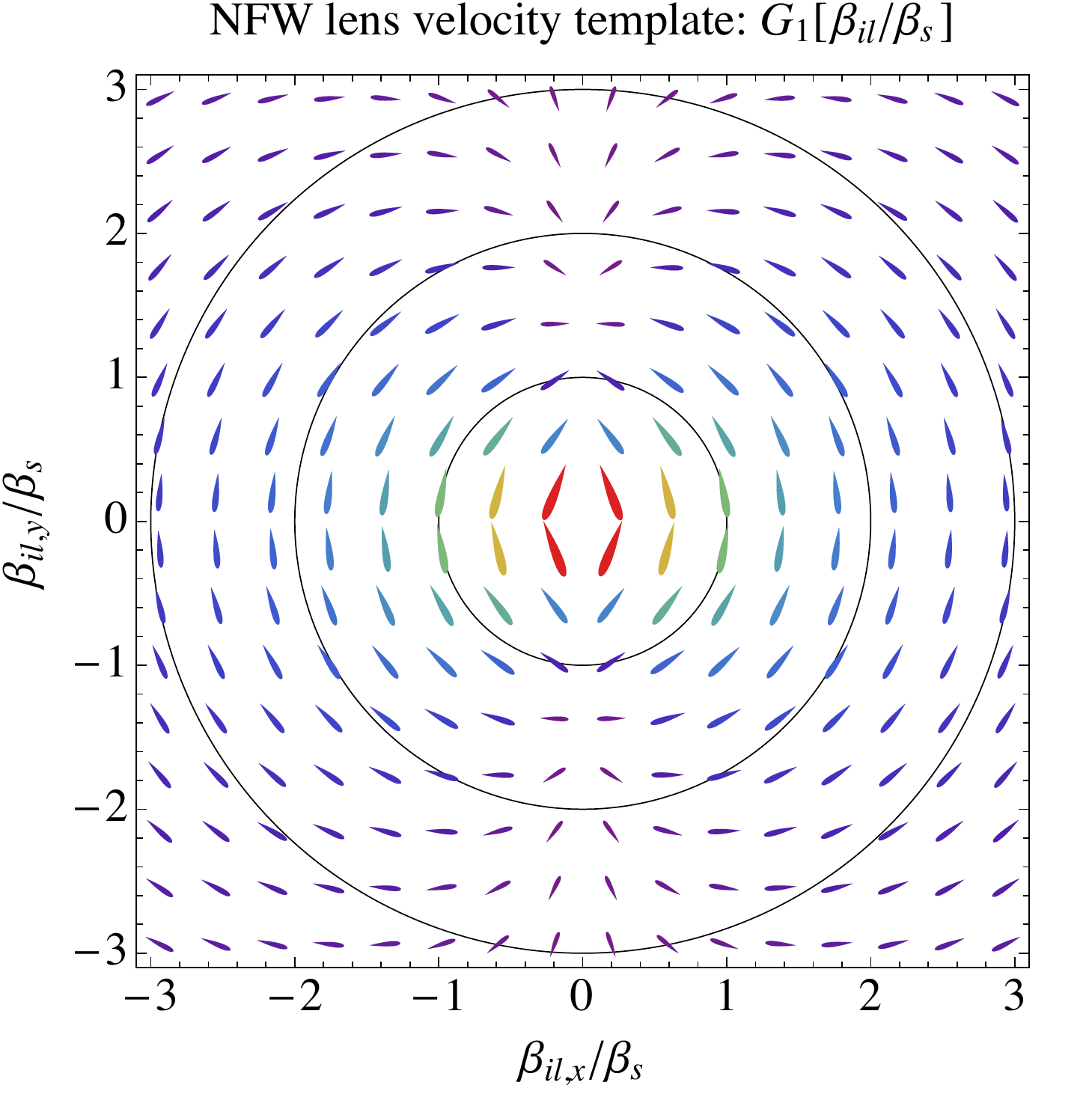}
\includegraphics[width=.45\textwidth,origin=0,angle=0]{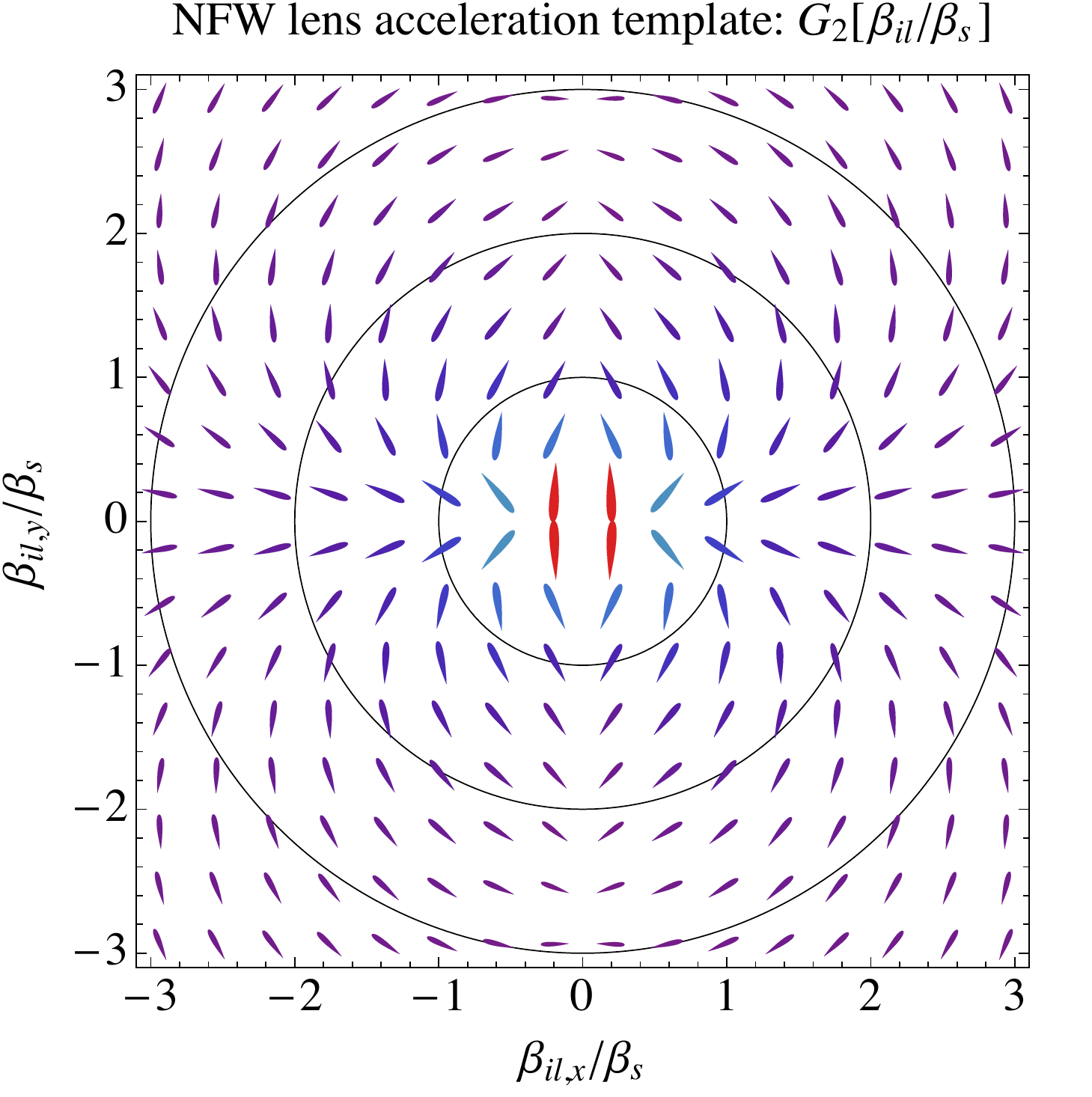}
\caption{Templates for NFW subhalos. \label{fig:templates} \textbf{Top left:} Azimuthally averaged square norm of the lensing template vectors for shifts in angle ($-\hat{\vect{\beta}}_{il} (\beta_s/\beta_{il})G_0[\beta_{il}/\beta_s]$), angular velocity ($\vect{G}_1[\beta_{il}/\beta_s]$), and angular acceleration ($\vect{G}_2[\beta_{il}/\beta_s]$) as a function of angular impact parameter $\beta_{il}$ in units of $\beta_s$. \textbf{Top right, bottom left, bottom right:} Vector templates for angular shifts in lensing angle, velocity, and acceleration, respectively, indicating both the direction and relative magnitude of the lensing correction. In all three panels, the lens is at the center of the plot by construction (where the angular impact parameter $\vect{\beta}_{il} = 0$), and assumed to move in the $+y$-direction, taking $\vect{v}_i = \vect{v}_\odot =0$ for illustrative purposes. In the velocity (acceleration) template, the depicted length of the template vectors is proportional to the fourth (eighth) root of their actual magnitude, to adjust for the large dynamic range. Black contours depict impact parameters at integer multiples of the angular size of the scale radius $\beta_s = r_s / D_l$.}
\end{figure} 

Finally, we are ready to discuss the lensing effects from an NFW subhalo. The position on the sky of a background source $i$ at a line-of-sight distance $D_i$ receives a gravitational lensing correction from a lens at distance $D_l$ and angular impact parameter vector $\vect{\beta}_{il} \equiv \vect{b}_{il}/D_l$ of:
\begin{align}
\Delta \vect{ \theta}_{il} = -\left(1 - \frac{D_l}{D_i} \right) \frac{4G_N M_s}{D_l \beta_{il}} \hat{\vect{\beta}}_{il} G_0\left[ \frac{\beta_{il}}{\beta_s} \right]
\end{align}
where we have defined the piecewise function:
\begin{align}
G_0[x] \equiv \frac{2}{\ln 4 -1}\left[\ln \frac{x}{2} + \begin{cases} \frac{1}{\sqrt{1-x^2}} \text{arccosh}(1/x) & x < 1 \\ \frac{1}{\sqrt{x^2-1}} \text{arccos}(1/x) & x > 1 \end{cases} \right].\label{eq:G0}
\end{align}
The function $G_0[x]$ scales as $x^2[-2\ln(x/2) - 1]/[2(\ln 4 -1)]$ in the limit of $x \to 0$, reaches $2(1-\ln 2)/(\ln 4-1) \approx 1.59$ at $x = 1$, and approaches logarithmic growth $[2/(\ln 4 - 1)]\ln[x/2]$ at large $x$. The angular velocity of the source also receives a correction if the impact parameter changes at a rate $\vect{v}_{il} \equiv \dot{\vect{b}}_{il}$:
\begin{align}
\Delta\dot{ \vect{\theta}}_{il} = \left(1 - \frac{D_l}{D_i} \right) \frac{4G_N M_s v_{il}}{D_l^2 \beta_s^2} \vect{G}_1\left[ \frac{\beta_{il}}{\beta_s}, \hat{\vect{\beta}}_{il}, \hat{\vect{v}}_{il} \right] \label{eq:NFWmu}
\end{align}
with a characteristic angular velocity profile function:
\begin{align}
\vect{G}_1 \left[x,\hat{\vect{\beta}},\hat{\vect{v}} \right] \equiv - \betahat (\betahat \cdot \vhat)  \frac{G_0'[x]}{x}  - \left[ \vhat - 2 \betahat ( \betahat \cdot \vhat ) \right] \frac{G_0[x]}{x^2}. \label{eq:G1}
\end{align}
Taking yet another time derivative, we get the angular acceleration shift:
\begin{align}
\Delta\ddot{ \vect{\theta}}_{il} = \left(1 - \frac{D_l}{D_i} \right) \frac{4G_N M_s v_{il}^2}{D_l^3 \beta_s^3} \vect{G}_2\left[ \frac{\beta_{il}}{\beta_s}, \hat{\vect{\beta}}_{il}, \hat{\vect{v}}_{il} \right] \label{eq:NFWalpha}
\end{align}
with an angular acceleration profile function of:
\begin{align}
\vect{G}_2 \left[x,\hat{\vect{\beta}},\hat{\vect{v}} \right] \equiv~& - \betahat (\betahat \cdot \vhat)^2  \frac{G_0''[x]}{x}  - \left[ 2 \vhat (\betahat \cdot \vhat) + \betahat \left(1-5(\betahat \cdot \vhat )^2 \right) \right] \frac{G_0'[x]}{x^2} \\
&+ \left[4 \vhat(\betahat \cdot \vhat) + 2 \betahat \left(1-4 (\betahat \cdot \vhat)^2 \right) \right] \frac{G_0[x]}{x^3}. \label{eq:G2}
\end{align}
The vector profiles for the lens-induced shifts in angular position, velocity, and acceleration are depicted in figure~\ref{fig:templates}. Also plotted, in the top left panel, are the azimuthally averaged square norms for the same three profiles.

\subsection{High-density subhalos}\label{sec:highdensityhalos}
One of the principal reasons that subhalos are hard to find is that they are low-density objects. The lens-induced angular velocity signal from an NFW subhalo scales as $\Delta \dot \theta_{il} \propto M_s/r_s^2 \propto \rho_s^{2/3} M_s^{1/3}$, while the induced acceleration scales as $\Delta \ddot \theta_{il} \propto M_s/r_s^3 \propto \rho_s$. As we saw in section~\ref{sec:subhalos}, for a nearly scale-invariant power spectrum of primordial perturbations, the median core density $\rho_s$ of the resultant subhalos varies only logarithmically with $M_s$ (smaller ones are only slightly denser, as the perturbations that seed them entered the horizon earlier). Due to the stochastic nature of the formation, accretion, and merger histories, there is a modest (log-normal) scatter about these median values for the density, but subhalos of much-higher-than-median densities should be exceedingly rare.

We stress that the above scenario presupposes a scale-invariant primordial power spectrum with an amplitude and constant spectral index extrapolated far beyond where they have been constrained in CMB~\cite{ade2016planck} and Lyman-$\alpha$~\cite{bird2011minimally} observations, which have so far only probed primordial fluctuations for wavenumbers smaller than $3~\text{Mpc}^{-1}$ (see gray exclusion regions of figure~\ref{fig:prim}). The \textit{Planck} experiment has set the tightest constraints on the primordial power, over the largest range of scales (in log space), between $10^{-4}~\text{Mpc}^{-1}$ and $0.5~\text{Mpc}^{-1}$. Parametrizing the power spectrum as 
\begin{align}
\mathcal{P}_\mathcal{R}(k) = \frac{k^3}{2\pi^2} \left| \mathcal{R}_k \right|^2 = A_s \left(\frac{k}{k_*} \right)^{\left[n_s-1+\frac{1}{2} \frac{\dd n_s}{\dd \ln k} \ln\frac{k}{k_*}+\frac{1}{6} \frac{\dd^2 n_s}{\dd (\ln k)^2} \left(\ln\frac{k}{k_*}\right)^2+\dots\right]} \label{eq:planck0}
\end{align}
with a pivot scale  $k_* = 0.05~\text{Mpc}^{-1}$, the \textit{Planck} $TT+\text{lowP}$ data set~\cite{ade2016planck} indicates that at 68\% CL, the scalar amplitude $A_s$ and spectral index $n_s$ are
\begin{align}
\ln(10^{10} A_s) = 3.089 \pm  0.036, \quad n_s = 0.9655 \pm 0.0062 \label{eq:planck1}
\end{align}
assuming a constant spectral index, i.e.~$0=\frac{\dd n_s}{\dd \ln k}=\frac{\dd^2 n_s}{\dd (\ln k)^2}$. When both the running and the running of the running of the scalar spectral index are allowed to float, the same data set gives the 68\%-CL ranges of:
\begin{align}
n_s = 0.9569 \pm 0.0077, \quad \frac{\dd n_s}{\dd \ln k} = 0.011^{+0.014}_{-0.013}, \quad \frac{\dd^2 n_s}{\dd (\ln k)^2} = 0.029^{+0.015}_{-0.016}. \label{eq:planck2}
\end{align}
(Including also the \textit{Planck} $TE,EE$ information slightly reduces the central values and error bars, but would lead to the same conclusions.) In figure~\ref{fig:prim}, we plot eq.~\ref{eq:planck0} for the central values of eq.~\ref{eq:planck1} as the black dashed line, and for the central values of eq.~\ref{eq:planck2} in dot-dashed yellow, assuming third- and higher-order logarithmic derivatives of $n_s$ are zero. The yellow band contains the parameters where both ${\dd n_s}/{\dd \ln k}$ and ${\dd^2 n_s}/{\dd (\ln k)^2}$ are simultaneously allowed to deviate by $1\sigma$. That extrapolation should not be taken too seriously, as higher-order effects would almost certainly kick in, and the deviation from a scale-independent $n_s$ is driven by the apparent power deficit at large scales, and is furthermore not statistically significant. Our point is that the \textit{Planck} data allow for a relatively large second logarithmic derivative of the spectral index, and that there could very well be enhanced fluctuations at smaller scales. Such enhancements can occur in a variety of inflationary models, in particular those where the inflaton potential flattens to a plateau towards the end of inflation~\cite{orlofsky2017inflationary}, or develops a saddle point, as happens naturally in hybrid models~\cite{garcia1996density}.

Given a general primordial curvature power spectrum $\mathcal{P}_\mathcal{R}(k)$, it is in principle possible to calculate the spectrum of halos, subhalos, subsubhalos,\dots at all masses. In practice, this is an exceedingly difficult problem, because a quantitative study of these nonlinear structures requires high-resolution simulations due to the inherently large dynamic range of time and length scales involved, and present-day computational resources are limited. Nevertheless, we will attempt to obtain at least a parametric analytic estimate, and leave their refinements to further work. We shall see that even on the most pessimistic end of plausible uncertainties, our techniques will still probe unconstrained parameter space of primordial fluctuations.

In the Press-Schechter formalism of spherical collapse~\cite{press1974formation}, when an overdense, spherical region of comoving radius $R$ becomes nonlinear and collapses, it will have an initial mass of $M_i(R) = (4\pi/3) \rho_c \Omega_\text{DM} R^3$, where $\rho_c$ is the critical energy density of the Universe today, and $\Omega_\text{DM}$ is its fractional DM abundance. Primordial fluctuations with comoving wavenumbers $k$ will seed overdense regions with $R \sim k^{-1}$. We will assume that the present-day core mass $M_s$ of the resultant (sub)halo (cfr.~\ref{eq:Ms}) is this initial mass $M_i$ up to a numerical prefactor $C_M$ which should only weakly depend on $k$:
\begin{align}
M_s(k) = C_M \frac{4\pi}{3} \rho_c \Omega_\text{DM} k^{-3} = C_M \frac{H_0^2}{2 G_N} \Omega_\text{DM} k^{-3}. \label{eq:Msk}
\end{align}
The collapse redshift $z_\text{coll}$ is defined as the redshift at which the fractional overdensity $\delta$ reaches $\delta_\text{coll} = (3/5)(3\pi/2)^{2/3}\approx 1.686$ in the linear theory. The ambient DM energy density at that time is $\rho_c \Omega_\text{DM} (1+z_\text{coll})^3$, and the collapsed object will virialize to a mean density about 200 times larger than that, with an even denser core. We estimate the resultant present-day ($z=0$) core density $\rho_s$ as defined in eq.~\ref{eq:rhoNFW} to be:
\begin{align}
\rho_s(z_\text{coll}) = C_\rho \rho_c \Omega_\text{DM} (1+z_\text{coll})^3 \left[1+b \log_{10} \frac{R_\text{sub}}{r_{200}^\text{MW}}\right]^3, \label{eq:rhoszcoll}
\end{align}
where the numerical prefactor $C_\rho$ should only have a weak dependence on $k$ and $z_\text{coll}$, and we take $b$ to be the same prefactor as in eq.~\ref{eq:c200moline}.

Let us define $\delta_\text{min}(k,z_\text{coll})$ as the minimum necessary fractional overdensity $\delta$ with comoving wavenumber $k$ at horizon crossing such that it would collapse by redshift $z_\text{coll}$. We refer to ref.~\cite{bringmann2012improved} for the $\delta_\text{min}$ calculation, for which we used $z_\text{eq} \approx 3250$ for the redshift of matter-radiation equality. We denote by $\sigma^2(k)$ the variance in fractional density inside a sphere of radius $k^{-1}$ at horizon crossing. If the power spectrum $\mathcal{P}_\mathcal{R}(k)$ is nearly scale-invariant in one $e$-fold around $k$, then it can be derived~\cite{bringmann2012improved} that:
\begin{align}
\sigma^2(k) \approx 0.91 \mathcal{P}_\mathcal{R}(k). \label{eq:sigmak}
\end{align}
With the above assumptions, the probability $\beta$ that a fluctuation of wavenumber $k$  collapses by redshift $z_\text{coll}$ is:
\begin{align}
\beta\left[k,\mathcal{P}_\mathcal{R}(k),z_\text{coll} \right] = \frac{2}{\sqrt{2\pi \sigma^2(k)}} \int_{\delta_\text{min}}^\infty \dd \delta \, e^{-\frac{\delta^2}{2\sigma^2(k)}}
\end{align}
if the power is equal to $\mathcal{P}_\mathcal{R}(k)$ around that scale and there are negligible nongaussianities.
Assuming that the survival probability of these structures is order unity, we then find that the abundance $\Omega^\mathcal{P}_\text{sub}$ in subhalos with core mass $M_s(k)$ from eq.~\ref{eq:Msk} and a density of at least $\rho_s(z_\text{coll})$ from eq.~\ref{eq:rhoszcoll} is: 
\begin{align}
\Omega^\mathcal{P}_\text{sub} \left[M_s, \rho_s\right] = \Omega_\text{DM} \beta\left[k,\mathcal{P}_\mathcal{R}(k),z_\text{coll} \right]. \label{eq:OmegaP}
\end{align}
In other words, given a power $\mathcal{P}_\mathcal{R}(k)$ near $k$, one can calculate the corresponding core mass, and the abundance of halos denser than a given $\rho_s$. 

We have parametrized our ignorance and unknown uncertainties in the coefficients $C_M$ and $C_\rho$. In figure~\ref{fig:prim}, we have assumed that $C_M = 1$ and $C_\rho = 6\times 10^2$. These values are not much more than educated guesses. Our choice of $C_M = 1$ is based on our assumptions that the halo cores do not significantly accrete after collapse and do not grow in mergers, but also that they do not lose mass due to tidal stripping (or that these effects cancel each other out). We will assume that a subhalo of median density corresponds to a $1\sigma$ overdensity ($\beta = 0.32$). A matching at a reference virial mass of $m_{200} = 10^{6}~M_\odot$ and \textit{Planck}'s measured spectrum of ref.~\cite{ade2016planck} to the results of ref.~\cite{moline2017characterization} then fixes $C_\rho$.\footnote{Matching at lower (higher) reference virial masses leads to values of $C_\rho$ that are up 50\% larger (smaller).} These coefficients likely also depend on the precise shape of the power spectrum, which will affect formation, merger, and assembly histories. Our assumed values should thus be understood to carry large systematic errors, even at the order-of-magnitude level. For reference, if $C_\rho$ were smaller (larger) by a factor of 10, our sensitivity in terms of primordial power would degrade (improve) roughly by a factor of $10^{2/3}$. Finally, there has been some speculation in the literature~\cite{bertschinger1985self,ricotti2009new,li2012new} that a spike in the power spectrum would produce subhalos with an inner cusp of $\rho(r) \propto r^{-9/4}$, although recent studies~\cite{delos2018ultracompact} find much shallower density profiles, not steeper than $\rho(r) \propto r^{-3/2}$. We shall conservatively assume NFW profiles with $\rho(r) \propto r^{-1}$ in the inner core.

In section~\ref{sec:senshalos}, we will find projected signal-to-noise-ratio functions for NFW-shaped subhalos of the form:
\begin{align}
\text{SNR}\left[M_s, \rho_s, \Omega_\text{sub}  \right].
\end{align}
The threshold sensitivity, e.g.~$\text{SNR}=1$, defines a 2D surface in the 3D space $\lbrace M_s, \rho_s, \Omega_\text{sub} \rbrace$. Alternatively, for a given $M_s$, it defines a detectable threshold $\delta \Omega_\text{sub}$ at that mass scale, which is a monotonically decreasing function of $\rho_s$. If, for the same $M_s$ and some $\mathcal{P}_\mathcal{R}(k)$, this function intersects the one from eq.~\ref{eq:OmegaP}---i.e.~$\delta \Omega_\text{sub} = \Omega^\mathcal{P}_\text{sub}$---for \emph{some} value of $\rho_s$, then that $\mathcal{P}_\mathcal{R}(k)$ is detectable at that scale. The function $\Omega^\mathcal{P}_\text{sub}$ is monotonically decreasing with $\rho_s$ and increasing with $\mathcal{P}_\mathcal{R}(k)$, so this defines a unique minimum detectable power. It is this threshold detectable power that is plotted in figure~\ref{fig:prim}, assuming $R_\text{sub} \approx 8~\kpc$. Those sensitivity curves are direct mappings of those in figure~\ref{fig:halosens} to be presented in section~\ref{sec:sensitivity}.

\subsection{Exotic objects}\label{sec:exoticobjects}
While extended halos are a standard feature of cold dark matter cosmology, compact objects are a generic class of DM objects that can arise from a variety of different physics processes. For a point-like object, the enclosed mass is the total mass, $M(b_{il}) = M_l$, and the time-domain lensing signals are
\begin{align}
\Delta \dot{\vect{\theta}}_{il} = \left( 1 - \frac{D_l}{D_i}\right) \frac{4G_N M_l v_{il}}{b_{il}^2} \vect{P}_1\left[\hat{\vect{v}}_{il}, \hat{\vect{b}}_{il} \right], \quad
\Delta \ddot{\vect{\theta}}_{il} = \left( 1 - \frac{D_l}{D_i}\right) \frac{4G_N M_l v_{il}^2}{b_{il}^3} \vect{P}_2\left[\hat{\vect{v}}_{il}, \hat{\vect{b}}_{il} \right],
\end{align}
which have dipole- and quadrupole-like vector profiles of
\begin{align}
\vect{P}_1\left[\hat{\vect{v}}, \hat{\vect{b}} \right] &\equiv - \hat{\vect{v}} + 2 \hat{\vect{b}} \left( \hat{\vect{v}} \cdot \hat{\vect{b}}\right), \quad
\vect{P}_2\left[\hat{\vect{v}}, \hat{\vect{b}} \right] &\equiv  4\hat{\vect{v}} \left( \hat{\vect{v}} \cdot \hat{\vect{b}}\right) + 2 \hat{\vect{b}}\left[1-4 \left( \hat{\vect{v}} \cdot \hat{\vect{b}}\right)^2\right].
\end{align}
The lens deflection angle $\Delta \vect{\theta}_{il}$ can be read off directly from eq.~\ref{eq:lensangle}.

\paragraph{Primordial black holes}
The canonical example of a dark compact object is a primordial black hole (PBH)~\cite{carr1975primordial}, which can arise after an $\mathcal{O}(1)$ density perturbation in the early Universe collapses quickly after entering the horizon. Such perturbations could arise if there is a spike or other increase in the primordial power spectrum at small scales, such as in hybrid inflation models~\cite{garcia1996density}.
Slightly smaller perturbations can collapse also during radiation domination into supermassive dark matter clumps (SDMCs) \cite{berezinsky2013formation}, which from our perspective would also be compact objects. Both PBHs and SDMCs seed dense halos that may also contribute, or even dominate, the lensing signatures. Translating our sensitivity projections to these objects depends on the form of their density profiles, and is left for future work.
However, large primordial fluctuations are constrained even at most of the small scales under consideration~\cite{chluba2012probing,josan2009generalized}.

\paragraph{Dark stars} 
Compact objects can form at later times through a variety of dissipative mechanisms and attractive dynamics. A concise review of many of these mechanisms may be found in ref.~\cite{giudice2016hunting}, some key results of which we summarize here.

Bosonic fields can condense into stars (see \cite{jetzer1992boson,liddle1992structure,schunck2003general,liebling2012dynamical} for reviews). Free-field configurations have a maximum mass
\begin{equation}
	M_\text{max}  \approx \frac{0.633}{G_N m} \approx 1.3 \times 10^5~M_{\odot} \left[\frac{10^{-15}~\text{eV}}{m}\right]
\end{equation}
and a relationship between their total mass $M$ and radius $R$
\begin{align}
	R \sim \frac{1}{G_N M m^2} \sim 10^{-4}~\text{pc} \left[\frac{1~M_\odot}{M}\right]\left[\frac{10^{-15}~\text{eV}}{m}\right]^2,\label{eq:massradius}
\end{align}
where $m$ is the mass of the scalar field~\cite{kaup1968klein}. If the field has a quartic interaction $\lambda$, then the maximum mass is~\cite{colpi1986boson}
\begin{equation}
	M_\text{max} \sim M_\odot \sqrt{\lambda} \left(\frac{30~\text{MeV}}{m}\right)^2.
\end{equation}
In principle, there are thus compact equilibrium solutions in a wide range of masses observable to our techniques, though whether plausible formation histories exist requires further study in most cases. One example with a motivated formation mechanism are so-called ``axion miniclusters'', which can be produced from axion isocurvature perturbations~\cite{hogan1988axion,kolb1993axion}, and may eventually form dark stars in equilibrium of the type in eq.~\ref{eq:massradius}. The resulting objects today would be relatively compact objects whose masses can span a large range~\cite{fairbairn2017structure}.

Fermionic dark matter can contain dissipative interactions in a variety of theoretical frameworks~\cite{berezhiani2004mirror, chacko2006natural,fan2013dark,fan2013double,foot2015dissipative}. In such scenarios, it is possible for dark matter to condense and form dark stars. In mirror matter models, these stars are analogous to our own, and can be supported by radiation or degeneracy pressures~\cite{kouvaris2015asymmetric}. Such strongly dissipative dark matter is generally constrained to make up only a few percent of the total dark matter abundance~\cite{cyr2014constraints}. 
In general, dissipative dark matter can cool like ordinary matter, and form a variety of compact structures that can lead to lensing signals~\cite{agrawal2017point,buckley2018collapsed}. It is possible that primordial dark matter clouds collapse and then fragment, yielding high-density ``subhalos'' akin to ordinary globular clusters but instead composed of dark compact objects.

Dark matter scattering is known to flatten cores of dwarf galaxies~\cite{spergel2000observational}, but at high values can lead to a ``gravithermal catastrophe'', forming very cuspy halos with black holes in their cores~\cite{balberg2002self,pollack2015supermassive}. Such phenomena are tightly constrained for the dominant dark matter component, but the collapse would still occur in the centers of dark matter halos even if only a subdominant fraction interacts strongly~\cite{pollack2015supermassive}.

\subsection{Outer Solar System planets}\label{sec:planets}
There may yet be undiscovered planets in wide orbits in our own Solar System. Although they could be reasonably bright in reflected sunlight or infrared emission, they are difficult to spot in long exposures because of their large proper motion and a-priori-unknown path. A planet with an orbital radius of 1000~AU would undergo parallax motion with an amplitude of order 0.1~deg and orbital motion of order 0.01~deg per year, numbers which scale with orbital radius as $1/R$ and $1/R^{3/2}$, respectively. Even a planet as small as 10 Earth masses can produce microarsecond-level lens deflections at impact parameters smaller than 0.1~AU, while the impact parameter would undergo changes of about 2~AU every 6 months, primarily due to Earth's motion. The small angular deflections of the fixed stars, in front of which this hypothetical planet would appear to travel, could be detected in aggregate, as we discuss in section~\ref{sec:planetnine}.

Intriguingly, there is evidence pointing to the existence of a planet with mass $M\sim 10~M_\oplus$ orbiting at $R\sim 1000~\text{AU}$. It was originally noted in ref.~\cite{trujilloplanet9} that objects with orbital semimajor axes greater than 150~AU and perihelia beyond Neptune had clustered perihelion arguments, while objects with smaller perihelia had random arguments. They argued that an interaction with a remote, super-Earth-mass object could produce such a clustering. Later, it was shown in ref.~\cite{batygin2016evidence} that the clustering was not only apparent in the properties of the perihelia, but also in the orbital planes of these objects. The combined effects have a low $p$-value of  $0.007\%$, providing statistical support for a super-Earth orbiting the Sun with a semimajor axis around 700~AU. Finally, ref.~\cite{becker2017evaluating} found that orbits of eight trans-Neptunian objects would naturally be destabilized by Neptune; their survival could be explained by the stabilizing influence of a planet with grossly similar characteristics to those preferred by refs.~\cite{trujilloplanet9}~and~\cite{batygin2016evidence}. 

Regardless of this set of anomalies, it is clear that observational astronomy has not yet satisfactorily explored regions beyond the Kuiper Belt. We think variable astrometry will become a valuable tool to determine the presence of massive objects in the outer reaches of the Solar System---well into the Oort Cloud and perhaps all the way to the Solar System's cosmographic boundary. 

%%%%%%%%%%%%%%%%%%%%%%%%%%%%%%%%%%%%%%%%%%%%%%%%%%%%%%%%%%%%%%%%%%%
\section{Signal observables}\label{sec:signal}

In this section, we introduce precise definitions of observables sensitive to time-domain gravitational lensing of background light sources. The basic idea of each signal observable is illustrated in figure~\ref{fig:signals}.
We propose three basic classes of signal observables, based on
\begin{itemize}
\item single sources: outliers and mono-blips for rare, isolated lensing events by point-like lenses (subsection~\ref{sec:outliers});
\item multiple sources: multi-blips for point-like lenses traversing large arcs on the sky (subsection~\ref{sec:outliers}), and templates for extended lenses (subsection~\ref{sec:templates});
\item global correlations: small-angle excesses in the two-point function of apparent velocities and accelerations, integrated over a large field of view (subsection~\ref{sec:correlations}).
\end{itemize}
We discuss mono- and multi-blips together in subsection~\ref{sec:outliers} as they can be naturally merged into one generalized ``blip observable'' of which mono-blips arise as a limiting case. 

\subsection{Outliers and blips}\label{sec:outliers}
In this subsection, we discuss signal observables appropriate for point-like lenses, i.e.~those with negligible spatial extent. At a minimum, this would apply to objects such as PBHs or dark stars but is relevant whenever the size $r_s$ of the object is smaller than the typical change in impact parameter over the lifetime $\tau$ of the astrometric mission:
\begin{align}
r_s \ll v_{il} \tau \approx 8.5 \times 10^{-4}~\pc \left[\frac{v_{il}}{\sigma_{v_l}} \right] \left[\frac{\tau}{5~\text{y}} \right]. \label{eq:pointcond}
\end{align}
For lensing by dark matter objects, we will often pick the local velocity dispersion of DM, $\sigma_{v_l} \approx 166~\kms$, as a typical rate of change in the impact parameter.
We further categorize this regime into two subregimes:
\begin{align}
\text{outliers:}~v_{il}\tau \lesssim b_{il}; \qquad \qquad \text{blips:~}~v_{il}\tau \gtrsim b_{il}. \label{eq:outlierblipcond}
\end{align}
In the ``outlier regime'', the fractional change in impact parameter over the astrometric mission lifetime $\tau$ can be neglected, whereas in the  ``blip'' regime, it is large.
One can always look for outliers; blips will only occur for sufficiently numerous source-lens pairs, such that one can expect to find a sufficiently small $b_{il}$ among all pairs.

\paragraph{Outliers}
Suppose the proper motion of a typical lens relative to the background stars is small enough such that it traverses an arc on the sky much smaller than the smallest angular separation between any source-lens pair, yet still large enough relative to the size of the lens to be effectively point-like,
\begin{align}
r_s \ll v_{il} \tau \lesssim \left\langle \min_{i,l} b_{il} \right\rangle.
\end{align}
In that case, the best local observables of gravitational lensing are outliers, namely anomalously large angular velocities or accelerations. 
In other words, the largest angular velocity among the sources relative to their expected noise, or the largest noise-weighted angular acceleration, namely:
\begin{align}
\mathcal{O}_{\mu} \equiv \max_i \frac{\dot{\vect{\theta}}_i^2}{2\sigma_{\mu,i}^2} \quad \text{and} \quad \mathcal{O}_{\alpha} \equiv \max_i \frac{\ddot{\vect{\theta}}_i^2}{2\sigma_{\alpha,i}^2}
\end{align}
are good test statistics to hunt for weak lensing by compact objects. 

The expected largest velocity outlier due to lensing by objects of mass $M_l$, uniformly distributed in space with energy density $\rho_l$, over a field of stars with constant angular number density $\Sigma_0$ and angular area $\Delta \Omega$, is:
\begin{align}
\left\langle \mathcal{O}_\mu \right\rangle &\simeq \left\langle \max_{i,l} \left[\left(1-\frac{D_i}{D_l}\right)\frac{4G_N M_l v_{il}}{b_{il}^2} \right]^2 \frac{1}{2\sigma_{\mu,i}^2} \right\rangle \simeq \frac{(4G_N M_l)^2 \sigma_{v_l}^2}{\sigma_{\mu,\text{eff}}^2} \left\lbrace\frac{1}{\left\langle \min_{i,l} b_{il}^2 \right\rangle } \right\rbrace^2 \\
&\simeq \frac{(4G_N M_l)^2 \sigma_{v_l}^2}{\sigma_{\mu,\text{eff}}^2} \left\lbrace\min\left[\frac{\pi \rho_l D_i \Sigma_0 \Delta \Omega}{M_l}, \frac{1}{\sigma_{v_l}^2 \tau^2},\frac{1}{r_s^2} \right]\right\rbrace^2. \label{eq:outliervelocity}
\end{align}
Above, we have ignored the $(1-D_l/D_i)^2$ factor, assumed that $\langle v_{il}^2 \rangle = 2\sigma_{v_l}^2$, and defined $\sigma_{\mu,\text{eff}}^2 \equiv \langle \sigma_{\mu,i}^{-2} \rangle^{-1}$. The minimum lens-source impact parameter over the whole field of sources and lenses is cut off at $\sigma_{v_l} \tau$ or at $r_s$, whichever is lower. For impact parameters less than $\sigma_{v_l} \tau$, one enters the blip regime, which is discussed below.  Completely analogously, one finds the expected largest acceleration outlier:
\begin{align}
\left\langle \mathcal{O}_\alpha \right\rangle \simeq 4 \frac{(4G_N M_l)^2 \sigma_{v_l}^4}{\sigma_{\alpha,\text{eff}}^2} \left\lbrace\min\left[\frac{\pi \rho_l D_i \Sigma_0 \Delta \Omega}{M_l}, \frac{1}{ \sigma_{v_l}^2 \tau^2}, \frac{1}{r_s^2} \right]\right\rbrace^{3}, \label{eq:outlieracceleration}
\end{align}
where $\sigma_{\alpha,\text{eff}}^2 \equiv \langle \sigma_{\alpha,i}^{-2} \rangle^{-2}$. 
To estimate the sensitivity, one has to model what the distributions of $\mathcal{O}_\mu$ and $\mathcal{O}_\alpha$ are. Unfortunately, these outlier observables are prone to systematics and heavy tails in their distribution, as we will discuss in section~\ref{sec:background}. For estimates, we will take $\left\langle \mathcal{O}_\mu \right\rangle = \left\langle \mathcal{O}_\alpha \right\rangle = 10^2$ as fiducial detection thresholds. The figures of merit that quantify the effective noise for $\mathcal{O}_\mu$ and $\mathcal{O}_\alpha$ are:
\begin{align}
\text{FOM}_{\mathcal{O}_\mu} = \frac{\sigma_{\mu,\text{eff}}}{D_i^{1/2} \Sigma_0^{1/2} \Delta \Omega^{1/2}}, \quad \text{FOM}_{\mathcal{O}_\alpha} = \frac{\sigma_{\alpha,\text{eff}}}{D_i^{1/2} \Sigma_0^{1/2} \Delta \Omega^{1/2}}, \label{eq:FOMO}
\end{align}
and are to be minimized.

\paragraph{Blips}
If the typical change in impact parameter is larger than the smallest expected separation between any source-lens pair (as well as the size $r_s$ of the lens), 
\begin{align}
\left\langle \min_{i,l} b_{il} \right\rangle  \lesssim v_{il} \tau , \quad r_s \ll v_{il} \tau,
\end{align}
one can observe the full nonlinear lensing displacement trajectory: a ``blip'' in the otherwise normal proper motion of a background source. Suppose one hypothesizes that there exists a lens with a linear path $\vect{x}_l(t)$ and $\vect{v}_l = \dot{\vect{x}}_l = \text{constant}$; to test whether this path is consistent with observations, we construct the blip test statistic: 
\begin{align}
\mathcal{B}\left[\vect{x}_l(t)\right] \equiv \sum_{i\in \Box} \frac{1}{\sigma_{\delta\theta,i}^2} \sum_n \delta \vect{\theta}_i\left(t_n\right) \cdot \Delta \vect{\theta}_{il}\left[\vect{x}_l(t_n)\right].
\end{align}
Here $\sum_{i\in \Box}$ signifies that one is to sum over all sources $i$ with a minimum impact parameter less than $v_{il} \tau / 2$ (the square box in figure~\ref{fig:signals}). The sum takes the angular position data residuals $\delta \vect{\theta}_i\left(t_n\right)$ at observation times $t_n = 0,f_\text{rep}^{-1}, 2 f_\text{rep}^{-1},\dots,\tau$ as input ($f_\text{rep}^{-1}$ is the typical time between observations), and takes the inner product with the lensing correction prediction $\Delta \vect{\theta}_{il}\left[\vect{x}_l(t_p)\right]$ of eq.~\ref{eq:lensangle} given the lens path $\vect{x}_l(t)$. We weight the terms in the sum by the inverse expected variance in residual angular position error $\sigma_{\delta\theta,i}^2$. This variance is the noise \emph{per epoch}, i.e.~at each of the individual observation times $t_n$.

If one guesses the lens path $\vect{x}_l(t)$ correctly, then the expectation value of the data residuals $\delta \vect{\theta}_i$ equals the lens-induced prediction. In that case, we have
\begin{align}
\left \langle \mathcal{B}\left[\vect{x}_l(t)\right] \right \rangle = \left\langle \sum_{i\in \Box} \frac{1}{\sigma_{\delta\theta,i}^2} \sum_n \Delta \theta_{il}^2\left[\vect{x}_l(t_n)\right] \right\rangle \simeq \left( \frac{4G_N M_l}{\sigma_{\delta\theta,\text{eff}}} \right)^2 \left\langle \sum_{i\in \Box} \frac{\pi f_\text{rep}}{v_{il} b_{il}^*} \right \rangle, \label{eq:blipexp}
\end{align}
where $b_{il}^*$ is the minimum impact parameter between the source $i$ and the lens path $\vect{x}_l(t)$, and
\begin{align}
\sigma_{\delta \theta,\text{eff}}^2 \equiv \left \langle \sigma_{\delta \theta,i}^{-2} \right\rangle^{-1} \label{eq:sigmatheta}
\end{align}
is the effective noise appropriately averaged over all stars.
For the second equality in eq.~\ref{eq:blipexp}, we have assumed negligible size $r_s \ll v_{il}\Delta \tau$, and marginalized over the possible discrete observation times taken at regular intervals $f_\text{rep}^{-1}$, as the signal is larger if the minimum impact parameter with the path occurs near an observation time $t_{n}$ rather than in between two observation times $t_{n}$ and $t_{n+1}$. It is straightforward to check that the noise power is $\text{Var}\, \mathcal{B}\left[\vect{x}_l(t)\right] = \left \langle \mathcal{B}\left[\vect{x}_l(t)\right] \right \rangle$, such that any lens path will have an expected signal to noise ratio of $\left \langle \mathcal{B}\left[\vect{x}_l(t)\right] \right \rangle^{1/2}$. There will generally be many lenses in the sky; the largest local signal to noise ratio produced by this ensemble of lenses is thus:
\begin{align}
\left \langle \max_l \text{SNR}_{\mathcal{B}[\vect{x}_l(t)]} \right\rangle \simeq \frac{4G_N M_l}{\sigma_{\delta \theta,\text{eff}}}  \left\langle \max_l \sum_{i\in \Box} \frac{\pi f_\text{rep}}{v_{il}  b_{il}^* } \right \rangle^{1/2}. \label{eq:blipSNR}
\end{align}
There are two ways in which the signal can be large compared to the noise. A lens can pass very close by a source (small $b_{il}^*$), causing a large ``mono-blip''. Alternatively, a lens can pass nearby many sources (large $\sum_{i \in \Box}$), generating a ``multi-blip'' signal. We illustrate the two behaviors in figure~\ref{fig:signals}. Typically, multi-blips will be better for nearby lenses (especially ones in our own Solar System), while mono-blips are more suitable for a rare population of point lenses (such that the closest one will still have a large line-of-sight distance).
We shall see that the noise figures of merit for mono- and multi-blips are roughly:
\begin{align}
\text{FOM}_{\mathcal{B}^\text{mono}} = \frac{\sigma_{\delta \theta, \text{eff}}}{D_i^{1/2} \Sigma_0^{1/2} \Delta \Omega^{1/2}}, \quad 
\text{FOM}_{\mathcal{B}^\text{multi}} = \frac{\sigma_{\delta \theta, \text{eff}}}{ \Sigma_0^{1/2} }. \label{eq:FOMB}
\end{align}
Stellar targets with the lowest FOMs will provide the best detection prospects.

\subsection{Templates}\label{sec:templates}

As we have seen in section~\ref{sec:summary}, the finite size of a gravitational lens generally suppresses the magnitude of the lensing corrections to stellar motions. However, part of this suppression can be recuperated by exploiting the fact that a finite-size lens can induce correlated lensing shifts in \emph{many stars at the same time}. 
One can test if a field of stellar motions is locally consistent with the hypothesis that there is a lens with a certain density profile, velocity, and angular size along the line of sight in that location on the sky.

We introduce a local test statistic $\mathcal{T}_\mu$, which takes stellar location and velocity data $\lbrace \vect{\theta}_i,\dot{\vect{\theta}}_i\rbrace$ as input, and tests for consistence with the existence of a lens with location $\vect{\theta}_t$, angular size $\beta_t$, and velocity direction $\hat{\vect{v}}_t$. We assume that in absence of lensing, the angular velocity $\dot{\vect{\theta}}_i$ of each star is a random variable drawn from an independent gaussian distribution with \emph{known} variance $\sigma_{\mu,i}^2$ (which depends on the stellar magnitude, color, and age; see section~\ref{sec:background}).
The test statistic $\mathcal{T}_\mu \left[\vect{\theta}_t, \beta_t, \hat{\vect{v}}_t \right]$ takes dot products of the observed velocities with the predicted stellar velocity profile $\vect{\mu}_t$ if there were a lens of angular size $\beta_t$ at $\vect{\theta}_t$ moving in a direction $\hat{\vect{v}}_t$, inversely weighted by the expected standard deviation:
\begin{align}
\mathcal{T}_\mu \left[\vect{\theta}_t, \beta_t, \hat{\vect{v}}_t \right] = \sum_i   \frac{\dot{\vect{\theta}}_i}{\sigma_{\mu,i}^{2} }\cdot \vect{\mu}_t \left[ \frac{\beta_{it}}{\beta_t}, \hat{ \vect{\beta}}_{it} , \hat{\vect{v}}_{t}\right] ; \quad \vect{\beta}_{it} \equiv \vect{\theta}_t -  \vect{\theta}_i.
\end{align}
For example, to perform a test for an NFW profile, one would take $\vect{\mu}_t = \vect{G}_1$ from eq.~\ref{eq:G1}. It can be shown that $\mathcal{T}_\mu$ is optimal (in the sense that it is a maximum-likelihood estimator) if the velocity noise in the stellar motions is expected to be spatially uniform, uncorrelated, and gaussian distributed. Analogously, one can also construct a test statistic based on an acceleration template $\vect{\alpha}_t$:
\begin{align}
\mathcal{T}_\alpha \left[ \vect{\theta}_t, \beta_t, \hat{\vect{v}}_t \right] = \sum_i  \frac{\ddot{\vect{\theta}}_i}{\sigma_{\alpha,i}^2} \cdot \vect{\alpha}_t \left[ \frac{\beta_{it}}{\beta_t}, \hat{ \vect{\beta}}_{it} , \hat{\vect{v}}_{t}\right],
\end{align}
where $\sigma^2_{\alpha,i}$ is the expected acceleration variance. For e.g.~an NFW subhalo, one would choose the template $\vect{\alpha}_t = \vect{G}_2$ from eq.~\ref{eq:G2}.
From here on, we will focus mostly on the velocity test statistic $\mathcal{T}_\mu$ rather than the acceleration test statistic $\mathcal{T}_\alpha$ as it will turn out to provide more promising prospects for detection, but all of the statements for $\mathcal{T}_\mu$ below have exact analogues for $\mathcal{T}_\alpha$.

In absence of lensing, one would expect no correlation of stellar angular velocities with any chosen lens template, so that $\langle \mathcal{T}_\mu \rangle = 0$. For concreteness, we will assume that the mean velocity is zero (or subtracted) $\langle \dot{\vect{\theta}}_i \rangle = 0$, that there is no covariance, i.e.~$\langle \dot{\vect{\theta}}_i \cdot \dot{\vect{\theta}}_j \rangle = 0$ for $i \neq j$, and that the variance $\langle \dot{\theta}_{i,x}^2 \rangle = \langle \dot{\theta}_{i,y}^2 \rangle = \sigma_{\mu,i}^2$ is independent of position but may depend on other characteristics such as apparent magnitude. In a spatially uniform field of stars with an angular number density of $\Sigma_0$, we can then compute the variance of $\mathcal{T}_\mu$:
\begin{align}
\text{Var} \, \mathcal{T}_\mu  =  \frac{\Sigma_0}{\sigma_{\mu,\text{eff}}^2} \int \dd^2 \beta_{it} \, \left| \vect{\mu}_t \left[ \frac{\beta_{it}}{\beta_t}, \hat{ \vect{\beta}}_{it} , \hat{\vect{v}}_{t}\right]  \right|^2; \quad \sigma_{\mu,\text{eff}}^2 \equiv \left\langle \sigma_{\mu,i}^{-2} \right\rangle^{-1}.
\end{align}
The quantity $\langle \sigma_{\mu,i}^{-2}\rangle$ should be read as the average inverse variance over the chosen stellar population; i.e.~$\sigma_{\mu,i}$ is not a random variable.
With the above assumptions, $\text{Var} \, \mathcal{T}_\mu $ should be independent of $\vect{\theta}_t$ and $\hat{\vect{v}}_t$ but not generally $\beta_t$. 

In the presence of a gravitational lens $l$, one expects that $\langle \dot{\vect{\theta}}_i \rangle = \Delta \dot{\vect{\theta}}_{il}$, so we can get nonzero template overlap:
\begin{align}
\langle \mathcal{T}_\mu \rangle &= \frac{\Sigma_0}{\sigma_{\mu,\text{eff}}^2} \int \dd^2 \beta_{it} \, \langle \dot{\vect{\theta}}_i \rangle \cdot\vect{\mu}_t \left[ \frac{\beta_{it}}{\beta_t}, \hat{ \vect{\beta}}_{it} , \hat{\vect{v}}_{t}\right] = \frac{\Sigma_0}{\sigma_{\mu,\text{eff}}^2}  \Delta \dot{\theta}_l \beta_t^2 \int \dd^2 \frac{\beta_{it}}{\beta_t} \, \left| \vect{\mu}_t \left[ \frac{\beta_{it}}{\beta_t}, \hat{ \vect{\beta}}_{it} , \hat{\vect{v}}_{t}\right]  \right|^2.
\end{align}
In the second equality we have assumed that there is a lens $l$ which perfectly matches the template in that location: $ \Delta \dot{\vect{\theta}}_{il}  =  \Delta \dot{\theta}_{l} \vect{\mu}_t\left[ {\beta_{it}}/{\beta_t}, \hat{ \vect{\beta}}_{it} , \hat{\vect{v}}_{t} \right] $, where $\Delta \dot{\theta}_{l}$ quantifies the size of the lens velocity correction.
As we have seen in section~\ref{sec:summary}, the lensing shift to the stellar velocity depends on the effective velocity of the light path's impact parameter, $\vect{v}_{il} = \vect{v}_l - (D_l / D_i) \vect{v}_i - (1-D_l / D_i) \vect{v}_\odot$, which is to be matched to the template velocity direction $\vect{v}_t$. 
The local signal-to-noise ratio for a perfectly matched template is then:
\begin{align}
\text{SNR}_{\mathcal{T}_\mu} = \frac{ \langle \mathcal{T}_\mu \rangle }{ \sqrt{\text{Var} \, \mathcal{T}_\mu } } = \frac{\Delta \dot{\theta}_l}{\sigma_{\mu,\text{eff}}} \sqrt{\Sigma_0 \beta_t^2}  \sqrt{ \int \dd^2 \frac{\beta_{it}}{\beta_t} \, \left| \vect{\mu}_t \left[ \frac{\beta_{it}}{\beta_t}, \hat{ \vect{\beta}}_{it} , \hat{\vect{v}}_{t}\right]  \right|^2} \label{eq:SNRTmu}
\end{align}
before accounting for look-elsewhere effects. If we choose a normalization such that $\vect{\mu}_t$ has $\mathcal{O}(1)$ norm over an angular area of $\mathcal{O}(\beta_t^2)$, one can expect a detection when the typical lens velocity shift $\Delta \dot{\theta}_l$ is larger than the \emph{averaged-down} noise, which is of order $\sigma_{\mu,\text{eff}} / \sqrt{\Sigma_0 \beta_t^2}$.

The figure of merit (FOM), the quantity that maximizes the local signal-to-noise ratio of the velocity template, is proportional to
\begin{align}
\text{FOM}_{\mathcal{T}_\mu} \equiv \frac{\sigma_{\mu,\text{eff}}}{\Sigma_0^{1/2} \Delta \Omega^{1/3}}, \label{eq:FOMTmu}
\end{align}
which itself is to be minimized. As we will discuss in section~\ref{sec:sensitivity}, the factor of angular area $\Delta \Omega$ arises because the maximum expected lens size is expected to be proportional to the third root of the angular area of the source target: $\langle \max_l \beta_s \rangle \propto \Delta \Omega^{1/3}$. In section~\ref{sec:background}, we discuss various promising source target populations which have a low $\text{FOM}_{\mathcal{T}_\mu}$.

\subsection{Correlations}\label{sec:correlations}

In some cases, even when localized lens shifts to stellar motion are too small to observe, it is still possible to measure the small-scale correlations that gravitational lensing induces on a field of stellar motions. For this purpose, we introduce a global test statistic $\mathcal{C}_\alpha$ that measures small-angle correlations in angular accelerations:
\begin{align}
\mathcal{C}_\alpha[\beta_-,\beta_+,\delta] = \frac{1}{2} \sum_{i \neq j} \frac{\ddot{\vect{\theta}}_i \cdot \ddot{\vect{\theta}}_j}{\sigma_{\alpha,i}^2\sigma_{\alpha,j}^2}  \frac{B[\beta_{ij}; \beta_-, \beta_+]}{\beta_{ij}^\delta} \label{eq:Calpha}
\end{align}
with $B[\beta_{ij}; \beta_-, \beta_+] = 1$ if $\beta_-< \beta_{ij} < \beta_+$ and zero otherwise, and $0 < \delta < 1$. This observable takes as input data all star pairs with $\beta_{ij} \equiv |\vect{\theta}_i - \vect{\theta}_j|$ between the angular scales of $\beta_-$ and $\beta_+$, and computes dot products of their accelerations inversely weighted by their expected acceleration noise and angular distance (raised to the power $\delta>0$). Likewise, we can construct a similar test statistic for correlations in angular velocities:
\begin{align}
\mathcal{C}_\mu[\beta_-,\beta_+,\delta] = \frac{1}{2} \sum_{i \neq j} \frac{\dot{\vect{\theta}}_i \cdot \dot{\vect{\theta}}_j}{\sigma_{\mu,i}^2 \sigma_{\mu,j}^2}  \frac{B[\beta_{ij}; \beta_-, \beta_+]}{\beta_{ij}^\delta}. \label{eq:Cmu}
\end{align}
Below, we will work out the details for $\mathcal{C}_\alpha$. The extension to $\mathcal{C}_\mu$ is obvious: accelerations are replaced by velocities, and $\alpha \leftrightarrow \mu$. We will keep the angular weighting exponent $\delta$ and the angular cutoffs $\beta_-$ and $\beta_+$ as general parameters for now, as their optimal values are somewhat model dependent; we will optimize over them in section~\ref{sec:sensitivity}. 

Let us first compute the expected variance of the correlation test statistic, in absence of lensing. Assuming that $\langle \ddot{\vect{\theta}}_i \cdot \ddot{\vect{\theta}}_j \rangle = 0$ for $i \neq j$, we immediately have that $\langle \mathcal{C}_\alpha \rangle = 0$. Taking $\langle \ddot{\theta}_{i,x}^2 \rangle = \langle \ddot{\theta}_{i,y}^2 \rangle = \sigma_{\alpha,i}^2$, we have
\begin{align}
\text{Var} \, \mathcal{C}_\alpha = \left \langle \sum_{i \neq j} \sigma_{\alpha,i}^{-2} \sigma_{\alpha,j}^{-2} \frac{B[\beta_{ij}; \beta_-, \beta_+]}{\beta_{ij}^{2\delta}} \right\rangle.
\end{align}
For $\beta_+$ smaller than the angular size of the region of interest, populated with stars of uniform angular number density $\Sigma_0$, this then becomes
\begin{align}
\text{Var} \, \mathcal{C}_\alpha = \frac{2\pi}{2-2\delta}  \frac{\Sigma_0^2 \Delta\Omega}{\sigma_{\alpha,\text{eff}}^4} \left[\beta_+^{2-2\delta}-\beta_-^{2-2\delta}\right], \quad \sigma_{\alpha,\text{eff}}^2 \equiv \left\langle \sigma_{\alpha,i}^{-2} \right\rangle^{-1}, \label{eq:VarCalpha}
\end{align}
for a region of angular area $\Delta \Omega$ with $N_0 = \Sigma_0 \Delta \Omega$ stars in it.

Gravitational lensing will cause correlated accelerations of nearby star pairs behind the lens. If the mutual separation between the star pairs $\beta_{ij}$ is smaller than either lensing impact parameter $\beta_{il}$ or $\beta_{jl}$, then the stellar acceleration lensing shifts will point in the same direction, as we illustrate in figure~\ref{fig:signals}. Explicitly,
\begin{align}
\langle \mathcal{C}_\alpha \rangle =  \frac{\pi\Sigma_0^2 \Delta \Omega}{\sigma_{\alpha,\text{eff}}^4} \int_{\beta_-}^{\beta_+} \dd \beta_{ij}\, \beta_{ij}^{1-\delta} \big \langle \ddot{\vect{\theta}}_i \cdot \ddot{\vect{\theta}}_j \big\rangle_{\beta_{ij}} \simeq  \frac{\pi \Sigma_0^2 \Delta \Omega}{\sigma_{\alpha,\text{eff}}^4} \int_{\beta_-}^{\beta_+} \dd \beta_{ij}\, \beta_{ij}^{1-\delta} \big \langle \sum_l \big| \Delta\ddot{\vect{\theta}}_{il}  \big|^2 \Theta(\beta_{il} - \beta_{ij} ) \big\rangle, \label{eq:Calphaexp}
\end{align}
where $\langle \ddot{\vect{\theta}}_i \cdot \ddot{\vect{\theta}}_j \rangle_{\beta_{ij}}$ is the expectation value of the dot product conditional on the two stars being separated by an angle $\beta_{ij}$.
To evaluate this quantity further, one would have to put in a specific spectrum of lenses, which we will do in section~\ref{sec:sensitivity}. However, we can already read off the figure of merit for maximizing the signal to noise ratio $\text{SNR}_{\mathcal{C}_\alpha} = \langle \mathcal{C}_\alpha \rangle / \sqrt{\text{Var}\, \mathcal{C}_\alpha }$ and the equivalent quantity $\text{FOM}_{\mathcal{C}_\mu}$ for correlated angular velocities:
\begin{align}
\text{FOM}_{\mathcal{C}_\alpha} \equiv \frac{\sigma_{\alpha,\text{eff}}^2}{\Sigma_0 \Delta \Omega^{1/2}}; \quad 
\text{FOM}_{\mathcal{C}_\mu}  \equiv \frac{\sigma_{\mu,\text{eff}}^2 }{\Sigma_0 \Delta \Omega^{1/2}} \label{eq:FOMC}
\end{align}
The relative scaling between $\Sigma_0$ and $\sigma_{\mu,\text{eff}}$ is the same as in eq.~\ref{eq:FOMTmu}, though the figure of merit for correlations is \emph{quadratically} sensitive to $\sigma_{\mu,\text{eff}}/\Sigma_0^{1/2}$, and will thus increase much faster with improved experimental sensitivity. The angular area scaling between the FOMs is also different.

%%%%%%%%%%%%%%%%%%%%%%%%%%%%%%%%%%%%%%%%%%%%%%%%%%%%%%%%%%%%%%%%%%%
\section{Backgrounds and noise}\label{sec:background}
To make robust predictions about the sensitivity of the techniques introduced in this work, we identify several potential sources of statistical and systematic noise. We will briefly describe the most promising astrometric surveys for time-domain gravitational lensing. We address how our signal observables can be safeguarded against mimicking the behavior of these backgrounds. Additional discrimination techniques are discussed in section~\ref{sec:sensitivity}.

In subsection~\ref{sec:instrument}, we summarize the instrumental limitations to the end-of-mission astrometric performance of the current \textit{Gaia} and future \textit{Theia} and SKA missions.
In subsections~\ref{sec:velocitynoise}~and~\ref{sec:accelerationnoise}, we examine non-instrumental noise sources, such as the effects of peculiar velocities in the stellar target environments, and accelerations in bound gravitational systems.
Adding the relevant noise contributions in quadrature, and using apparent magnitudes from the \textit{Gaia} DR1 data set~\cite{van2017gaia}, we ultimately obtain rough numerical estimates for the effective per-epoch positional precision $\sigma_{\delta \theta,\text{eff}}$ (see eq.~\ref{eq:sigmatheta}), average angular velocity noise $\sigma_{\mu,\text{eff}}$ (below eq.~\ref{eq:outliervelocity}), and average acceleration noise $\sigma_{\alpha,\text{eff}}$ (eq.~\ref{eq:outlieracceleration}). We estimate the quantities that enter in the figures of merit for several populations of background light sources, and list them in table~\ref{tab:prec}.

\begin{table}[h]
\begin{center}
\addtolength{\tabcolsep}{4pt} %
\begin{tabular}{
 l *{1}{S[table-format=-2.4,table-space-text-post=*]} *{1}{S[table-format=-2.3,table-space-text-post=*]} *{1}{S[table-format=-3.1,table-space-text-post=*]} *{1}{S[table-format=-3.0,table-space-text-post=*]} *{1}{S[table-format=-2.1,table-space-text-post=*]}
}
\toprule
   & {$\Sigma_0$} & {$\Delta \Omega$} & {$\sigma_{\delta\theta,\text{eff}}$} & {$\sigma_{\mu,\text{eff}}$} & {$\sigma_{\alpha,\text{eff}}$} \\
   & {$10^9~\text{rad}^{-2}$} & {$\text{rad}^2$} & {$\muas$} & {$\muasy$} & {$\muasyy$} \\
\midrule
    LMC  & 0.88 & 0.009  & 212 & 205 & 27.2  \\
                          & 8.8   & 0.009                  &  4.4 & 127 &  0.9 \\
	SMC  & 0.71 & 0.002 & 211 & 192 & 27.1  \\
    					   & 7.1 & 0.002 & 4.4  & 105 & 0.9 \\
    Disk & 4.6 & 0.2 & 73.7 & {---} & 9.4  \\
    					  & 46 &  0.2 & 1.9 & {---}  & 0.4 \\
    QSO  &  0.0001 & {$4\pi$} & 100 & 10 & 1 \\
     					& 0.01 & {$4\pi$}  & 10 & 1 & 0.1  \\
\bottomrule
\end{tabular}
\end{center}
\caption{Numerical estimates for the angular number density of sources $\Sigma_0$, angular area $\Delta \Omega$, effective standard errors for per-epoch position $\sigma_{\delta\theta ,\text{eff}}$, average angular velocity $\sigma_{\mu,\text{eff}}$, and average angular acceleration $\sigma_{\alpha,\text{eff}}$. These quantities are estimated for several background source populations: the Large and Small Magellanic Clouds (LMC, SMC), the Galactic Disk, and quasi-stellar objects (QSO). The first line for each  source population is indicative of the parameters for the \textit{Gaia} astrometric survey; the second line contains optimistic estimates for future surveys such as \textit{Theia} or SKA.} \label{tab:prec}
\end{table}

\subsection{Instrumental precision}\label{sec:instrument}

\begin{figure}[t]
\centering
\includegraphics[width=0.9\textwidth]{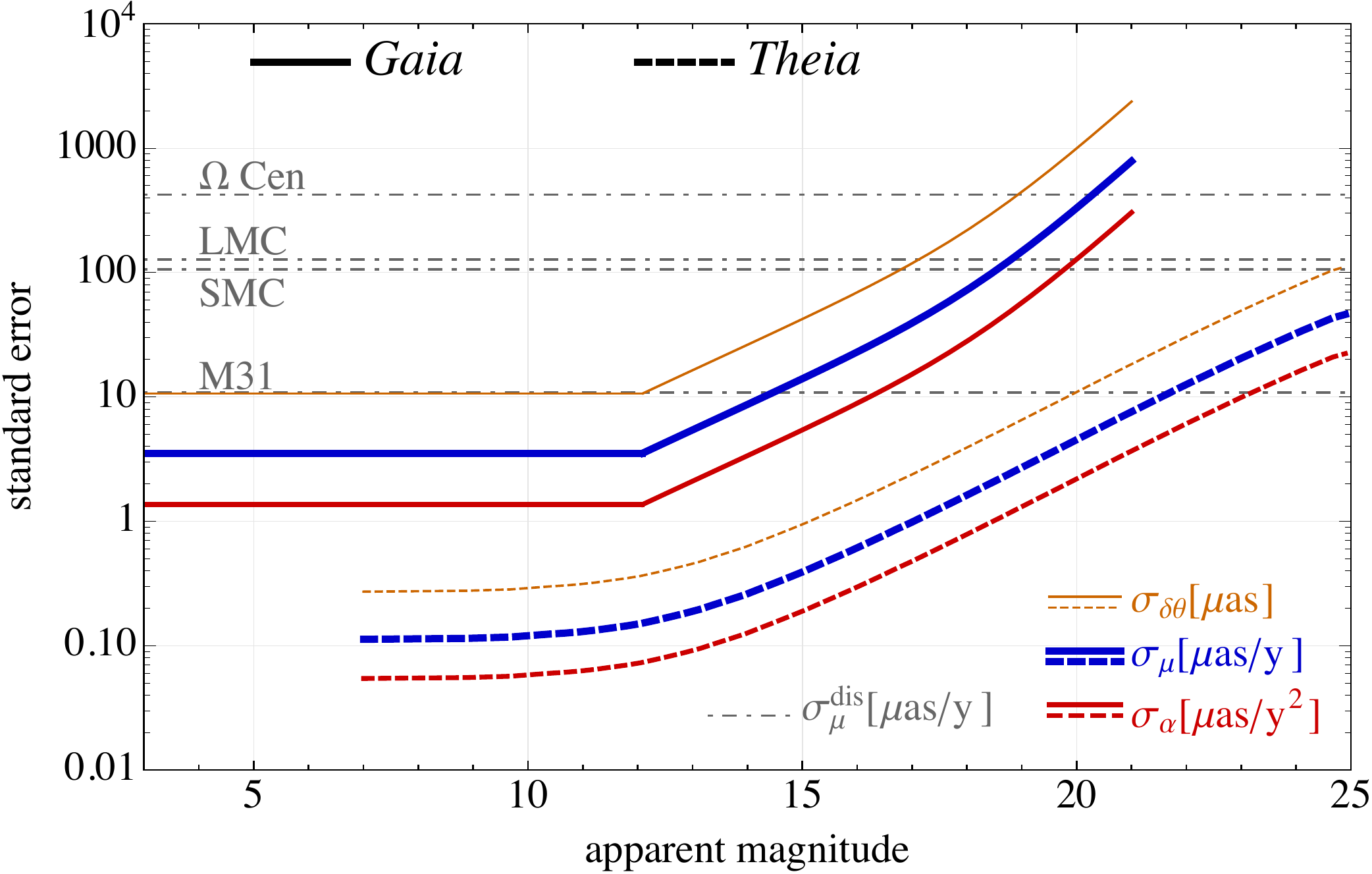}
\caption{Projected end-of-mission standard errors for per-epoch position $\sigma_{\delta \theta}$ (orange, thin), average angular velocity $\sigma_{\mu}$ (blue, thick), and average angular acceleration $\sigma_{\alpha}$ (red) in each dimension (e.g.~longitude and latitude) for mission times of $\tau=5~\text{y}$ (\textit{Gaia}, solid lines) and $\tau=4~\text{y}$ (\textit{Theia}, dashed lines). The standard errors are plotted as a function of apparent magnitude in the G band for \textit{Gaia} and in the R band for \textit{Theia}. Projections of the average proper motion precision $\sigma_\mu$ were taken from refs.~\cite{prusti2016gaia, boehm2017theia}, and rescaled using eq.~\ref{eq:noiseeffects} for $\sigma_{\delta \theta}$ and $\sigma_\alpha$. Also shown in dot-dashed gray are the intrinsic angular stellar velocity dispersions $\sigma_\mu^\text{dis}$ for the globular cluster Omega Centauri ($\Omega$ Cen), the Small and Large Magellanic Clouds (SMC, LMC), and Andromeda (M31).}
\label{fig:precision}
\end{figure}

\paragraph{Space observatories}
The \textit{Gaia} satellite was successfully launched at the end of 2013 and started taking scientific data in mid-2014. \textit{Gaia} is a spacecraft that slowly rotates around an axis perpendicular to the lines of sight of its two telescopes, which are mounted at a fixed relative angle. By measuring transit times of point sources across its focal planes, it will provide astrometry of unprecedented precision for an extensive catalog of 1.3 billion objects, in addition to photometric and spectroscopic measurements. The impending second data release will include 2D average positions $\vect{\theta}_i$ and average proper motions $\dot{\vect{\theta}}_i$, as well as line-of-sight distance $D_i$ (via parallax) for this uncatalogued. The nominal mission time was to be $\tau =  5~\text{y}$ but may be extended to $\tau =  9~\text{y}$.
\textit{Gaia} measures light in the G band (330--1050~nm), which covers the visible wavelengths, over a range of apparent magnitudes $\text{G} = 3$ to $\text{G} = 20.7$. According to the performance assessment of ref.~\cite{prusti2016gaia}, \textit{Gaia} will reach a sky-averaged proper motion standard error of $\sigma_\mu \approx 7~\muasy$ after the nominal 5-year mission time for bright stars in the range $3 \lesssim \text{G} \lesssim 12$.  The few stars that are brighter than this require special treatment, and the precision degrades for point sources fainter than $\text{G} = 12$ due to photon shot noise. We have plotted these predicted standard errors as solid lines in figure~\ref{fig:precision}.\footnote{There is a weak dependence on wavelength due to diffraction; for concreteness, we assume a chromatic index of $\text{V}-\text{I} = 0.7$ in figure~\ref{fig:precision}.} The per-epoch precision $\sigma_{\delta \theta}$ on the position of a point source is a known function of apparent magnitude and by construction does not depend on the mission time $\tau$. The average proper motion standard error $\sigma_\mu$ decreases (improves) as $\tau^{-3/2} f_\text{rep}^{-1/2}$, while that of the average angular acceleration scales as $\tau^{-5/2} f_\text{rep}^{-1/2}$. The frequency of observations (epochs) is about $f_\text{rep} \approx 14 / \text{y}$. If the mission time is extended from the nominal 5~years to 9~years, then $\sigma_\mu$ and $\sigma_\alpha$ will improve by factors of $0.41$ and $0.23$, respectively, from the ones plotted in figure~\ref{fig:precision}.

The planned \textit{Theia} mission, the natural successor to \textit{Gaia}, is to exploit technological improvements in image sensors and lessons learned from \textit{Gaia}, and promises another one or two orders of magnitude in improvement of relative astrometric precision~\cite{boehm2017theia}. It will cover nearly the same set of wavelengths (the R band, 300--1000nm), but unlike \textit{Gaia}, which constructed an unbiased uncatalogued with its ``stare-everywhere'' strategy, \textit{Theia} is set to be a more dedicated instrument, with a ``point-and-stare'' strategy. For our purposes, this flexibility offers significant advantages, as more observation time can be spent on targets that are more interesting from an astrometric lensing perspective.

A deep survey on a relatively small field of stars can allow for significant reductions in the figures of merit of several observables because of decreased statistical noise. 
Statistically-limited precisions $\sigma$ scale like $f_\text{rep}^{-1/2}$. 
Hence, on top of the improved technological capabilities, targeted observation strategies may dramatically improve statistics-limited observables such as acceleration correlations and blips. In figure~\ref{fig:precision}, we show the instrumental sensitivity as a function of R-band apparent magnitude, assuming a total observation time of 1000~hours per object spread evenly over a 4-year mission lifetime~\cite{boehm2017theia}. 

\paragraph{Long-baseline radio interferometry}
Another future survey with great potential for astrometry is the Square Kilometer Array (SKA), bound to become the world's largest and most sensitive radio telescope to date. It has a broad set of scientific goals, for which it will incrementally add instruments in three frequency bands: 70--300~MHz, 0.3--10~GHz, and 10--25+~GHz, with the objective to have a fully operational array below 10~GHz completed by 2020.

SKA should have the capability of determining the position of bright celestial objects in the radio band down to the microarcsecond level. The main limitation to the precision will likely come from tropospheric refraction, which after modeling should make a $10~\muas$ accuracy attainable for the long-baseline operating mode at 8~GHz~\cite{fomalont2004microarcsecond} for the brightest radio sources. After 10~years of operation, this would already translate to a proper motion error on the order of $1~\muasy$; SKA will likely run for much longer than a decade, with corresponding improvements.

SKA is bound to see an enormous number of extragalactic objects emitting at radio frequencies. Ref.~\cite{jarvis2015cosmology} presented estimates based on the simulations of \cite{wilman2008semi} for the predicted number of detected sources as a function of redshift and flux density threshold. At 1~GHz, a cumulative angular number density of $7 \times 10^7$ ($6\times 10^8$) is expected for a flux density threshold of $5~\mu\text{Jy}$ ($100~\text{nJy}$), providing a plenitude of sources for our purposes. Most of those radio sources are star-forming or starburst galaxies, but they also include a number density of $1.4\times 10^7$ ($1.3\times 10^8$) of radio-quiet and radio-loud quasars for a $5~\mu\text{Jy}$ ($100~\text{nJy}$) threshold. We therefore project that in the far-future, a full-sky survey may approach parameters of $\Sigma_0 \sim 10^7$ and $\sigma_{\mu,\text{eff}} \sim 1~\muasy$.

\subsection{Peculiar velocities}\label{sec:velocitynoise}

The majority of stars in the Milky Way have a velocity dispersion $\sigma_{v_i}$ that translates into an intrinsic angular velocity noise $\sigma_\mu^\text{dis} = \sigma_{v_i} / D_i$ that dwarfs the instrumental precision of the same quantity. After all, observatories such as \textit{Gaia} are meant to map out the kinematic structure of the Galaxy. For velocity-based observables, the best sensitivity will come from targets that are far away, because even a low-dispersion object such as the globular cluster Omega Centauri, which has $\sigma_{v_i}\sim 10~\text{km}/\text{s}$ and $D_i \approx 5~\text{kpc}$, has undesirably large intrinsic proper motion noise of $\sigma_\mu^\text{dis} \sim 400~\muasy$ (see figure~\ref{fig:precision}). More distant objects such as the Andromeda Galaxy (M31), which has a velocity dispersion of $\sigma_{v_i}\sim 40~\text{km}/\text{s}$ for its disk stars and is located at $D_i \approx 800~\text{kpc}$, has a much lower intrinsic angular velocity noise of $\sigma_\mu^\text{dis} \sim 10~\muasy$, though many of its stars are too faint to see. Stellar populations exhibit velocity correlations, e.g.~coherent rotational velocities, among their constituents, which is why we only find use for the velocity correlation observable $\mathcal{C}_\mu$ in extra-galactic source populations such as quasars. For the same reason, the template velocity observable $\mathcal{T}_\mu$ is only robust on angular scales much smaller than that of the stellar target, and ultimately will provide the best sensitivity on extra-galactic source targets.

\paragraph{Magellanic Clouds}
The kinematic structure of the Large Magellanic Cloud (LMC) has been studied through observations of tracer populations of stars. The magnitude of the velocity dispersion appears to be correlated with the age of the stellar populations, ranging from $6~\text{km}/\text{s}$ for young stars and $30~\text{km}/\text{s}$ for old ones~\cite{gyuk2000self}. These measurements indicate disk-like kinematics, as a much higher dispersion of $50~\text{km}/\text{s}$ would be expected in a stellar halo. There is evidence for a kinematically hot and metal-poor old halo in the inner regions of the LMC, where 43 RR Lyrae stars exhibit typical dispersions of $53\pm 10~\text{km}/\text{s}$~\cite{minniti2003kinematic}, but this old population constitutes only 2\% of the LMC mass. The bulk of the LMC disk probably consists of intermediate-age carbon stars with a dispersion of  $20\pm 0.5~\text{km}/\text{s}$~\cite{van2006large}. We adopt a representative value of $\sigma^\text{dis}_{v_i} = 30~\text{km}/\text{sec}$ in our sensitivity estimates, which translates to an intrinsic angular velocity noise of $\sigma_\mu^\text{dis} \approx 127~\muasy$ given the LMC's line-of-sight distance $D_i \approx 50~\text{kpc}$.

The Small Magellanic Cloud (SMC) appears to be largely supported by dispersion rather than rotation. A sample of 2046 red giant stars was found to have a velocity dispersion of $27.6\pm 0.5~\text{km}/\text{s}$, in rough agreement with figures for stars of very different ages~\cite{harris2006spectroscopic}. For example, a sample of radial velocities on 2045 young stars of OBA spectral type was found to have a dispersion of $30~\text{km}/\text{s}$ also~\cite{evans2008kinematics}. We therefore choose $\sigma^\text{dis}_{v_i} = 30~\text{km}/\text{sec}$ for the SMC as well, corresponding to $\sigma_\mu^\text{dis} \approx 105~\muasy$ given the SMC's distance of 60~kpc.

\paragraph{Quasi-Stellar Objects (QSOs)}
Very long baseline interferometry has been used to determine the positions of extragalactic radio sources---mostly QSOs---to define the International Celestial Reference Frame (ICRF)~\cite{ma1998international}. A key concept that allows for the practical realization of a celestial coordinate system is their stationarity: these extragalactic radio sources are, owing to their large distance, the objects that best approximate fixed reference points on the sky. Their light centroids do undergo some proper motion, however, as they are violent and active regions at the centers of galaxies. 

The Very Long Baseline Array has measured the positions of four compact radio sources in close proximity on the sky, using phase referencing over multiple frequency bands~\cite{fomalont2011position}. These high-resolution measurements had an instrumental accuracy of $20~\muas$. While there was apparent motion in some of the emission regions, the radio cores themselves could be identified, and appeared stable up to the $20~\muas$ instrumental precision (set by atmospheric propagation effects) over the one-year observation period. 

At this level of astrometric precision, the secular aberration drift, caused by the acceleration of the Solar System's acceleration towards the Galactic Center, should also be accounted for. This drift produces a dipole pattern in the apparent proper motion field of extragalactic objects. The effect was confirmed for the first time in proper motion data on 555~radio sources in ref.~\cite{titov2011vlbi}, yielding a maximum amplitude of $6.4\pm 1.5~\muasy$, consistent with theoretical expectations. This drift can be modeled~\cite{titov2013secular}, and is easily distinguishable from gravitational lensing effects.

\subsection{Peculiar accelerations}\label{sec:accelerationnoise}
As astrometry improves, there is a critical question as to whether there is a limit to the precision in acceleration observables. Given the recent growth in knowledge about exoplanets, an obvious source of intrinsic noise could come from Jovian planets. Accelerations from planets with orbital periods below the mission time should be easily excluded because the periodicity would be detectable, but those in wide orbits could induce an apparently linear acceleration on the host star.
This acceleration is easily calculated: for a star $i$ of mass $m_i$ with a lower-mass companion with mass $m_c$ in a circular orbit of period $T$, at line-of-sight distance $D_i$, the resulting acceleration is:
\begin{align}
\ddot{\theta}_{ic} = \frac{G_N^{1/3}}{D_i} \frac{m_c}{m_i^{2/3}} \left(\frac{2\pi}{T} \right)^{4/3} \approx 0.18~\muasyy \left[\frac{10~\kpc}{D_i}\right] \left[\frac{m_c}{10^{-3} M_\odot}\right]\left[\frac{M_\odot}{m_i}\right]^{2/3} \left[\frac{10~\text{y}}{T}\right]^{4/3}.
\end{align}
Such a small acceleration is below \textit{Gaia}'s threshold, but could potentially limit \textit{Theia}'s acceleration sensitivity for nearby, low-mass stars if super-Jupiter planets are common around them. This effect points to the tremendous capabilities of \textit{Theia} in the search for exoplanets, but for our purposes it acts as a source of uncorrelated, stochastic noise that should be added in quadrature to the instrumental noise.

There are physical, correlated accelerations to consider as well. Gravitational attraction from small-scale structures in the Milky Way can be eliminated by vetoing star pairs whose physical separation is small in both angle and line-of-sight distance. Still, there are large-scale correlated acceleration fields toward the Galactic Disk and Center. Accelerations toward the disk can be calculated as
\begin{align}
\alpha_{d}[D_i, R, z] \simeq \frac{G_N \Sigma_{d,0}}{2D_i} e^{-\frac{R}{R_d}} \left[1-e^{-\frac{|z|}{z_d}}\right] \approx 2 \times 10^{-6}\muasyy \left[\frac{10~\kpc}{D_i}\right]\left[\frac{e^{-\frac{R}{R_d}}}{e^{-\frac{R_\odot}{R_d}}}\right] \left[1-e^{-\frac{|z|}{z_d}}\right], 
\end{align}
and are too small to be observed. Accelerations toward the Galactic Center are at most one order of magnitude larger in absolute terms, but are suppressed by a small angle for disk stars, and are thus similarly far below instrumental threshold. Even if they were not, they could in principle be distinguished from a lensing signal, which would manifest itself at small angular separations, for example by changing the angular cutoffs $\beta_-$ and $\beta_+$ in eq.~\ref{eq:Calpha}.

%%%%%%%%%%%%%%%%%%%%%%%%%%%%%%%%%%%%%%%%%%%%%%%%%%%%%%%%%%%%%%%%%%%
\section{Sensitivity projections}\label{sec:sensitivity}
There is a wide variety of objects for which we could work out the sensitivity of our techniques. For brevity, we will consider two baseline scenarios: NFW subhalos and point masses. They are representative objects for lens targets that are ``soft'' and ``hard'': for soft lenses, the signal to noise ratio is driven by their aggregate effect on a large number of background sources (e.g.~NFW subhalos), while for hard lenses it is dominated by short-distance effects (e.g.~point-like lenses). Spherically symmetric lenses with inner density profiles $\rho(r) \propto r^{-\gamma}$ generally fall in the hard-lens category when $\gamma \ge 2$, and in the soft-lens category when $\gamma < 2$. Our soft-lens observables are optimized for NFW subhalos with $\gamma=1$, though if the true population of lenses were to deviate from this scaling behavior, our sensitivity estimates would not change dramatically. We leave a calculation of the projected sensitivity to lenses with generalized density profiles to future work. 

Likewise, there is a wide variety of luminous sources that we can target to apply these techniques. As we have already discussed, these sources have a broad range of properties, including the individual angular precisions, the density of the sources, intrinsic noise, and intrinsic correlations. For concreteness, we list here our benchmark parameters and from where they come. For our single-source observables, $\mathcal{B}^\text{mono}$, $\mathcal{O}_\mu$, and $\mathcal{O}_\alpha$, we will describe the near-future sensitivity of \textit{Gaia}, while our long-term estimates are for \textit{Theia}. For the multi-source observable $\mathcal{T}_\mu$, we base our near-term sensitivity on \textit{Gaia} observations of the LMC and SMC, while the best long-term sensitivity will come from quasar observations by SKA or comparable radio observatories. For the multi-source observable $\mathcal{C}_\mu$, we base our near-term sensitivity on \textit{Gaia} observations of quasars, while our long-term sensitivity is based on radio observations of quasars. We do not attempt to use the Magellanic Clouds for $\mathcal{C}_\mu$, as the presence of intrinsic correlations would yield spurious signals that are difficult to disentangle from lensing. Finally, for the multi-source observables $\mathcal{C}_\alpha$ and $\mathcal{B}^\text{multi}$, we use \textit{Gaia} observations of Milky Way stars for our near-term projections, and \textit{Theia} observations of stars for our longer-term projections. 

We will show explicit quantitative reaches for our observables in terms of \emph{local} signal to noise ratios, specifically $\text{SNR}=1$ for $\mathcal{T}_\mu$, $\mathcal{C}_\mu$, $\mathcal{C}_\alpha$, and $\mathcal{B}$, and $\text{SNR}=10$ for $\mathcal{O}_\mu$ and $\mathcal{O}_\alpha$. 
While these are transparent descriptors of sensitivity, one must take care of the look-elsewhere effect in a proper analysis of a limit or positive signal. For those, the \emph{global} SNR is the relevant quantity, whose threshold may be chosen to be 2(5) for e.g.~a $2\sigma$ limit (a $5\sigma$ signal). The global SNR is smaller than the local SNR by a trials factor, which for gaussian noise is roughly:
\begin{align}
\text{SNR}^\text{global} \simeq \frac{\text{SNR}^\text{local}}{\sqrt{1+\ln N_\text{trial}}}, \label{eq:globalSNR}
\end{align}
$N_\text{trial}$ is the number of independent trials that had to be done to find the largest local signal, which in some cases can be quite large.

For the outlier observables $\mathcal{O}_\alpha$ and $\mathcal{O}_\mu$, the number of trials is most simply the total number of sources. For the full \textit{Gaia} catalog of $10^9$ stars, the denominator of eq.~\ref{eq:globalSNR} is 4.7. However, as mentioned before, acceleration outliers likely have nongaussian tails, and velocity outliers have an incredulity cutoff (e.g.~a few times the escape velocity in the Galaxy), so a more careful analysis would be in order once the real data set becomes available. For mono-blips, $N_\text{trial}$ is the number of sources times the number of tentative lens paths tested for each star, a reasonable number for which is $4\pi (f_\text{rep} \tau) \sim 10^3$; the total $N_\text{trial} = 10^{12}$ leads to a trials suppression factor of 5.4 in eq.~\ref{eq:globalSNR}. Correlations need not have trials factors at all, as they are summed over the whole field of view, and their control parameters $\delta$, $\beta_-$, and $\beta_+$ have predetermined optimal values.

The trials factor calculation for multi-source observables is more involved. For the velocity template $\mathcal{T}_\mu$, it depends on the minimum and maximum angular scales considered for the template size $\beta_t$, and the number of template positions $\vect{\theta}_t$ considered per angular size $\beta_t$. The number of trial directions $\hat{\vect{v}}_t$ should be 2, e.g.~longitude and latitude, for each $\beta_t$ and $\vect{\theta}_t$. If the centroid $\vect{\theta}_t$ is off by approximately $e \beta_s$ from a real NFW lens, then the local SNR drops by about $1/e$ (the precise value depends on whether the template is off parallel or perpendicular to $\hat{\vect{v}}_t$). For a given angular scale $\beta_t$ and $\hat{\vect{v}}_t$, we need $\Delta \Omega/(e \beta_t)^2$ trials to cover an angular area $\Delta \Omega$. A trial template with the correct position and direction but incorrect size $\beta_t$ gives a template statistic $\sim 1/e$ below optimality when its size is off by $\sim 1/e$. If we wish to consider templates with angular scales between  $\beta_t^\text{min}$ and $\beta_t^\text{max}$, then we have a total number of trials $N_\text{trial} \sim \sum_{n=0}^{\ln \beta_t^\text{max}/\beta_t^\text{min}} 2\Delta \Omega/(e^{n+1} \beta_t)^2 \sim 2\Delta \Omega/ (\beta_t^\text{min})^2$. For $\beta_t^\text{min} = 1~\text{deg}$ and a full-sky $\Delta \Omega = 4\pi$ analysis, we have $N_\text{trial} \sim 10^5$, and a look-elsewhere suppression factor of 3.5 in the global SNR. 

For multi-blip lensing searches for Solar System planets, the number of trials is the number of planetary paths tested. To a good approximation, their apparent motion is determined by Earth's, so the number of paths is the number of tentative initial positions of the hypothetical planet. Without losing a significant fraction of the signal, one needs to test for $N_\text{trial} \sim \Sigma_0 \Delta \Omega$ initial locations, where $\Sigma_0$ is the angular number density of stars over the tested area $\Delta \Omega$, which can be less than $4\pi$ given some priors from auxiliary methods. For a fiducial $\Sigma_0 = 10^8$ and $\Delta \Omega = 0.1$, we have $N_\text{trial} \sim 10^7$ and a trial suppression factor of 4.1.  

\subsection{Subhalos} \label{sec:senshalos}
\begin{figure}[t]
\centering 
\includegraphics[width=.78\textwidth,trim=0 26 0 0,clip]{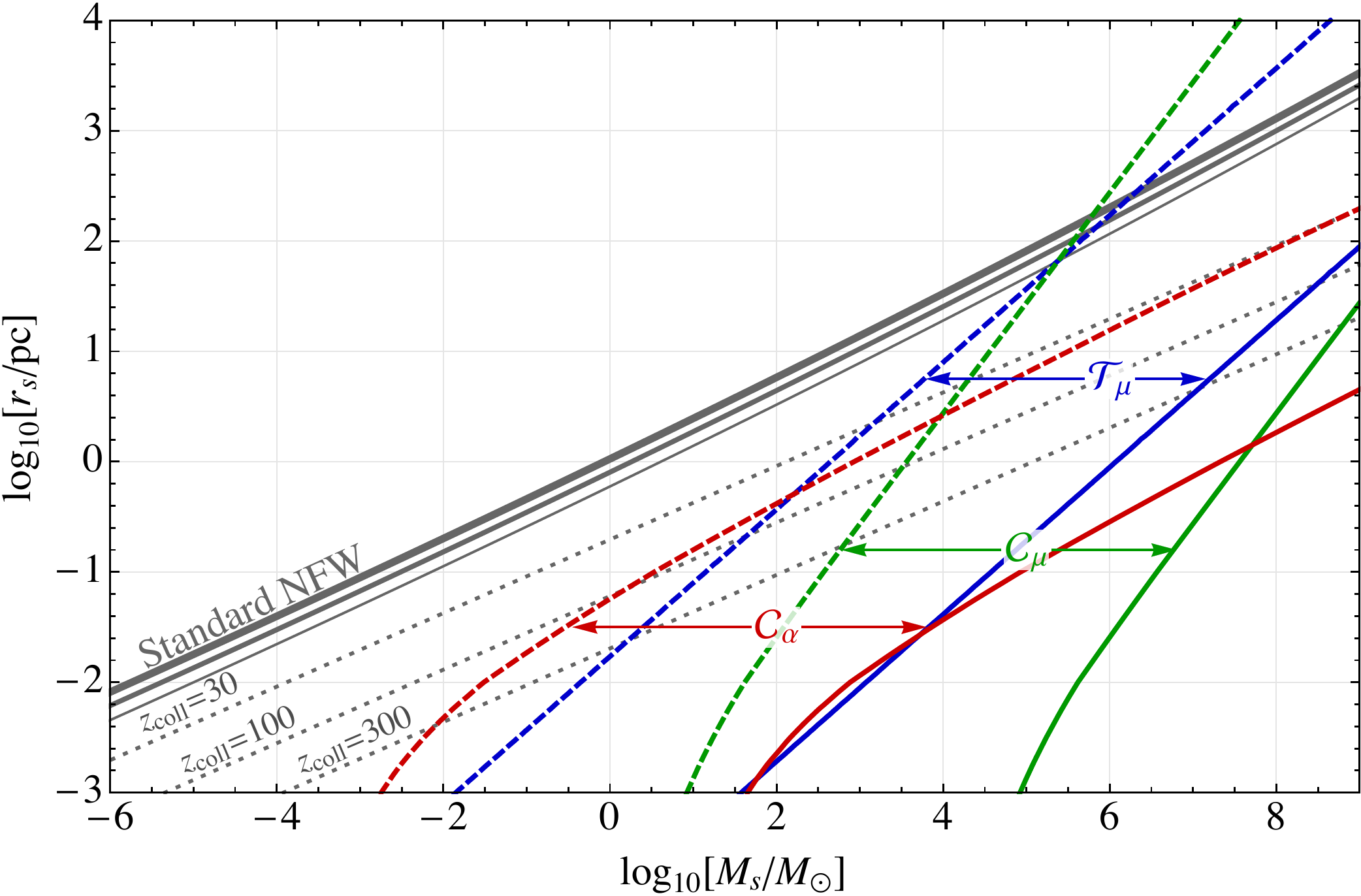}
\includegraphics[width=.78\textwidth,trim=0 0 0 0,clip]{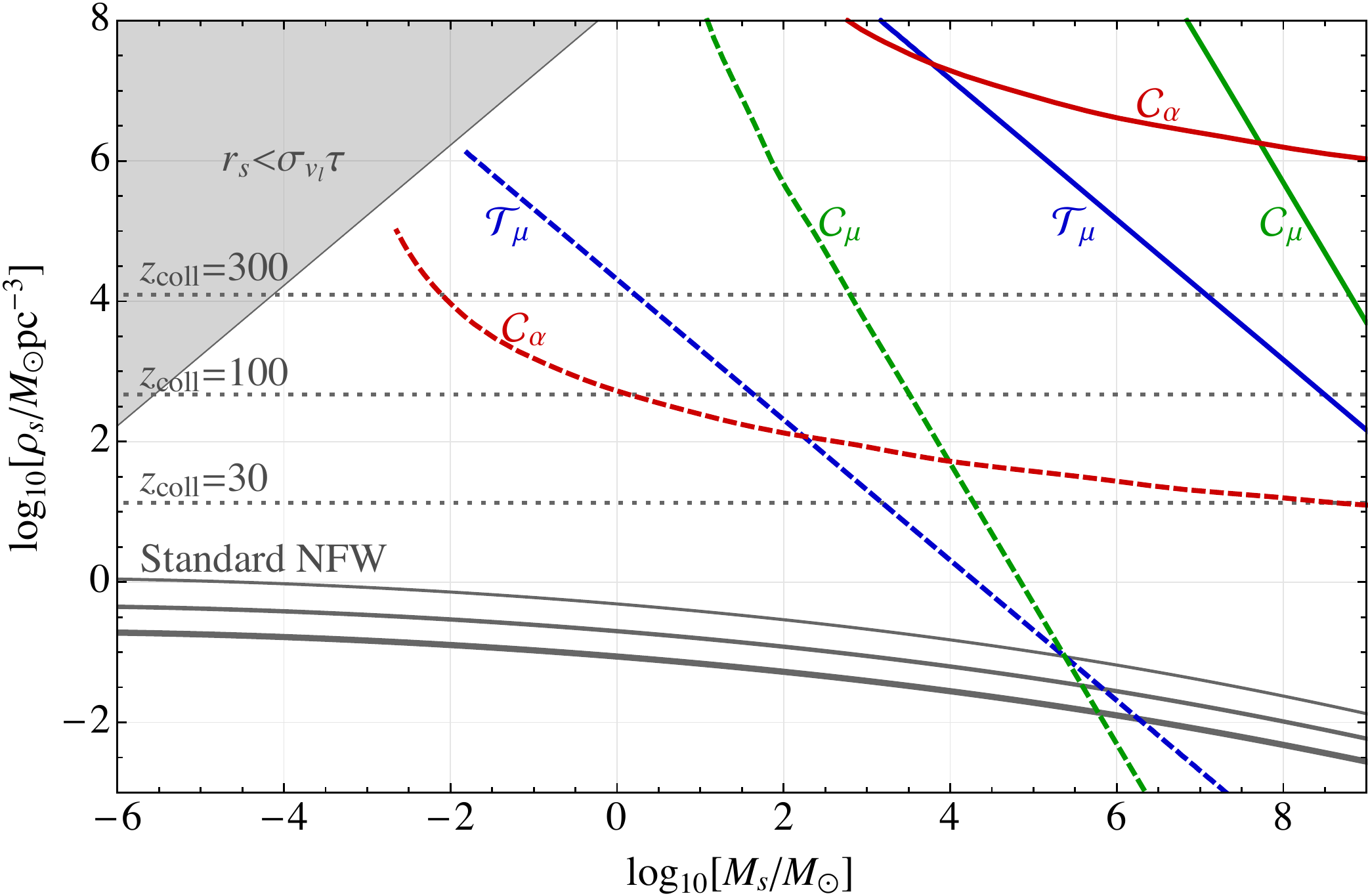}
\caption{Sensitivity projections for NFW subhalos as a function of core mass $M_s$. 
Vertical axes are the scale radius $r_s$ (\textbf{Top}) and density $\rho_s = M_s / [r_s^3 16\pi (\ln 2 -1/2)]$ (\textbf{Bottom}). At fixed mass, more compact objects (smaller $r_s$, larger $\rho_s$) are easier to see. 
The blue solid (dashed) curves show the local $\text{SNR}=1$ curve for the template velocity test statistic $\mathcal{T}_\mu$, assuming $\sigma_{\mu,\text{eff}} = 200(1)~\muasy$, $N_0 = 10^7 (10^8)$, and $\Delta \Omega = 0.01 (4\pi)$, representative of \textit{Gaia} observations toward the Magellanic Clouds (SKA radioastrometry of quasars). The green solid (dashed) curves show the global $\mathcal{C}_\mu$ velocity correlation test statistic sensitivity for $\sigma_{\mu,\text{eff}} = 10 (1)~\muasy$, $N_0 = 10^6(10^8)$, and $\Delta \Omega = 4\pi$, representative of near-future (far-future) astrometric observations of quasars in the radio and visible bands. The red solid (dashed) curve depicts the global $\text{SNR}=1$ sensitivity for acceleration correlations $\mathcal{C}_\alpha$, assuming $\sigma_{\alpha,\text{eff}} = 10 (0.1)~\muasyy$, $N_0 = 10^9(10^{10})$, and $\Delta \Omega = 0.2$ for \textit{Gaia} (\textit{Theia}) observations of Galactic disk stars. Also shown in solid gray is the ``standard'' NFW subhalo median relation between $M_s$ and $r_s$ from ref.~\cite{moline2017characterization} for three subhalo distances $R_\text{sub} = \lbrace 240,10,5\rbrace~\kpc$ away from the Galactic Center (closer ones are denser), as well as a rough estimates (in dotted gray) for the scale radius and density for nonstandard collapse redshifts $z_\text{coll}$. }
\label{fig:halosens}
\end{figure}

In this subsection, we will first work out the expected sensitivity to NFW subhalos assuming that their volume number density $n_l$ is uniform, that their mass distribution is a delta function centered at the core mass $M_s$, and that they all have the same scale radius $r_s$. We take the lens energy density $\rho_l$ to saturate the local dark matter energy density $\rho_{\text{DM},\odot} \approx 0.4~\GeV~\cm^{-3} \approx 0.0106~M_\odot~\pc^{-3}$, i.e.~$\Omega_\text{sub}(M_s') = \delta(M_s'-M_s)$. Concretely, we assume a volume number density of:
\begin{align}
n_l = \frac{\rho_l}{20 M_s}. \label{eq:nl}
\end{align}
The factor of 20 is to account for a typical ratio between the original virial mass \emph{before} tidal stripping and the core mass, cfr.~figure~\ref{fig:NFWs}. We assume a gaussian velocity distribution $f(\vect{v}) = (2 \pi \sigma_{v_l}^2 )^{-3/2} \exp(- \vect{v}^2 / 2\sigma_{v_l}^2)$ with $\sigma_{v_l} \approx 5.5\times 10^{-4} \approx 166 ~\kms$. We will comment on how these sensitivity curves can be convoluted with extended mass distributions.

\subsubsection{Velocity templates}
Following the discussion around eq.~\ref{eq:SNRTmu}, it is optimal to take the velocity template $\vect{\mu}_t$ equal to the expected velocity profile of the lens, which for an NFW subhalo is $\vect{G}_1$ from eq.~\ref{eq:G1}. When the template position, angular scale, and lensing velocity direction all match those of the lens, i.e.~$\vect{\theta}_t = \vect{\theta}_l$, $\beta_t = \beta_s$, and $\hat{\vect{v}}_{il} = \hat{\vect{v}}_t$, we can expect a template SNR of:
\begin{align}\label{eq:SNRTemp1}
\text{SNR}_{\mathcal{T}_\mu}\left[ D_l , v_{il}\right] = C_1 \left(1-\frac{D_l}{D_i}\right) \frac{4 G_N M_s v_{il}}{r_s D_l} \sqrt{\frac{\Sigma_0}{\sigma_{\mu,\text{eff}}^2}}
\end{align}
where we used eqs.~\ref{eq:SNRTmu},~\ref{eq:NFWmu},~and~\ref{eq:FOMTmu}. We also defined $C_1 = [\int \dd^2 (\beta_{il}/\beta_t)|\vect{G}_1|^2]^{1/2}$; we have $C_1 \approx 5.55$ but it is already $1.19,~2.83,~3.62,~4.32$ when the radial integral is cut off at $\beta_{il}/\beta_t = 0.1,~0.5,~1,~2$, respectively, so most of the information of the template is contained at scales $\beta_{il} \sim \beta_t$. For simplicity, let us ignore the Sun's velocity relative to the line of sight as well as the average velocity of the stellar field to the line of sight, such that $\langle v_{il} \rangle = \langle v_l \rangle = \sqrt{\pi/2} \sigma_{v_l}$. Most of the lens-to-lens variation in terms of signal to noise ratio comes from the variation in the line-of-sight distance $D_l$, the closest giving the largest angular size and thus the best potential SNR. For uniformly distributed halos, we expect the closest halo to be located at:
\begin{align}
\left\langle \min_l D_l \right \rangle = \left( \frac{3 }{n_l \Delta \Omega} \right)^{1/3} \approx 18~\kpc \left[\frac{M_s}{10^7 M_\odot} \frac{1}{\Omega_\text{sub}} \frac{0.01}{\Delta \Omega} \right]^{1/3}.
\end{align}
To a good approximation, we expect the largest (local) signal-to-noise ratio to be one where the lens has $D_l = \langle \min_l D_l  \rangle$ and $v_{il} = \sqrt{\pi/2} \sigma_{v_l}$ :
\begin{align}
\left \langle \max_l \text{SNR}_{\mathcal{T}_\mu} \right\rangle  & \simeq \frac{\pi^{1/2} C_1 }{2^{1/2} 3^{1/3}} \frac{4 G_N M_s \sigma_{v_l}}{r_s} \left( n_l \Delta \Omega \right)^{1/3} \sqrt{\frac{\Sigma_0}{\sigma_{\mu,\text{eff}}^2}} \label{eq:SNRmumax}\\
& \approx 0.4~\Omega_\text{sub}^{1/3} \left[\frac{M_s}{10^7 M_\odot}\right]^{2/3} \left[\frac{10~\pc}{r_s}\right] \left[ \frac{N_0}{10^7}\right]^{1/2} \left[\frac{0.01}{\Delta \Omega} \right]^{1/6} \left[ \frac{200~\muasy}{\sigma_{\mu,\text{eff}}}\right]. \nonumber
\end{align}
Above, we have furthermore assumed that $\sigma_{v_l} = 166~\kms$. We see that the velocity template is sensitive to the most massive halos, since approximate scale invariance implies constant density $\rho_s$ and $r_s \propto M_s^{1/3}$, such that $\langle  \max_l \text{SNR}_{\mu} \rangle \propto M_s^{1/3}$. Equation~\ref{eq:SNRmumax} is just the \emph{expected} largest, local SNR: it is possible one could get ``lucky'' with a faster-than-normal $v_{il}$ or closer-than-expected smallest $D_l$. So far, we have also been ignoring any scatter in the scale radius $r_s$ at fixed scale mass $M_s$: a further signal boost can be obtained if any of the nearby lenses is smaller than the median value of $r_s$.

In figure~\ref{fig:halosens}, we show $\text{SNR}=1$ contours for the template velocity observable in blue. The solid line represents our projection for what \textit{Gaia} can achieve at the end of its mission: we took $\sigma_{\mu,\text{eff}} = 200~\muasy$, $N_0 = 10^7$, and $\Delta \Omega = 0.01$, representative of its combined observations of the Magellanic Clouds. The blue dashed line is a more optimistic projection for $\sigma_{\mu,\text{eff}} = 1~\muasy$, $N_0 = 10^8$, and $\Delta \Omega = 4\pi$, indicative of \textit{Theia}'s and/or SKA's potential after a decade of quasar observations. It can be seen in figure~\ref{fig:halosens} that \textit{Gaia} will have sensitivity potential for a swath of dilute dark matter structures, though likely not those in standard scenarios. We think it is likely that future astrometric missions such as \textit{Theia} and radio interferometers such as SKA will be able to discover dark matter substructure that is truly ``dark'', i.e.~subhalos containing a negligible amount of stars and gas, perhaps with core masses as low as $10^6~M_\odot$. 

The discovery potential with velocity templates is not particularly sensitive to the substructure fraction or its mass function, due to weak scaling of the signal to noise ratio with abundance $\text{SNR}_{\mathcal{T}_\mu} \propto \Omega_\text{sub}^{1/3}$. Suppose the subhalo spectrum consists of exclusive\footnote{Not counting subsubhalos and subsubsubhalos, which can only boost the sensitivity.} structures with core mass function $\dd \Omega_\text{sub} / \dd (\log M_s)$. If we were to take this function to be constant over 12 decades (27 $e$-folds), then the sensitivity in the heaviest-mass $e$-fold bin is reduced only by a factor of $3 = 27^{1/3}$ compared to the case where all of the substructure were in that mass, while the smaller mass bins can contribute also, albeit less so.

If a tentative signal is seen, there are several handles that may be used to discriminate a real lensing signal from a spurious one, the most likely culprit being a correlated rotational velocity field in e.g.~the Magellanic Clouds. Obviously, one should first try to model and subtract this motion. The characteristic angular scale of this motion is a few degrees, much larger than the angular scales of lensing subhalos to which \textit{Gaia} would be sensitive. With \textit{Gaia}, the threshold sensitivity to a $\Omega_\text{sub}=1$ population of $M_s = 10^7~M_\odot$ is about $r_s = 4~\text{pc}$ (see figure~\ref{fig:halosens}), whose closest subhalo would have an angular scale of about $0.01~\text{deg}$, several hundred times smaller than the SMC or LMC. Let us assume that after rotational velocity subtraction procedure, there is still a tentative maximum $\mathcal{T}_\mu$ at a template location $\vect{\theta}_t^*$, size $\beta_t^*$, and direction $\hat{\vect{v}}_t^*$. A subset of tests that a real signal should pass, and that are difficult to pass simultaneously by a correlated stellar velocity field, are: (1) the SNR in every quadrant around $\vect{\theta}_t^*$ should be equal; (2) the SNR should decrease as $\beta_t/\beta_t^*$ for $\beta_t \ll \beta_t^*$ and as $\beta_t^*/\beta_t$ for $\beta_t \gg \beta_t^*$; (3) the SNR should scale as $\hat{\vect{v}}_t \cdot \hat{\vect{v}}_t^*$. The feasibility of these tests merits further study, and should become possible after \textit{Gaia}'s upcoming data release. 

\subsubsection{Velocity correlations}
The velocity template is aimed at local measurements of a single lens on a stellar velocity field. However, the lens with e.g.~the second- and third-largest effects (relative to the noise) are expected to have comparable velocity shifts and angular sizes, and there is a potential to measure the global effect of this lens population with velocity correlations. Stellar populations in the Milky Way are expected to exhibit intrinsic velocity correlations, and they will be difficult to disentangle from lens-induced ones. However, extragalactic objects such as quasars should have independent proper motions, and are thus a prime target for the observable $\mathcal{C}_\mu$.

\begin{figure}[t]
\centering 
\includegraphics[height=.32\textwidth,trim=0 0 0 0,clip]{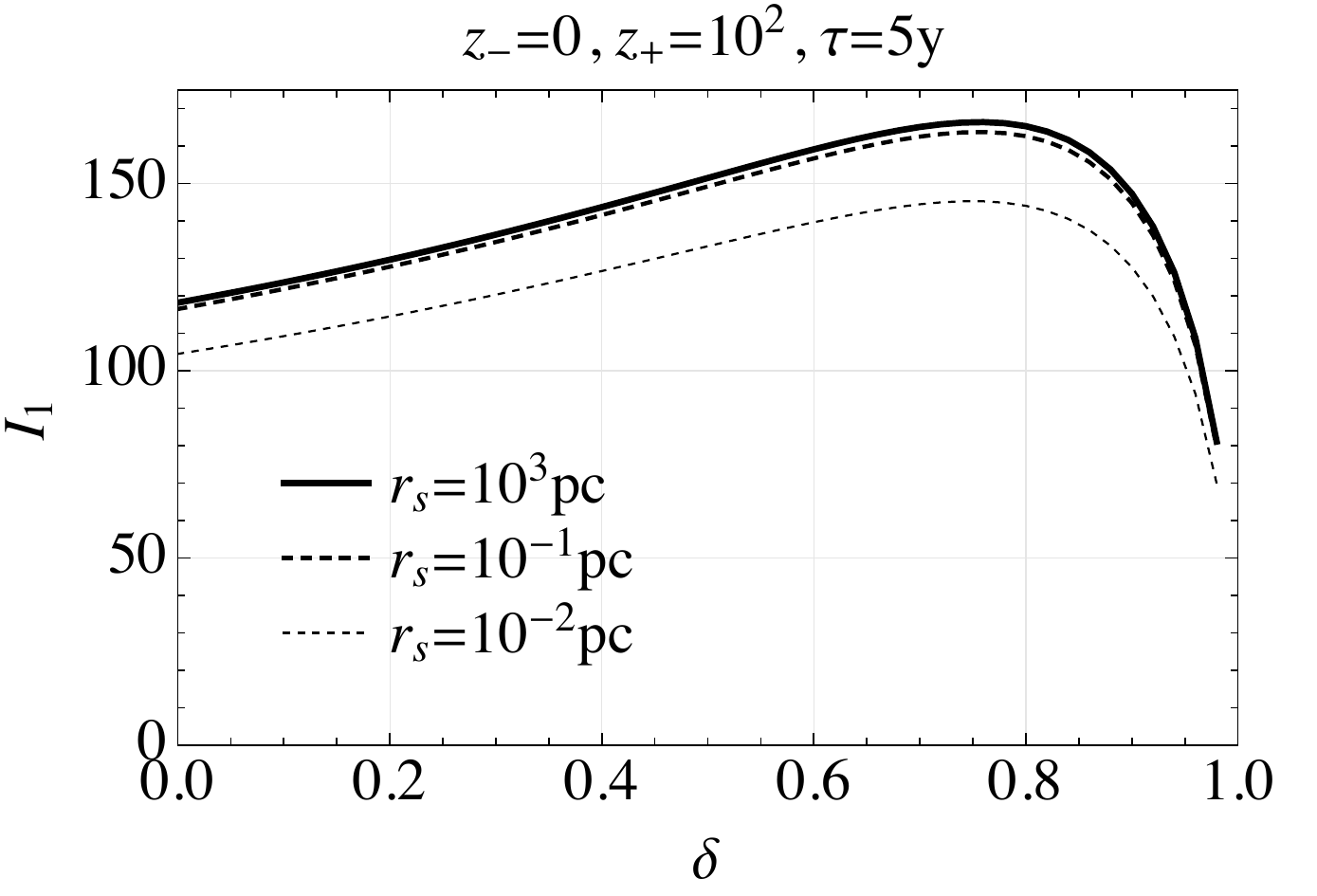}\hspace{1.2em}
\includegraphics[height=.32\textwidth,trim=0 0 0 0,clip]{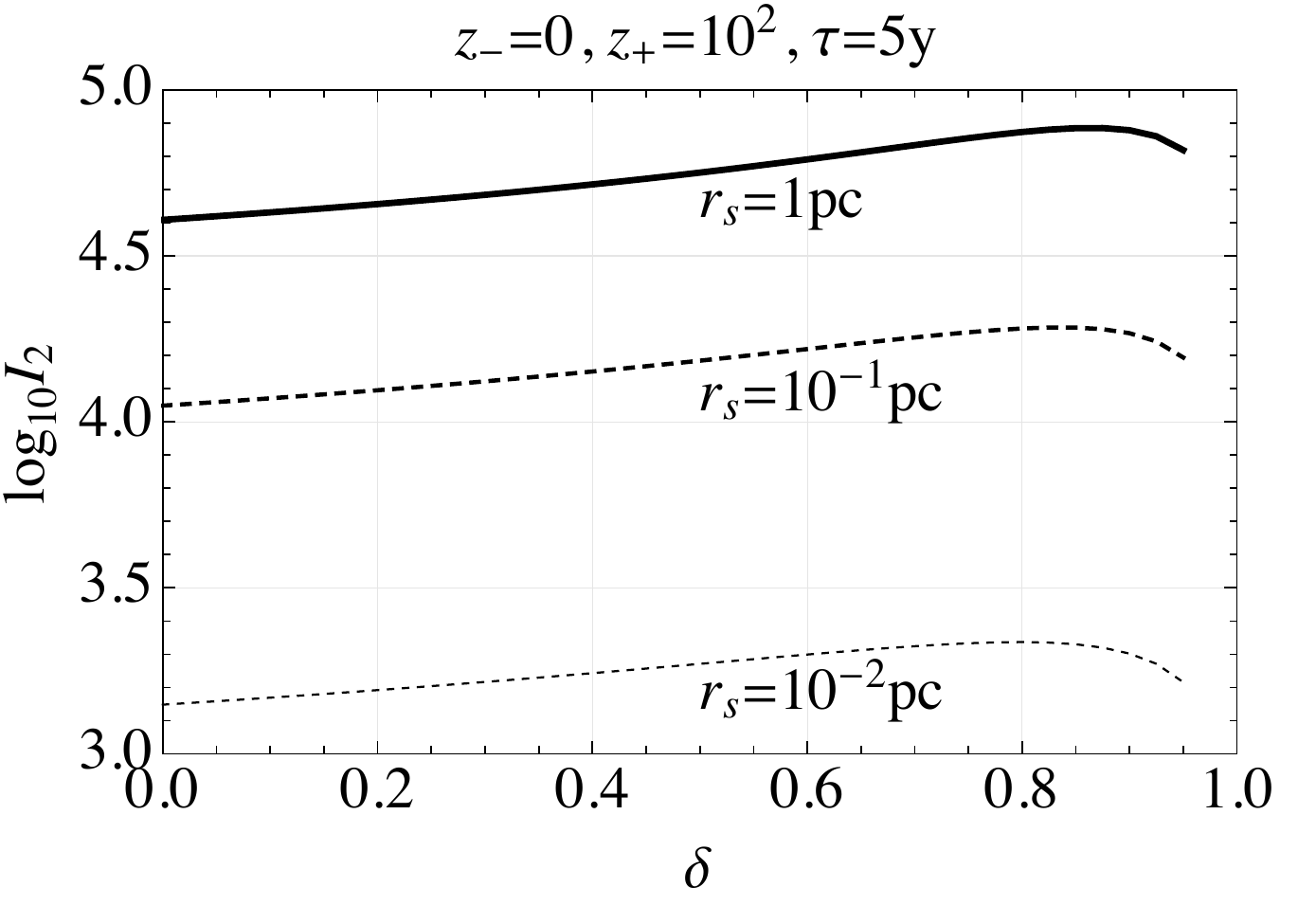}\\
\includegraphics[height=.32\textwidth,trim=0 0 0 0,clip]{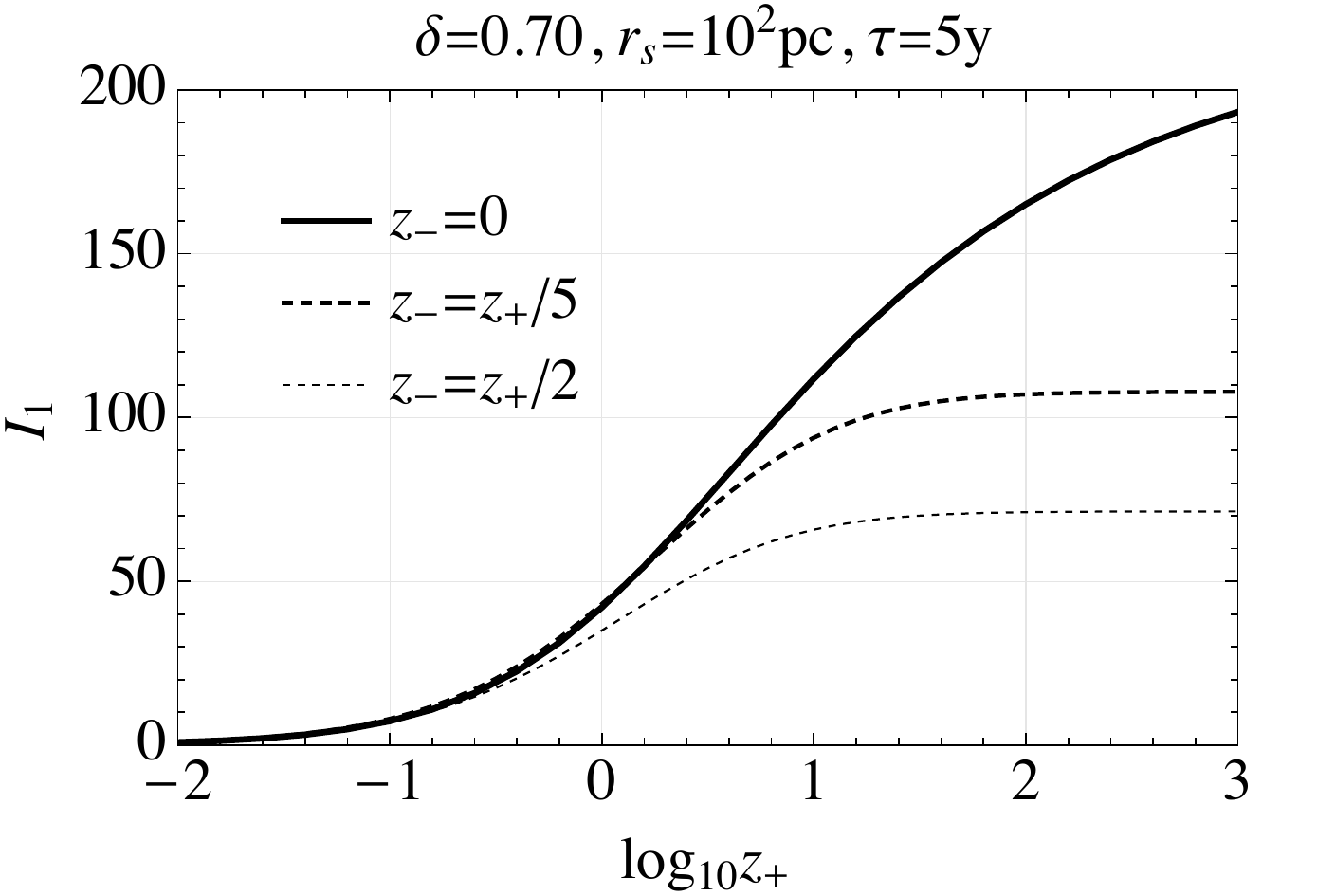}
\includegraphics[height=.32\textwidth,trim=0 0 0 0,clip]{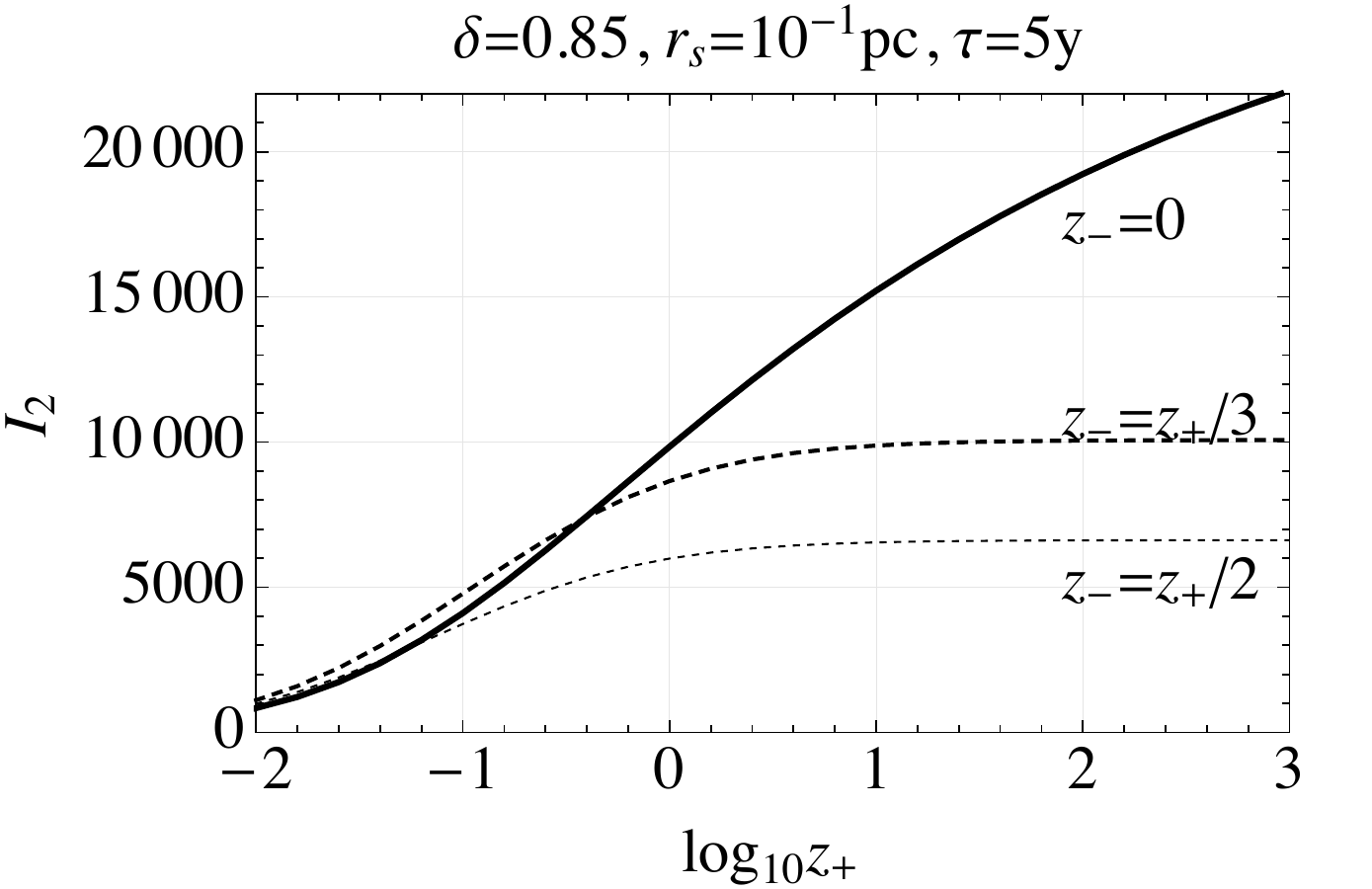}\\
\includegraphics[height=.32\textwidth,trim=0 0 0 0,clip]{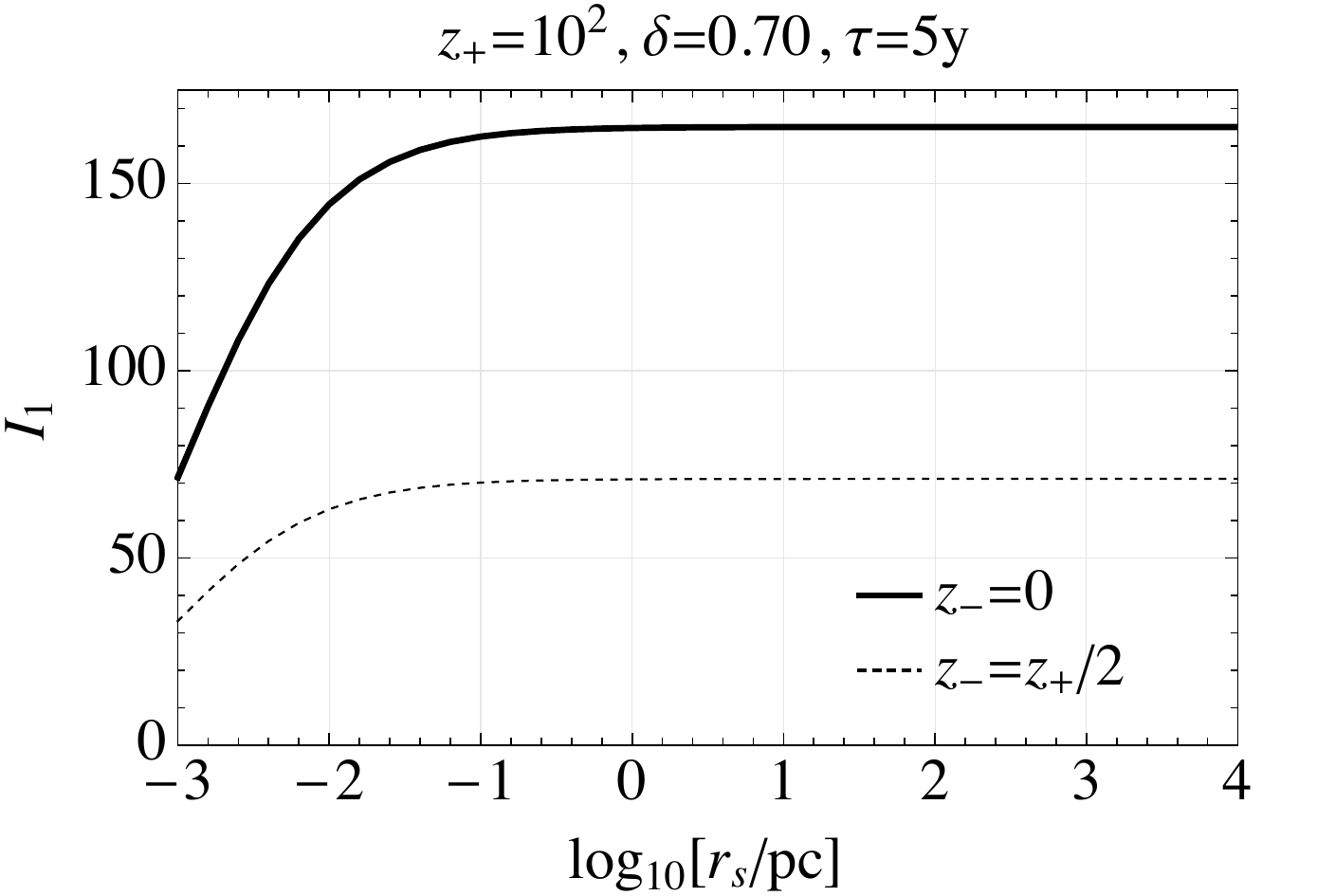}\hspace{1.5em}
\includegraphics[height=.32\textwidth,trim=0 0 0 0,clip]{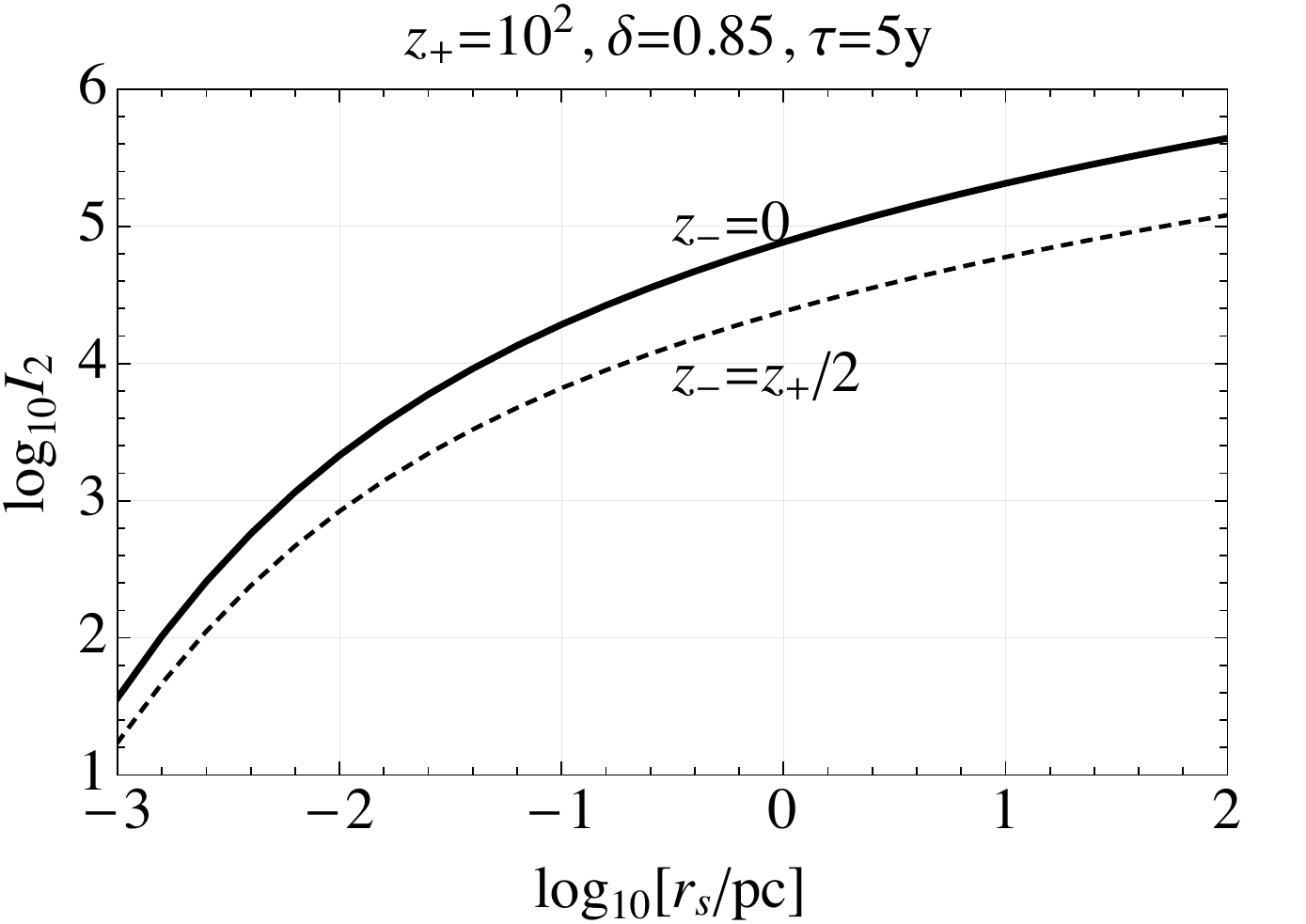}
\caption{Plots of the integral function $I_1[z_-,z_+,\delta_,\epsilon_s]$ relevant for velocity correlations as a function of $\delta$ (\textbf{Top left}), $z_+$ (\textbf{Middle left}), and $r_s$ (\textbf{Bottom left}). \textbf{Right:} Analogous plots for the integral function $I_2$ relevant for correlated accelerations. We assumed $\epsilon_s = \sigma_{v_l} \tau / r_s$ with $\sigma_{v_l} = 166~\text{km}/\text{s}$ and $\tau = 5~\text{y}$.}
\label{fig:Ifun}
\end{figure}

To estimate the power of the correlated velocity test statistic $\mathcal{C}_\mu$ of eq.~\ref{eq:Cmu}, we first need to evaluate the expected square velocity correction from all lenses with angular impact parameter larger than $\beta_{ij}$:
\begin{align}
&\left \langle \sum_l \left| \Delta\dot{\vect{\theta}}_{il}  \right|^2 \Theta(\beta_{il} - \beta_{ij} ) \right\rangle \label{eq:mu2expNFW}\\
& {\hspace{3em}} =  \int_0^{D_i} \dd D_l \, n_l D_l^2 \left( 
1-\frac{D_l}{D_i} \right)^2  \left(\frac{4 G_N M_s v_{il}}{r_s^2}\right)^2 \beta_s^2 \int_{x_{il}>\frac{\beta_{ij}}{\beta_s} + \epsilon_s } \dd^2 x_{il} \,  \left| \vect{G}_1[x_{il},\hat{\vect{x}}_{il},\hat{\vect{v}}_{il} ]\right|^2. \nonumber
\end{align}
%We have summed up the contributions from all lenses with angular impact parameters larger than $\beta_{ij}$.
We imposed a short-distance cutoff of $\sigma_{v_l} \tau$ on the minimum impact parameter due to the motion of the lenses on the sky over the integration time $\tau$, represented by the parameter $\epsilon_s$:
\begin{align}
\epsilon_s \equiv \frac{\sigma_{v_l} \tau}{r_s}.
\end{align}
Since we will be considering extragalactic sources, we can safely drop the factor of $(1-D_l/D_i)^2$ in eq.~\ref{eq:mu2expNFW}. For simplicity, we will take $n_l$ to be a constant as discussed above eq.~\ref{eq:nl} for $D_l < D_\text{max}$, and to vanish for $D_l > D_\text{max}$. A reasonable value for the maximum line-of-sight distance up to which lenses could appear at sufficient number density would be to take $D_\text{max} = 50~\kpc$; our results will only depend logarithmically on this choice.
Using the identity $r_s = \beta_s D_l$, and using the results of section~\ref{sec:correlations} to compute $\langle \mathcal{C}_\mu \rangle$ and $\text{Var}\, \mathcal{C}_\mu$, we find a signal to noise ratio of:
\begin{align}
\text{SNR}_{\mathcal{C}_\mu} = \sqrt{\frac{\pi}{2}} \sqrt{\Sigma_0^2 \Delta \Omega} \left( n_l r_s^3\right) \left(\frac{4G_N M_s v_{il}}{r_s^2 \sigma_{\mu,\text{eff}}} \right)^2 I_1\left[ \frac{\beta_-}{r_s/D_\text{max}}, \frac{\beta_+}{r_s/D_\text{max}}, \delta, \epsilon_s \right] , \label{eq:SNRCmuNFW}
\end{align}
where we have defined the function $I_1$:
\begin{align}
I_1\left[z_-, z_+, \delta,\epsilon_s \right] \equiv \sqrt{\frac{2-2\delta}{z_+^{2-2\delta} - z_-^{2-2\delta}}} \int_{z_-}^{z_+}\dd z \, z^{1-\delta} \int_0^1\dd y \, \int_{x> zy+\epsilon_s} \dd^2 x \, \left| \vect{G}_1[x,\hat{\vect{x}},\hat{\vect{v}} ]\right|^2. \label{eq:I1}
\end{align}
We plot the function $I_1$ in the left panels of figure~\ref{fig:Ifun} as a function of $z_- \equiv \beta_-/(r_s/D_\text{max})$, $z_+\equiv \beta_+/(r_s/D_\text{max})$, the angular weighting exponent $\delta$, and the scale radius $r_s$, assuming a fixed $\sigma_{v_l} = 166~\text{km}/\text{s}$ and $\tau =  5~\text{y}$. It is optimal to take $z_- = 0$, $\delta \approx 0.7$, and $z_+$ as large as possible, although it is only logarithmically increasing for $z_+ \gtrsim 10$. The control variables $z_-$, $z_+$, and $\delta$ can serve as handles for rejecting systematic backgrounds, as well discriminating between different gravitational lens profiles.

Parametrically, the signal to noise ratio can be read as $\sqrt{\text{\# of source pairs}}$ $\times$ [fractional volume of space in subhalos] $\times$ [typical velocity divided by typical noise$]^2$ $\times$ [enhancement function from close encounters $=I_1]$. As a conservative estimate, we take $v_{il}^2$ equal to its expectation value assuming only the lenses move $\langle v_{il}^2 \rangle = \langle v_{l}^2 \rangle = 2 \sigma_{v_l}^2$. 
Our sensitivity projections for $\mathcal{C_\mu}$ at unit signal to noise are depicted in green in figure~\ref{fig:halosens}. We picked $z_- = 0$, $z_+ = 10^2$,  $\delta = 0.70$, $\epsilon_s = \sigma_{v_l} (5~\text{y}) / r_s$, and $\Delta \Omega = 4\pi$ throughout. The solid line assumes $N_0 = 10^6$ and $\sigma_{\mu,\text{eff}} = 10~\muasy$, while the dashed line is for the more futuristic parameters $N_0 = 10^8$ and $\sigma_{\mu,\text{eff}} = 1~\muasy$. These parameters are plausible current and future goals for optical space-based observatories (such as \textit{Gaia} and \textit{Theia}) and radio interferometers. We think these future astrometric missions have discovery potential for dark matter structures with cores lighter than a million solar masses, using the observables $\mathcal{C}_\mu$ and $\mathcal{T}_\mu$, as we discussed in the previous subsection. The minimum detectable core mass with $\mathcal{C}_\mu$ scales as $\sigma_{\mu,\text{eff}}^2/\Sigma_0$, where as it scales more slowly, as $(\sigma_{\mu,\text{eff}}^2/\Sigma_0)^{1/2}$, for $\mathcal{T}_\mu$.
The correlated velocity test statistic depends more strongly on the substructure fraction than the velocity template technique, and thus its reach will be diminished for extended mass functions with significant support at small subhalo masses, where it is less sensitive. 

\subsubsection{Acceleration correlations}

Gravitational lensing by NFW subhalos can produce spatial correlations in the stellar acceleration field, as we discussed around eq.~\ref{eq:Calphaexp}. If we plug in the expected combined acceleration of eq.~\ref{eq:NFWalpha}, we find: 
\begin{align}
&\left \langle \sum_l \left| \Delta\ddot{\vect{\theta}}_{il}  \right|^2 \Theta(\beta_{il} - \beta_{ij} ) \right\rangle \label{eq:alpha2expNFW}\\
& {\hspace{3em}} = \int_0^{D_i} \dd D_l \,  n_l  D_l^2 \left( 
1-\frac{D_l}{D_i} \right)^2 \left(\frac{4 G_N M_s v_{il}^2}{r_s^3}\right)^2 \beta_s^2 \int_{x_{il}>\frac{\beta_{ij}}{\beta_s} + \epsilon_s } \dd^2 x_{il} \,  \left| \vect{G}_2[x_{il},\hat{\vect{x}}_{il},\hat{\vect{v}}_{il} ]\right|^2. \nonumber
\end{align}
We can combine eqs.~\ref{eq:alpha2expNFW}, \ref{eq:Calphaexp}, and \ref{eq:VarCalpha} to compute the signal to noise ratio:
\begin{align}
\text{SNR}_{\mathcal{C}_\alpha} = \sqrt{\frac{\pi}{2}} \sqrt{\Sigma_0^2 \Delta \Omega} \left( n_l r_s^3\right) \left(\frac{4G_N M_s v_{il}^2}{r_s^3 \sigma_{\alpha,\text{eff}}} \right)^2 I_2\left[ \frac{\beta_-}{r_s/D_i},\frac{\beta_+}{r_s/D_i}, \delta , \epsilon_s\right]. \label{eq:SNRCalphaNFW}
\end{align}
The definition of the function $I_2$ is the same as that of $I_1$ in eq.~\ref{eq:I1} except for the profile replacement $\vect{G}_1 \leftrightarrow \vect{G}_2$, and the evaluation of the line-of-sight integral, which is now cut off by the distance $D_i$ to the nearest of the star pairs, instead of by the distance $D_\text{max}$ where the DM density drops significantly:
\begin{align}
I_2\left[z_-, z_+, \delta,\epsilon_s \right] \equiv \sqrt{\frac{2-2\delta}{z_+^{2-2\delta} - z_-^{2-2\delta}}} \int_{z_-}^{z_+}\dd z \, z^{1-\delta} \int_0^1\dd y \, (1-y)^2 \int_{x> zy+\epsilon_s} \dd^2 x \, \left| \vect{G}_2[x,\hat{\vect{x}},\hat{\vect{v}} ]\right|^2. \label{eq:I2}
\end{align}
We plot the function $I_2$ in the right panels of figure~\ref{fig:Ifun}. It is optimal to take $\delta \approx 0.85$. Again $z_- = 0$ is optimal and $z_+$ is to be as large as possible, although for $z_+ \gtrsim 10$ growth is only logarithmic (or slower). Just like for $\mathcal{C}_\mu$, the variables $z_-$, $z_+$, and $\delta$ can serve as discrimination handles. Spurious correlated accelerations can be rejected by not including stars with similar distances $D_i \simeq D_j$, and instead only summing over ``fake double stars'', which appear nearby (in angle) but have a widely different line-of-sight distances.

Heuristically, the signal to noise ratio is again proportional to the square root of the number of source pairs, the fractional volume of space occupied by subhalos, and the typical acceleration correction (at the scale radius) divided by the typical acceleration noise. The enhancement function $I_2$ tends to be much larger than $I_1$ for the NFW profile, however.
We conservatively take $v_{il}^4$ to be equal to its expectation value assuming only the lenses move $\langle v_{il}^4 \rangle = \langle v_{l}^4 \rangle = 8 \sigma_{v_l}^4$. 
Our resulting sensitivity projections for $\mathcal{C}_\alpha$, at signal to noise ratio equal to unity, are depicted in red in figure~\ref{fig:halosens}. We assume $z_- = 0$, $z_+ = 10^2$, $\delta = 0.85$, and $\epsilon_s = \sigma_{v_l} (5~\text{y}) / r_s$ throughout. The solid curve is for parameters of $N_0 = 10^9$, $\sigma_{\alpha,\text{eff}} = 5~\muasyy$, $\Delta \Omega = 0.2$, representative of what \textit{Gaia} can achieve if the mission is moderately extended from its nominal 5~y mission time. The dashed red curve bumps these numbers up to the more optimistic $N_0 = 10^{10}$ and $\sigma_{\alpha,\text{eff}} = 0.1~\muasyy$, indicative of \textit{Theia}'s potential for observations of disk stars. 

The observable $\mathcal{C}_\alpha$ has a roughly scale-invariant signal to noise ratio for constant-density subhalos: it is nearly equally sensitive to the smallest subhalos as it is to the largest---over 10 orders of magnitude in mass---if the subhalos obey the scaling $M_s \propto r_s^3$. Extended, scale-invariant mass distributions should therefore not reduce its discovery power. Finally, the next layer of substructure, namely subsubhalos, may also contribute to the signal, an effect we have not included yet in our above analysis.

\subsection{Compact objects}\label{sec:senscompact}
Dark matter clumps smaller than $10^{-3}~\pc \sim 200~\text{AU} \sim 3~\times 10^{10}~\text{km}$ are effectively point-like lenses for our purposes, as their typical motion over a multi-year mission is larger than their size, cfr.~eq.~\ref{eq:pointcond}. The appropriate signal observables to use are thus those for point-like lenses: outliers and blips.
We will perform a sensitivity analysis for objects of vanishing size; the extension to finite-size objects is left to future work. We will assume that all compact objects are uniformly distributed with energy density $\rho_l$ near the Solar system's location in the Galaxy. We will take their mass function to be a delta function, i.e.~all lenses have the same mass $M_l$, so that the natural parameter space is the $M_l$--$\rho_l$ plane. Generalizations to extended mass distributions are left to future work. 

\begin{figure}[t]
\centering 
\includegraphics[width=.98\textwidth,origin=0,angle=0]{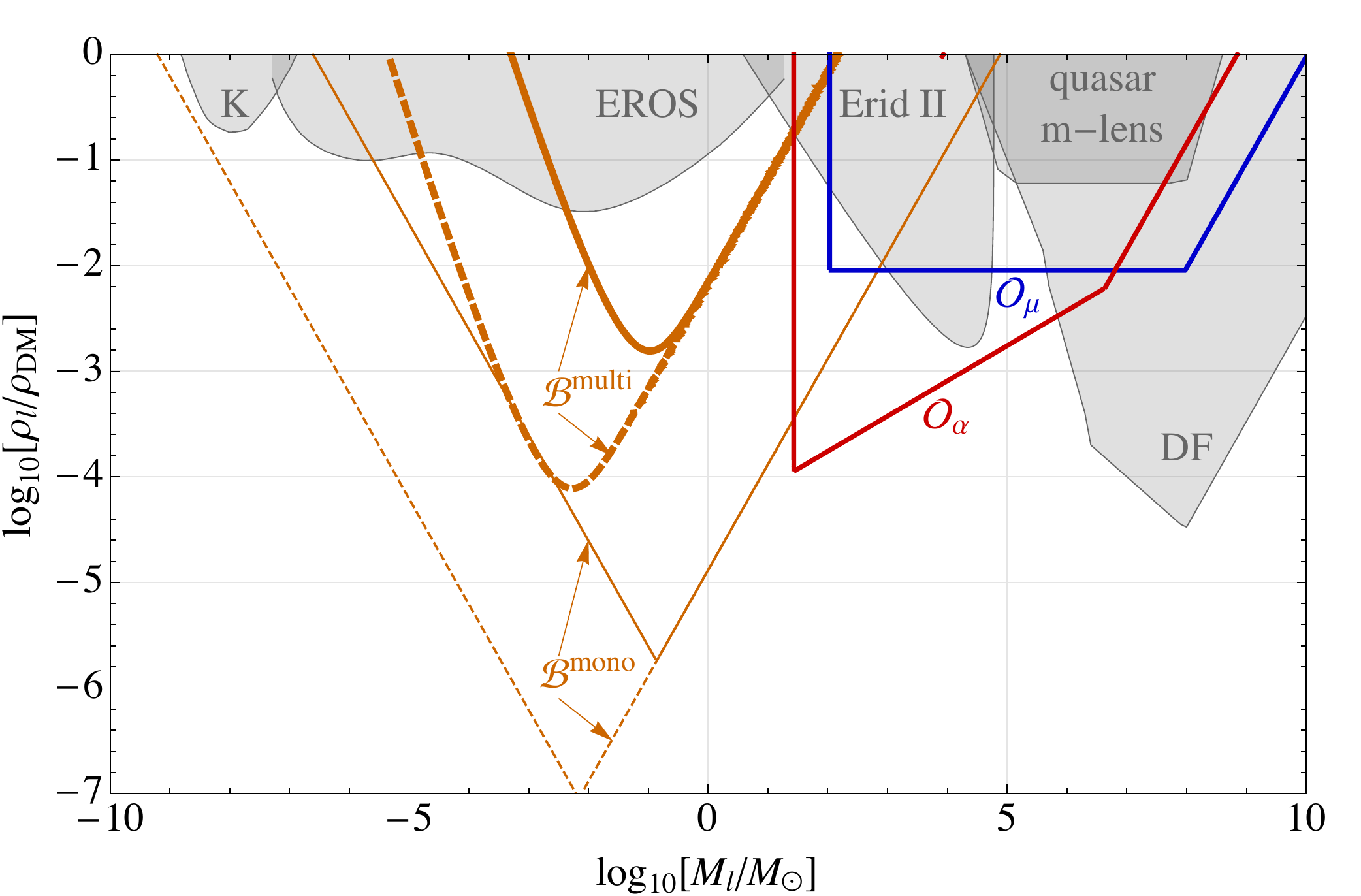}
\caption{Projected sensitivity of point-like lensing searches for compact objects of mass $M_l$ and dark-matter fraction $\rho_l/\rho_{DM}$. In thin and thick solid orange, we show unit signal to noise lines using mono- and multi-blip searches; blue dashed and red dotted lines depict fiducial sensitivity from outlier velocity and acceleration events. The projections indicated by solid lines are representative of the end-of-mission capabilities of \textit{Gaia}, with parameters as specified in section~\ref{sec:senscompact}. The dashed lines are futuristic projections for blip searches with a survey such as \textit{Theia}. Also shown by the gray regions, from left to right, are constraints from Kepler and EROS microlensing, the survival of the Eridanus II cluster, quasar millilensing, and dynamical friction.} \label{fig:machosens} 
\end{figure}

\paragraph{Outliers} Close approaches of point-like lenses can cause outlier velocities or accelerations, measured by the observables $\mathcal{O}_\mu$ and $\mathcal{O}_\alpha$ in eqs.~\ref{eq:outliervelocity} and \ref{eq:outliervelocity}, respectively. They will generally provide better sensitivity than blip observables when the smallest source-lens impact parameter is still larger than the motion of the lens on the sky, which is expected to be the case when:
\begin{align}
\left \langle \min_{i,l} b_{il} \right\rangle = \sqrt{\frac{M_l}{\pi \rho_l D_i \Sigma_0 \Delta \Omega}} \gtrsim \sigma_{v_l} \tau. \label{eq:outlierregime}
\end{align}
In figure~\ref{fig:machosens}, we plot the sensitivity of an outlier velocity search in solid blue for $\Delta \Omega = 0.01$, $N_0 = 10^{7}$, $D_i = 50~\kpc$, $\sigma_{\mu,\text{eff}} = 200~\muasy$, and $\tau = 5~\text{y}$, which is representative of \textit{Gaia} parameters confined to the patch of sky containing the Magellanic Clouds. In solid red, we show the sensitivity of an outlier acceleration search, for which we took $\Delta \Omega = 0.2$, $N_0 = 10^9$, $D_i = 10~\kpc$, $\sigma_{\alpha,\text{eff}} = 5~\muasyy$, and $\tau = 5~\text{y}$, representative of the set of disk stars in the \textit{Gaia} data. Lines represent curves where $\left \langle \mathcal{O}_\mu \right\rangle = 10^2$ and $\left\langle \mathcal{O}_\alpha \right\rangle = 10^2$, a large threshold chosen to partially overcome expected systematics from hyper-velocity stars and wide, massive binary systems, which we discussed in sections~\ref{sec:velocitynoise} and \ref{sec:accelerationnoise}, respectively. In the regime of eq.~\ref{eq:outlierregime}, the lowest detectable $\rho_l$ is mass-independent for $\mathcal{O}_\mu$ and scales as $M_l^{-1/3}$ for $\mathcal{O}_\alpha$. At very high masses, the sensitivity degrades because one simply does not expect to find a lens within a cone of solid angle $\Delta \Omega$ and distance $D_i$. 

\paragraph{Blips} At low masses where the inequality~\ref{eq:outlierregime} is violated, blip searches are more sensitive. The largest mono-blip SNR will come from the lens with a path $\vect{x}_l[t]$ whose smallest expected impact parameter $b_{il}^*$ with a source $i$ is:
\begin{align}
\left \langle \min_{i,l} b^*_{il} \right\rangle = \frac{M_l}{\sigma_{v_l} \tau \rho_l D_i \Sigma_0 \Delta \Omega} \lesssim \sigma_{v_l} \tau. \label{eq:minbstar}
\end{align}
Ignoring the size $r_s$ of the object, setting $v_{il} = \sigma_{v_l}$, and marginalizing over the discrete observation times $t_n$ in eq.~\ref{eq:blipSNR}, then gives the mono-blip SNR:
\begin{align}
\text{SNR}_{\mathcal{B}}^\text{mono} \simeq \frac{4G_N M_l}{\sigma_{\delta \theta,\text{eff}}} \sqrt{\frac{\pi f_\text{rep} }{\sigma_{v_l} } \frac{1}{\left \langle \min_{i,l} b^*_{il} \right\rangle }}
\end{align}
where it is understood that the minimum $b_{il}^*$ is that of eq.~\ref{eq:minbstar} above. At low masses, the smallest detectable dark matter abundance $\rho_l$ scales as $M_l^{-1}$. We plot the $\text{SNR}=1$ line in this low-mass regime as the thin orange curve in figure~\ref{fig:machosens} for  $\Delta \Omega = 0.2$, $N_0 = 10^9$, $D_i = 10~\kpc$, $\sigma_{\delta\theta,\text{eff}} = 100~\muas$, $\tau = 5~\text{y}$, and $f_\text{rep} = 14/\text{y}$, representative of the disk stars sample in \textit{Gaia}. At high masses, the sensitivity to small $\rho_l$ degrades because one does not expect to find a mono-blip (however small); in this high-mass regime, we plot the line where the inequality~\ref{eq:minbstar} is saturated. The sensitivity on the low-mass end may improve with a survey like \textit{Theia}: in figure~\ref{fig:machosens}, we show in dashed the mono-blip sensitivity also for $N_0  = 10^{10}$ and $\sigma_{\delta\theta,\text{eff}} = 5~\muas$.

Multi-blip signals will occur if the number density of lenses is so high that the smallest line-of-sight distance among all lenses
\begin{align}
\left\langle \min_{l} D_l \right\rangle = \left(\frac{3M_l}{\rho_l \Delta \Omega}\right)^{1/3} \label{eq:minDl}
\end{align}
is so small that the typical motion is larger than the typical angular separation between stars:
\begin{align}
\frac{\sigma_{v_l} \tau}{\left\langle \min_{l} D_l \right\rangle} \gtrsim \frac{1}{\sqrt{\Sigma_0}}.
\end{align}
The largest multi-blip SNR is expected to come from the closest lens:
\begin{align}
\text{SNR}_{\mathcal{B}}^\text{multi} \simeq \frac{4G_N M_l}{\sigma_{\delta \theta,\text{eff}}} \sqrt{\frac{\pi f_\text{rep}}{\sigma_{v_l} } \frac{4\Sigma_0 \sigma_{v_l}\tau}{\left\langle \min_{l} D_l \right\rangle} \ln \left[\frac{\Sigma_0^{1/2} \sigma_{v_l} \tau}{\left\langle \min_{l} D_l \right\rangle} \right]   }
\end{align}
with $\left\langle \min_{l} D_l \right\rangle$ from eq.~\ref{eq:minDl}. In figure~\ref{fig:machosens}, we show the $ \text{SNR}_{\mathcal{B}}^\text{multi} = 1$ curves in thick orange for the same parameters as the mono-blip lines. While not as sensitive as mono-blips to ultra-low abundances, multi-blips offer a potentially spectacular signature, where correlated anomalous motion could be seen at different times in nearly collinear background sources.
We note that blip searches are especially sensitive to objects in a several-decade range around $1~M_\odot$, where mergers or close orbits of these objects may be visible also in current and future gravitational wave observatories.

In figure~\ref{fig:machosens}, we depict in gray the existing exclusion regions in the $M_l$--$\rho_l$ plane of compact objects. Photometric microlensing observations by Kepler~\cite{griest2014experimental} and EROS~\cite{tisserand2007limits} constrain compact objects below $10~M_{\odot}$, while millilensing of quasars~\cite{wilkinson2001limits} excludes high-mass ones. The kinematic constraints coming from the nondisruption of the Eridanus II star cluster~\cite{brandt2016constraints}, and from dynamical friction due to the ``granularity'' of dark matter~\cite{carr1999dynamical}, are also shown. If the compact objects are primordial black holes, there is an additional constraint at high masses because their accretion-induced radiation would alter the ionization history of the Universe, and affect CMB anisotropies~\cite{ali2017cosmic}.

\subsection{Planet Nine} \label{sec:planetnine}

Multi-blip astrometric lensing signatures are significantly enhanced for more nearby objects, which naturally have smaller impact parameters $b_{il}^*$---increasing the signal---and trace out larger arcs on the sky, thus moving in front of a larger number of background sources---averaging down the noise. For compact dark matter objects in the MW halo, we have seen that the best sensitivity comes from mono-blip rather than multi-blip events, primarily because it is statistically unlikely to find a nearby lens if the density is low.
However, if dark objects are gravitationally bound to the outer Solar System, they may have their motion ``gravitationally imaged'' by a long string of multi-blip events, with a typical angular deflection scale of:
\begin{align}
\Delta \theta_{il} = \frac{4 G M_l}{b^*_{il}} \approx 0.024~\muas\left[\frac{M_l}{M_\oplus}\right]\left[\frac{\text{AU}}{b_{il}} \right].
\end{align}
We shall argue that the correlated deflection of many background sources and occasional close approaches $b^*_{il}\ll \text{AU}$ allows the multi-blip method to tease out this signal from below the noise. This technique would open up a discovery potential for Solar System planets beyond the Kuiper Belt, where they would appear so faint as to have escaped detection so far.

The apparent motion of objects in the outer Solar System is dominated by the motion of Earth, with the impact parameter changing by $2~\text{AU}$ every half-year (for an average impact parameter speed of $\langle v_{il} \rangle = 4~\text{AU}/\text{y}$) roughly over the same patch of the sky, behind which we can expect to find a number of ``blipped'' background sources $N^{\mathcal{B}}_{0}$ equal to
\begin{align}
N_{0}^{\mathcal{B}} \simeq \frac{ 4\Sigma_0 \text{AU}^2}{R_l^2} \approx 400~\left[\frac{\Sigma_0}{10^8}\right]\left[\frac{1000~\text{AU}}{R_l}\right]^2,
\end{align}
where $R_l$ is the distance of the planet to the Sun at the present time. For simplicity, we assumed the apparent motion is linear, i.e.~the orbit of the planet lies in the ecliptic plane, though this is by no means crucial. In our analytic estimates, we can safely ignore the motion of the lensing planet itself ($R_l \gg \text{AU}$) as well as those of the background stars ($R_l / D_i \simeq 0$). We also neglect small fractional changes in the line-of-sight distance as the Earth goes around the Sun, and, in the case of a satellite mission, the observer being located at e.g.~Lagrange point L2 instead of Earth. These details will not affect the gross sensitivity estimates, but should be taken into account in the actual analysis. With all of the above assumptions, we find that
\begin{align}
\left\langle \sum_{i \in \Box} \frac{1}{b^*_{il}} \right\rangle \simeq \frac{\tau}{\text{y}/2} \frac{4 \Sigma_0 \text{AU}}{R_l^2} \ln\left[ \frac{4\Sigma_0 \text{AU}^2}{R_l^2} \right],
\end{align}
essentially saying that after 6 months, the effective impact parameter $\langle \sum_{i\in \Box} 1/b^*_{il}\rangle^{-1}$ is equal to $\text{AU} /(N_{0}^\mathcal{B} \ln N_{0}^\mathcal{B})$, and linearly decreases for longer observation times. Plugging this into eq.~\ref{eq:blipSNR}, we find a signal to noise ratio of:
\begin{align}
\text{SNR}_{\mathcal{B}_l}^\text{multi} \simeq \frac{4 G_N M_l}{R_l} \frac{\sqrt{f_\text{rep} \tau }}{\sigma_{\delta \theta,\text{eff}}} \sqrt{2\pi \Sigma_0 \ln\left[ \frac{4\Sigma_0 \text{AU}^2}{R_l^2} \right]}.
\end{align}
Parametrically, we can interpret this number as [the planet's Schwarzschild radius divided by the orbital radius] $\times$~[the end-of-mission positional precision$]^{-1}$ $\times$~[the angular number density of sources$]^{1/2}$, up to a logarithmic factor. The gravitational lensing signal thus scales as $M_l / R_l$, in contrast to the signal from reflected sunlight, which scales as $M_l^{2/3} / R_l^4$ at fixed density and albedo. The infrared power from the planet's own blackbody emission is likely considerably higher (but much less certain) than the reflected power, but also scales down fast, as $1/R_l^2$. This distance scaling of the lensing effect is thus much more favorable to find massive objects in the outer reaches of our Solar System.

\begin{figure}[t]
\centering 
\includegraphics[width=.9\textwidth,origin=0,angle=0]{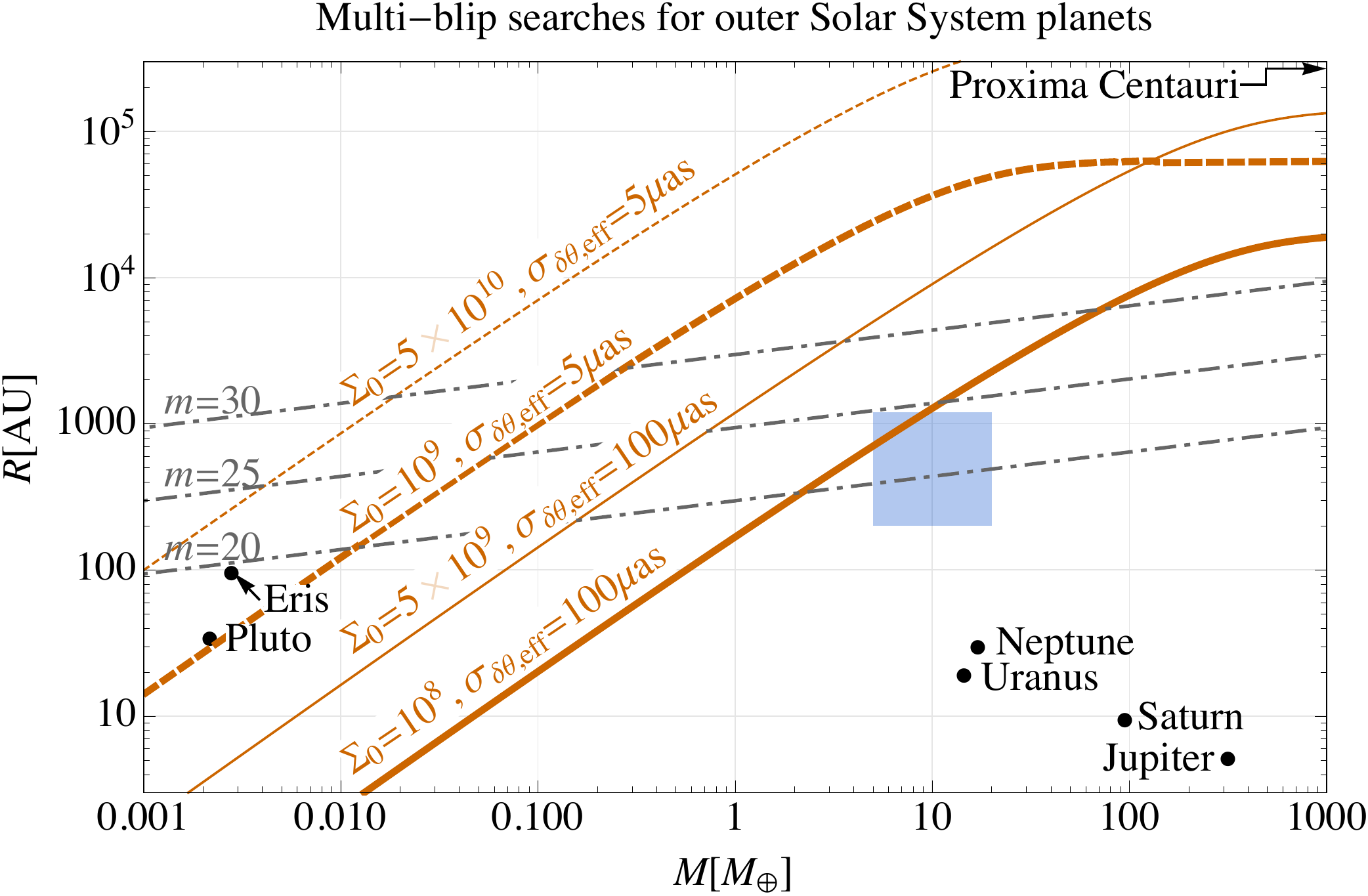}
\caption{Sensitivity to multi-blip astrometric events caused by planets in the outer Solar System as a function of their mass $M$ in Earth-mass units $M_\oplus$ and distance $R$ in astronomical units (AU). In solid thick orange, we plot the $\text{SNR}=1$ curve assuming $N_0 = 1.3 \times 10^9$ uniformly distributed point sources over the whole sky, angular position precision of $\sigma_{\delta \theta,\text{eff}} = 100~\muas$, $f_\text{rep} = 14/\text{y}$, and $\tau = 5~\text{y}$, representative of full-sky \textit{Gaia} sensitivity at the end of its mission. In thin solid orange, we plot the same curve but for an angular source number density approximating that of the MW disk, $\Sigma_0 = 5 \times 10^9$ . Finally, another factor of about 60 improvement in observable radius can be gained for a precision of $5~\muas$ and angular number densities that are 10 times higher, as shown by the thick and thin dashed curves. Gray dot-dashed lines show contours of apparent magnitude in the visible band from reflected sunlight, assuming the planet has the same average density and albedo as Neptune. The blue shaded region at 5--20~$M_\oplus$ and 200--1,200~AU roughly indicates the preferred parameters of a planet that could explain the statistical clustering and survival of several Kuiper-Belt objects. Also shown are the four known outermost planets, and the two heaviest known dwarf planets (black dots).} \label{fig:planetnine} 
\end{figure}

In figure~\ref{fig:planetnine}, we plot the sensitivity projections for \textit{Gaia}-like parameters in thick solid orange for when the planet is in front of a field of background sources with average angular number density, and in thin solid orange for when the planet is in front of a dense star field such as the Galactic disk, a situation which may not be that unlikely for the hypothesized Planet Nine of refs.~\cite{trujilloplanet9,batygin2016evidence,becker2017evaluating} since it is weakly preferred that its aphelion points in that direction. In dashed orange, we show the sensitivity at unit signal to noise for a mission such as \textit{Theia} with a positional precision improvement by a factor of 20, and an angular number density of $10^{9}$ and $5 \times 10^{10}$, respectively. We also show contours of equal apparent magnitude of reflected sunlight, illustrating the different mass and distance scaling, assuming the planet has Neptune's mass density and albedo. We note that our newly proposed method for finding planets is complementary to direct imaging methods, in the sense that traditional imaging is easier against relatively void patches of sky and difficult against dense star fields due to the large associated backgrounds, while exactly the opposite is true for our method, which is more powerful the denser the field of background stars. Finally, it would be interesting to calibrate the multi-blip method on known planets; astrometric lensing by Jupiter and Saturn has been considered previously in the context of high-precision tests of general relativity~\cite{crosta2006microarcsecond, kopeikin2007gravitational}.

\section{Conclusions}\label{sec:conclusions}

Precision astrometry has the potential to usher in the field of halometry via weak gravitational lensing. The new generation of astrometric surveys bring with them two dramatic changes. The first is the unprecedented precision and accuracy to which the locations and proper motions of luminous objects can be measured. The second is the sheer number of cataloged objects. 
In this context, we have considered weak gravitational lensing signatures from dark compact objects and extended structures inside the Milky Way. We proposed a set of techniques to measure these minuscule effects, and have estimated the potential discovery reach to nonluminous objects---dark matter and planets---with the ongoing \textit{Gaia} mission, and with the planned surveys of \textit{Theia} and the Square Kilometer Array. Our methods are applicable to any set of observations with sufficient precision and number of observed luminous sources. 

We find, as have previous authors, that lensing signals of typical objects are too small to be observed, despite the impressive precision of modern observatories. The impact parameter of the light path from the visible source to the gravitational lens is typically too large---in the case of compact objects---or the lens is too diffuse---in the case of standard NFW subhalos. However, with a large number of sources, one can look for rare or aggregate signals, opening the possibility for halometry of completely dark structures at small length scales.

The idea of using weak lensing to detect objects in time-domain astrometry has a significant history. Previous work has focused on what we call ``mono-blips'', rare events where the impact parameter of a source and lens changes by $\mathcal{O}(1)$ over the course of a mission~\cite{dominik2000astrometric,belokurov2002astrometric,erickcek2011astrometric}. We have revisited this observable, and shown that current and planned astrometric surveys can use it to detect an exceedingly rare population of compact objects, well below existing limits. We have also constructed a new ``multi-blip'' observable, which generalizes this idea to the case where a lens transits past many luminous sources. While offering a weaker discovery potential compared to mono-blips for Galactic dark matter, it preferentially picks out \emph{nearby} compact objects and provides a robust and spectacular signature. The proximity of the lenses that could be discovered in this way means that follow-up astronomical observations are in a better position to investigate their nature.

The multi-blip observable may have a profound impact on mapping out the contents of the outer Solar System. Objects located beyond the Kuiper Belt are hard to spot directly due to their small apparent brightness and large proper motion. On the other hand, their large parallax will cause objects located even as far as in the Oort Cloud to traverse past several background stars. The combined lensing effect of these multiple transits can provide new detection prospects for (dwarf) planets more massive than the Moon. Using our proposed technique, future surveys such as \textit{Theia} will have the capability to detect \textit{any} object heavier than a few Earth masses all the way to the cosmographic boundary of the Solar System.

We have also moved beyond these dramatic lensing events to consider observables for situations in which even the smallest impact parameter changes only by a small fractional amount over the course of an astrometric survey. In such cases, it is economical to package the results in terms of time derivatives of the lensing angle, $\Delta \dot \theta$ and $\Delta \ddot \theta$. Massive point-like lenses may be so rare that close angular approaches with a star are not expected to occur. Nonetheless, their relative motion may produce ``outlier'' velocities and accelerations far beyond reasonable expectations. Outlier observables are more prone to systematics, but may be used to set stringent limits on very massive lenses.

Diffuse objects such as subhalos---even relatively dense ones---produce much smaller apparent angular velocities and accelerations of light sources in their background. However, these effects will present themselves on many sources in a correlated way, either because any one lens may have a large angular size and eclipse many background sources, or because extended structures together may cover a large fraction of the sky. We have developed a number of approaches to leverage the large number of lensed sources to bring these signals above the noise. ``Templates'' are localized observables that compare an expected lens-induced velocity (acceleration) field to the observed velocities (accelerations) of the background light sources. ``Correlation'' observables measure the degree to which velocities or accelerations of nearby (in angle) light sources are correlated over a large field of view.

\textit{Gaia}'s observations of the Magellanic Clouds will already probe new parameter space of subhalos, using velocity templates. Although these stellar targets are distant, their density of stars is high, and their large distances suppress the intrinsic proper motion of their constituent stars. Conventional NFW subhalos are beyond \textit{Gaia}'s reach, but $10^6$--$10^8~M_\odot$ subhalos that collapsed at high redshift could be observed if sufficiently copious. These templates are inherently local measurements, allowing for follow-up observations and a build-up in the lensing signal as new astrometric surveys get off the ground. Quasars will likely prove to be the ultimate targets for velocity templates, given their stable locations and uncorrelated intrinsic motions.
Future surveys such as \textit{Theia} and SKA may achieve $\muasy$ precision on millions of quasar proper motions, enabling them to detect conventional NFW subhalos as light as $10^6~M_\odot$. Putative cosmic ray signals from subhalos, such as those claimed by ref.~\cite{bertoni2016gamma} would be directly testable. Known dwarf galaxies will provide attractive first targets. More massive halos, such as those of the Large Magellanic Cloud, and even extragalactic ones such as Andromeda and Virgo, may be probed at high precision.

We have introduced two correlation observables that measure the degree to which angular velocities and accelerations of nearby sources are aligned with one another. 
Lens-induced velocity correlations on stellar sources are difficult to detect because they exhibit intrinsic correlations that can be challenging to subtract. Quasars have no such irreducible systematics, and their lens-induced velocity correlations provide a comparable threshold sensitivity to velocity templates for NFW subhalos, while the velocity correlation signal for lens populations above threshold scales up much faster than with the template approach.
Correlated accelerations have a much lower irreducible background, and can be studied with Milky Way stars. \textit{Theia}'s sensitivity is likely just short of the standard NFW subhalo spectrum, but will provide---over a huge dynamic range---a discovery reach to denser subhalos, which may be produced from an increase in the power of primordial curvature fluctuations at small scales. %, the sensitivity is short of NFW halos, but can probe dense small halos at scales where even moderate scale dependence of primordial power, integrated over a large dynamic range, could provide the necessary fluctuations to be observable.

Our work is the first foray into time-domain astrometric lensing of multiple light sources, and opens up several avenues for future inquiry. It would be interesting to work out projections for generalized radial density profiles, for nonspherically-symmetric objects such as dark matter disks, filaments, and strings, and for generalized mass distributions. All of the estimates in this work are approximate analytic expressions, so it would be worthwhile to check their validity on mock simulations or early data releases of \textit{Gaia}. Finally, a careful study of possible systematic effects, discrimination methods, and contamination from lensing by baryonic matter is in order.

Now that detection of completely dark objects in the Milky Way will become more likely, it is paramount to study their expected properties. Dedicated simulations should attempt to predict the subhalo mass functions and concentration distributions for general primordial power spectra, as well as for the standard, scale-invariant power spectrum. Particularly pressing issues are the survival probability of microhalos in the presence of baryons (and other tidal disruption effects), and the fractal nature of dark matter substructure (what is the expected spectrum of subsubhalos in a given subhalo?). These are difficult questions, but we hope that our work gives a renewed impetus to answer them.

Finally, we repeat our call on astrometric surveys to release their full time-series data as soon as possible, to make many of the analyses presented here a possibility. There is an outside chance that evidence for a new planet in our Solar System is already lurking in the current \textit{Gaia} data set. At the very least, we encourage the collaborations to fit for blips and average acceleration, and to make this derivative information publicly available. Looking to the future, we hope that planned astrometric surveys deliberate their potential to do halometry when deciding on their observational strategies.

The techniques and estimated sensitivities laid out in this work present the start of a set of opportunities to learn about the small-scale structure of our Universe and the outer regions of our very own Solar System. A full implementation of our methods in a realistic data set will no doubt involve a great deal of careful thought and unexpected issues. Nonetheless, it is clear that there is a tremendous amount of previously untapped information in astrometric catalogs, and that the observables introduced here are a powerful means of accessing it. The unfolding era of precision astrometry thus promises to be an exciting one for halometry.

\acknowledgments
We thank Yacine Ali-Haimoud, Asimina Arvanitaki, Masha Baryakhtar, Vasily Belokurov, Liang Dai, Adrienne Erickcek, David Gerdes, David Hogg, Junwu Huang, Derek Inman, Alexander Kaurov, Matthew Kleban, Robert Lasenby, Mariangela Lisanti, Aaron Pierce, Ben Safdi, Oren Slone, Jeremy Tinker, Tejaswi Venumadhav, and Matthias Zaldarriaga for helpful discussions. KVT recognizes funding by the AMIAS. NW and KVT are supported by the National Science Foundation under grant PHY-1620727. The work of NW is supported by the Simons Foundation. This work has made use of data from the European Space Agency (ESA) mission {\it Gaia} (\url{https://www.cosmos.esa.int/gaia}), processed by the {\it Gaia} Data Processing and Analysis Consortium (DPAC,
\url{https://www.cosmos.esa.int/web/gaia/dpac/consortium}). Funding for the DPAC has been provided by national institutions, in particular the institutions participating in the {\it Gaia} Multilateral Agreement.

%\paragraph{Note added.} This is also a good position for notes added after the paper has been written.

\bibliographystyle{JHEP}
\bibliography{halometry}
\end{document}